\documentclass[aps,prb,twocolumn,superscriptaddress,10pt]{revtex4-2}
\pdfoutput=1 
\usepackage[utf8]{inputenc}
\usepackage[english]{babel}
\usepackage[T1]{fontenc}
\usepackage[pdftex]{graphicx}
\usepackage{amsmath,amsthm,amssymb,mathtools,braket}
\usepackage{xcolor}
\usepackage{dcolumn}
\usepackage{bbm}
\usepackage{bbold}
\usepackage[normalem]{ulem}
\usepackage[colorlinks=true,linkcolor=blue,citecolor=blue,urlcolor=blue,unicode]{hyperref}
\usepackage{footmisc}
\usepackage{comment}

\begin{document}
\title{State preparation and detection for quantum simulation of particle collisions}
\author{Federica Maria Surace}
\affiliation{Universität Innsbruck, Institut für Theoretische Physik, Technikerstraße 21a, 6020 Innsbruck, Austria}
\affiliation{School of Physics, Trinity College Dublin, Dublin 2, Ireland}
\affiliation{Department of Physics and Institute for Quantum Information and Matter,
California Institute of Technology, Pasadena, California 91125, USA}
\author{Sary Bseiso}
\affiliation{Department of Applied Physics, University of Michigan, Ann Arbor, Michigan 48109, USA}
\author{John Preskill}
\affiliation{Department of Physics and Institute for Quantum Information and Matter,
California Institute of Technology, Pasadena, California 91125, USA}

\begin{abstract}
Simulating the real-time dynamics of particle collisions is a promising application of quantum simulators, because classical methods such as tensor networks struggle to capture the highly entangled states generated in high-energy scattering. Realizing such simulations requires both the preparation of incoming wave packets and the detection of the outgoing scattering products. In this work, we propose protocols that address both challenges on programmable analog and digital quantum simulation platforms.
Our state-preparation scheme uses a weakly coupled auxiliary qubit --- or, more generally, a customized local quench --- to inject a single quasiparticle with well-defined momentum. 
Because it relies only on conservation of energy, this scheme requires no fine-tuning or prior knowledge about particle eigenstates, making it robust against errors in calibration and implementation. The momenta of scattering products are then extracted, using only local measurements, from the interference pattern that arises when particles are reflected at the system's boundary. We validate our protocols through numerical simulations, first in a simple single-particle model and subsequently in two interacting many-body systems: a Rydberg atom chain and an Ising chain in a mixed field. We demonstrate how high-energy regimes, necessary to access inelastic scattering processes, can be reached through an adiabatic ramp, and how the wave packet shape can be optimized by spatially modulating the Hamiltonian. Finally, we show how the protocol can be generalized to systems with more than one spatial dimension.
Our proposal provides a versatile approach to the quantum simulation of scattering phenomena, and is compatible with several quantum simulation platforms that are already experimentally available.

\end{abstract}

\maketitle

\section{Introduction}

Quantum simulators have emerged as a powerful approach for simulating the dynamics of quantum many-body systems, competing directly with state-of-the-art classical computational techniques \cite{Daley_2022,eisert2025mind}. A promising application is the simulation of scattering processes, akin to those occurring in particle colliders \cite{jordan2011quantum,Jordan2012,preskill2018simulating,Bauer2023,Wang2024,DiMeglio2024,bauer2025efficient,Burbano2026,Hardy2026,Barata2026}. In this setting, the theory of interest is modeled as a lattice Hamiltonian, and the particles are the low-energy excitations (quasiparticles) carrying well-defined momentum. Quantum simulators can reproduce the collision of two such particles in real time, probing not only the scattering matrix (which encodes the probabilities of the final asymptotic states), but also the state of the system during the collision or shortly after. This capability opens the door to studying phenomena such as the emergence of hydrodynamics, thermalization, and aspects of heavy-ion collisions in controlled quantum many-body systems \cite{BergesQCD,Busza2018}.

Quantum simulators may become particularly advantageous in the high-energy regime, where the energy is much larger than the mass gap and inelastic channels with many particles become accessible. Since the entanglement entropy generated in a scattering event is expected to increase with the number of particles produced, such processes are challenging to simulate using tensor-network methods \cite{milsted2022collisions}. While tensor networks have proven highly successful at low energies \cite{Vanderstraeten2014,VanDamme2021,Rigobello2021,Karpov2022,milsted2022collisions,Belyansky2024,Jha2025,Papaefstathiou2025,pavevsic2026scattering}, their applicability is expected to deteriorate as the collision energy and the number of particles increase. Furthermore, although tensor-network methods have achieved remarkable success in one dimension, extending them to two dimensions is substantially more demanding \cite{pavevsic2026scattering}, and they have not yet been demonstrated for scattering problems in three dimensions. Quantum simulators therefore offer a potential route toward exploring scattering regimes that are difficult to access with existing classical methods.

Simulating scattering on a quantum simulator requires the following steps \cite{jordan2011quantum,Jordan2012,preskill2018simulating}: (i) preparing an initial state consisting of two particle wave packets traveling toward one another, (ii) evolving this state with the model Hamiltonian until after the particles have collided, (iii) measuring observables to extract relevant properties of the final state.
Assuming that the desired Hamiltonian can be implemented, either with analog approaches or with Trotterization in a digital quantum computer, the preparation of the initial state and characterization of the final state remain key challenges. Several methods have been proposed to address these tasks \cite{jordan2011quantum,surace2021scattering,Barata2021,Belyansky2024,Turco2024,Farrell2024,Su2024,Bennewitz2025simulatingmeson,Turco2025,joshi2025probing,ingoldby2025real,Abel2025,lee2026studying,morgavi2026preparation,zemlevskiy2026exclusive}, particularly for digital quantum simulators. Some of these approaches have been implemented in the first digital quantum simulations of particle collisions \cite{davoudi2024scattering,Chai2025fermionicwavepacket,Zemlevskiy2025,farrell2025digital,schuhmacher2025observation}. These works demonstrated an impressive degree of control over wave packet preparation. However, they also highlighted the difficulty of maintaining high fidelity over the long evolution times required to observe inelastic scattering, due to multiple sources of error. This motivates the development of robust scattering protocols for analog quantum simulators, where the time evolution can be implemented directly, without Trotter errors.

\begin{figure*}
    \centering
    \includegraphics[width=\linewidth]{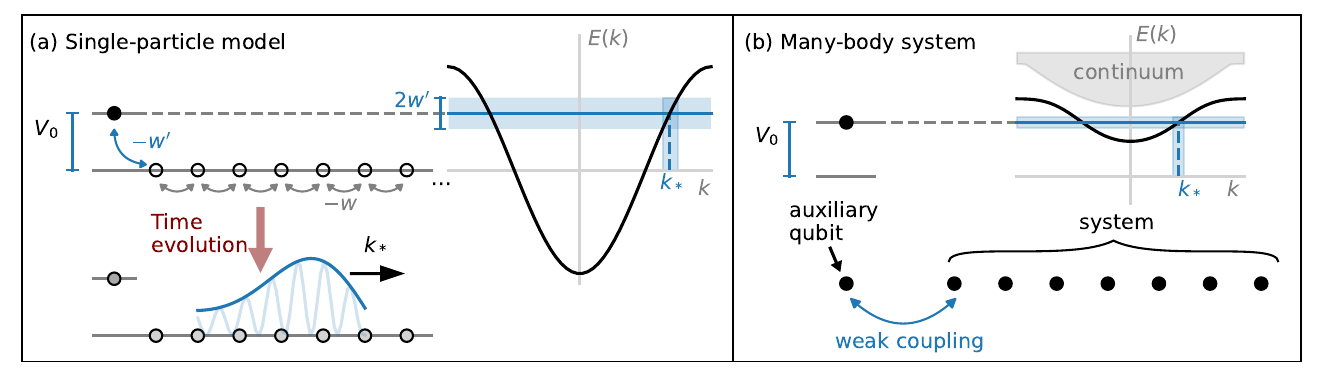}
    \caption{(a) Wave packet preparation protocol in the single-particle model. The particle is initialized in the $j=0$ site, with energy $V_0=E(k_*)$. It is coupled to the rest of the chain with a (small) hopping amplitude $w'$. As the system evolves, a wave packet with momentum $k_*$ is produced. (b) Wave packet preparation protocol in the many-body system. An auxiliary qubit with energy gap $V_0=E(k_*)$ is weakly coupled to the rest of the system.  }
    \label{fig:scheme}
\end{figure*}

In this work, we propose a method for state preparation and detection that can be used to simulate collisions on several different platforms and is particularly suited for analog simulators. The wave packet is prepared by coupling one boundary site (in 1D) or a corner site (in 2D) of the system to an auxiliary qubit, initialized in the excited state, which then decays by emitting an excitation into the system (Fig.~\ref{fig:scheme}). More generally, the same effect can be achieved by designing a local quench which injects the energy needed for the target excitation, while maintaining a small energy variance. The method we propose has the significant advantage of being robust to imperfections or errors in the calibration of the model's parameters, since it relies exclusively on the conservation of energy and conservation of energy variance, and does not require fine-tuning or prior knowledge of the energy eigenstate representing the particle. The momenta of the final states are then detected, using only local measurements, by observing the interference pattern when the particles are reflected at the system's boundary.

Our state preparation protocol shares  similarities with the well-studied model of a superconducting qubit or quantum dot (here, the auxiliary qubit) coupled to a waveguide (the system) \cite{Roy2017}. In the context of wave packet preparation for particle collisions, related protocols have been suggested in Refs.~\cite{surace2021scattering,Belyansky2024}, but have not been verified in a quantum many-body system. Here, we demonstrate the preparation of particle wave packets first in a simple single-particle model, using exact time evolution, and then in two interacting many-body systems (a Rydberg atom chain and a mixed-field Ising chain) using tensor-network simulations. By combining this protocol with an adiabatic ramp and a ``quantum slide'' (in which the Hamiltonian is spatially modulated), we show how to prepare approximately Gaussian wave packets in a regime where inelastic scattering is kinematically allowed.

The paper is structured as follows. In Section \ref{sec:SP} we focus on the analytically tractable single-particle model for a pedagogical illustration of the state preparation and detection protocols. In Section \ref{sec:MB} we move on to the case of a generic many-body system, here exemplified by a Rydberg atom chain, demonstrating all the steps needed to simulate elastic scattering. In Section \ref{sec:inelastic} we show how to use adiabatic ramps to probe inelastic scattering, focusing on a model with multiple distinct single-particle excitations (a mixed-field quantum Ising chain). In Section \ref{sec:slide} we demonstrate a method to control the shape of the wave packet using a quantum slide, and discuss its underlying theory. In Section \ref{sec:2D} we demonstrate the wave packet preparation in a two-dimensional single-particle model and discuss prospects for simulating scattering in more than one spatial dimension.

\section{Single-particle model}
\label{sec:SP}

To illustrate the protocols for state preparation and detection, we start with the textbook model of a single particle hopping in a one-dimensional lattice (tight-binding model).
This model captures the general idea of the protocol and is simple enough that we can have analytic understanding, giving us the tools to later approach the case of a full many-body Hamiltonian.

The Hilbert space of a single particle in a one-dimensional lattice is spanned by the states $\ket{j}$ with $j\in \mathbb{Z}$ labeling the position of the particle. We will consider a model where the particle can hop between neighboring sites, with a hopping amplitude $w$:

\begin{equation}
\label{eq:Htb}
    H=-w\sum_{j} (\ket{j}\bra{j+1}+\text{H.c.}).
\end{equation}

The Hamiltonian Eq.~(\ref{eq:Htb}) is easily diagonalized in momentum space. The eigenvectors $\ket{\phi_k}$ are plane waves $\ket{\phi_k}=\sum_je^{ikj}\ket{j}$ labeled by the lattice momentum $k\in [0, 2\pi)$ with energy $E(k)=-2w\cos k$.

Our first goal is to prepare a wave packet, i.e., a state with a narrow momentum distribution around a value $k_*$ of choice, and having a finite width in real space. An example is a Gaussian wave packet of the form $\ket{\psi_G}=\mathcal N^{-1}\sum_j e^{-\alpha(j-j_0)^2+ikj}\ket{j}$, where $\alpha^{-1/2}$ represents the spatial width of the wave packet and $\mathcal N$ is a normalization constant. Gaussian wave packets are particularly useful because they minimize the uncertainty product between position and momentum, yielding the most localized states for a given momentum uncertainty. We will more generally consider wave packets with non-Gaussian shapes, provided that they are narrow in momentum space and sufficiently localized in real space. We want to prepare such states starting from a simple initial state -- a particle localized on the first site of a semi-infinite one-dimensional lattice -- by time evolving using a Hamiltonian with locally tuned parameters. As will be discussed in Sec.~\ref{sec:MB}, this protocol will mimic the evolution after a local quench in a quantum simulator. To study a collision in a one-dimensional system, one can use this protocol to ``inject'' a particle from each edge of a finite chain.

Our second goal is to find a protocol for detecting the momentum of a wave packet. To mimic the measurements that are typically done in an experimental setup such as an array of neutral atoms, of superconducting qubits, or a trapped ion system, we will assume that the observer can measure in a local basis. In our single-particle model, this corresponds to measuring the position of the particle, i.e., having access to the probability distribution $P(j)=|\braket{j|\Psi}|^2$ for a state $\ket{\Psi}$, while direct measurements in the momentum basis are not allowed.

\subsection{State preparation}

Our state preparation scheme for a wave packet moving from the left edge of a chain uses site-dependent hopping amplitudes and on-site potentials in a small region near the edge. As the simplest setup, we now consider a scheme where we tune the Hamiltonian parameters on the first site of a semi-infinite chain. We will later consider a more refined scheme, where the local parameters are modulated in a larger region (Sec.~\ref{sec:slide}). The Hamiltonian for the simple case reads

\begin{align}
    H_\mathrm{prep}=&V_0\ket{0}\bra{0}-w'(\ket{0}\bra{1}+\text{H.c.})\nonumber\\
    &+\sum_{j>0} (-w\ket{j}\bra{j+1}+\text{H.c.}),\label{eq:Hwv}
\end{align}
where we add a potential $V_0$ at the first site $j=0$, and we change the hopping parameter between first two sites to $w'$. The system is initialized in the state $\ket{\psi(t=0)}=\ket{0}$, i.e., with the particle localized at site $j=0$. In the evolution under $H_\mathrm{prep}$, the particle will hop and eventually move away from the edge. The crucial ingredient for our protocol is that the energy and energy variance are conserved in the evolution, i.e.,

\begin{equation}
\braket{\psi(t)|H_\mathrm{prep}|\psi(t)}=\braket{0|H_\mathrm{prep}|0}=V_0,
\end{equation}
\begin{equation}
\sigma^2_{H_\mathrm{prep}}(t) =\braket{0|H_\mathrm{prep}^2|0}-V_0^2=(w')^2.
\end{equation}

\begin{figure}[h]
    \centering
    \includegraphics[width=\linewidth]{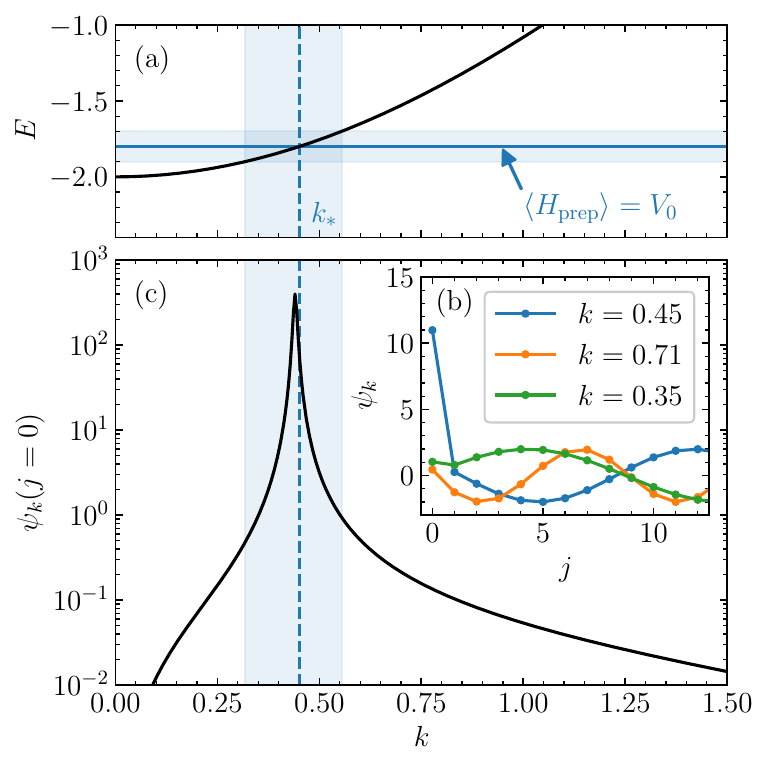}
    \caption{Scheme for preparation of a wave packet with momentum $k_*$ in the single-particle model. Parameters: $w=1, w'=0.1, V_0=-1.8$. (a) Bulk spectrum of the single-particle model $E(k)=-2w\cos k$ (black line). The conservation of energy and of energy variance defines an energy window $V_0\pm w'$ (horizontal blue-shaded region) of modes that can be excited in the time evolution under $H_\mathrm{prep}$, starting from the initial state $\ket{\psi(t=0)}=\ket{0}$. This energy window corresponds to a small range in momenta, centered around $k=k_*=\cos^{-1}(-V_0/2w)$ (vertical blue-shaded region). (b) Wavefunctions $\psi_k(j)$ of the eigenstates of $H_\mathrm{prep}$ for a few values of momenta. For $k\approx k_*$ the eigenstate $\psi_k$ has large amplitude on the first site. (c) Overlap $\psi_k(j=0)$ between the initial state $\ket{0}$ and the eigenstate $\ket{\psi_k}$ as a function of $k$. The distribution is peaked around $k_*$.}
    \label{fig:spec_1}
\end{figure}

Conservation of energy and energy variance strongly constrains the modes that can be excited in the bulk of the system, once the particle has moved away from the edge: We expect that only modes in an energy window $\approx V_0\pm w'$ will be excited. As shown in Fig.~\ref{fig:spec_1}a, by choosing $V_0=E(k_*)=-2w\cos k_*$ and $w'\ll w$, this small energy window corresponds to a small range of momenta around $k_*$.

\subsubsection{Spectrum}
We now make the above statement more precise by solving for the spectrum of $H_\mathrm{prep}$, and showing that only eigenstates with $k\approx k_*$ have large overlap with the initial state $\ket{\psi(t=0)}=\ket{0}$.

The eigenstates of $H_\mathrm{prep}$ in a semi-infinite chain are superpositions of an incoming plane wave with momentum $-k$ and an outgoing plane wave with momentum $+k$ and energy $E(k)=-2w\cos k$. Their exact form (including the phase shift) is found by imposing the boundary condition at the edge $j=0$ (see App.~\ref{app:sp})

\begin{equation}
    \psi_k(j)\equiv\braket{j|\psi_k}= \begin{dcases}
        2\sin (kj+z_k) & \text{ for }j>0\\
        2\dfrac{w}{w'}\sin z_k & \text{ for } j=0.
    \end{dcases}
    \label{eq:psik}
\end{equation}
where $k \in [0, \pi]$ and 

\begin{equation}
\label{eq:tanz}
    z_k =\tan^{-1}\left(\frac{(w')^2 \sin k}{wV_0-[(w')^2-2w^2] \cos k }\right).
\end{equation}

The eigenstates for some values of $k$ are plotted in Fig.~\ref{fig:spec_1}b. 
From Eq.~(\ref{eq:psik}) we immediately find that the eigenstate's overlap $|\psi_k(j=0)|$ with the initial state is largest when $z_k= \pi/2$, i.e., so that $\cos k = wV_0/[(w')^2-2w^2]$. In the limit $w'\ll w$, this condition corresponds to the one predicted by energy conservation $E(k_*)=V_0$ (Fig.~\ref{fig:spec_1}c).

\subsubsection{Time evolution}
\begin{figure}[t]
    \centering
    \includegraphics[width=\linewidth]{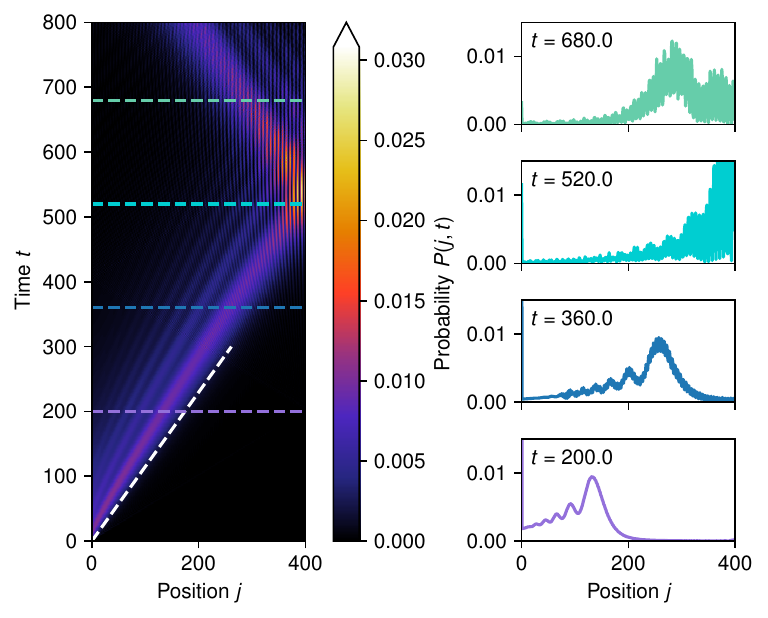}
    \caption{Left: Real-space probability distribution $P(j,t)$, obtained from the time evolution of an initial state with a particle localized on the first site (same parameters as in Fig.~\ref{fig:spec_1}). Right: The same spatial profile at selected fixed times $t$, indicated in the left panel by horizontal dashed lines. The wave packet propagates with the predicted group velocity $v(k_*)$, as indicated by the white dashed line, and is later reflected at the boundary.}
    \label{fig:single_x}
\end{figure}

We now study the time-evolved state $\ket{\psi(t)}=e^{-iH_{\mathrm{prep}}t}\ket{0}$, demonstrating the formation of a wave packet with momentum $k_*$. The real-space probability distribution $P(j,t)=|\braket{j|\psi(t)}|^2$ is plotted in Fig.~\ref{fig:single_x}.

A localized pulse is emitted from the boundary site $j=0$ at time $t=0$ and propagates into the bulk with the characteristic group velocity $v(k_*)=(dE/dk)|_{k_*}=2w\sin(k_*)$, corresponding to the target momentum $k_*$, indicated by the white dashed line. The resulting wave packet is clearly non-Gaussian, exhibiting pronounced oscillations as well as a long asymmetric tail. Nevertheless, it is sharply peaked in momentum space, as anticipated from our spectral analysis and as we will further illustrate in the next subsection. In the present simulation we consider a finite system of fixed size $L=400$; consequently, once the wave packet reaches the opposite boundary it undergoes reflection.

\subsection{State detection}
As shown in the previous subsection, the group velocity of the wave packet can be inferred from the local probability distribution, which in turn allows one to determine the momentum, assuming the dispersion relation is known. While this approach is useful in many cases, it becomes less practical when the velocity varies little with $k$, as, for example, in the relativistic regime. Here, we illustrate an alternative method that, like the previous one, relies solely on measurements of the local probability distribution, and does not require previous knowledge about the dispersion relation.

As shown in Fig.~\ref{fig:single_x}, when the wave packet is reflected at the boundary, the real-space probability distribution exhibits pronounced oscillations. These oscillations arise from the interference between the incoming wave packet, with momentum $k_*$, and the reflected wave packet, with momentum $-k_*$. The resulting interference pattern is $|e^{ik_*j}+e^{-ik_*j}e^{i\phi}|^2\propto \cos(2k_*j-\phi)$, where $\phi$ is the phase shift acquired upon reflection. The momentum $k_*$ can then be extracted from the Fourier transform of the probability distribution $P(j,t)$ with respect to the space coordinate $j$:
\begin{equation}
\label{eq:Pk}
    \mathcal P(q,t)= \sum_{j=1}^{L-1} P(j,t)e^{iqj}.
\end{equation}

Since we are interested in the interference fringes on the right end of the chain, in the definition of $\mathcal P(q,t)$ we omit the first site ($j=0$), thereby obtaining a cleaner signal.
Note that 
$\mathcal P(q,t)$ differs from the actual momentum-space probability distribution, which requires {\it first} Fourier transforming $\psi(j,t)$ and {\it then} taking the modulus squared. This distinction is important: the true momentum-space distribution cannot be accessed through single-site measurements alone, whereas $\mathcal P(q,t)$ can be reconstructed directly from measurements of the local probability $P(j,t)$.

\begin{figure}
    \centering
    \includegraphics[width=\linewidth]{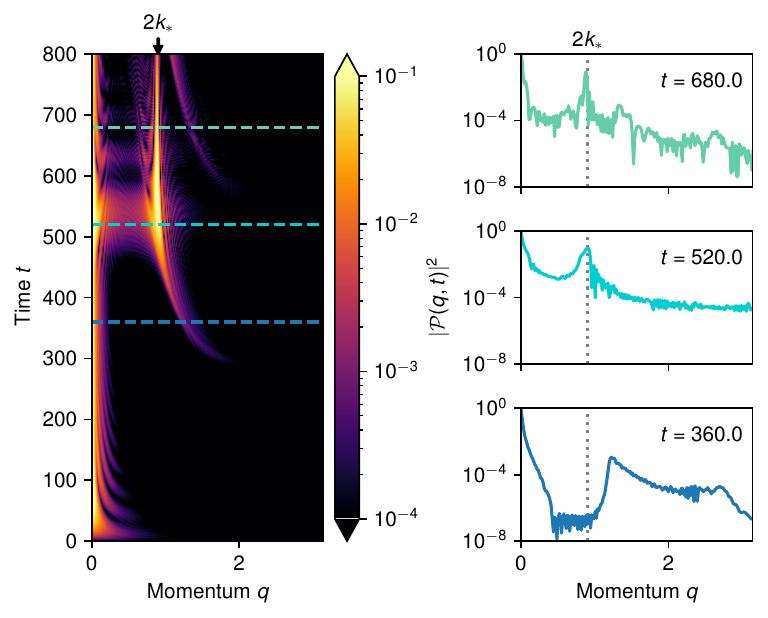}
    \caption{Left: Absolute value squared of $\mathcal P(q,t)$, defined in Eq.~(\ref{eq:Pk}) as the Fourier transform of the real-space probability distribution with respect to the spatial coordinate. Since $|\mathcal P(q,t)|^2=|\mathcal P(-q,t)|^2=|\mathcal P(2\pi-q,t)|^2$, we restrict to the range $q\in [0,\pi]$. Right: The same function $|\mathcal P(q,t)|^2$ at selected fixed times $t$, indicated in the left panel by horizontal dashed lines. The function $|\mathcal P(q,t)|^2$ develops a peak at $q=2k_*$ when the wave packet is reflected at the boundary, originating from the interference between the incoming and reflected wave packets. }
    \label{fig:single_k}
\end{figure}

The function $|\mathcal P(q,t)|^2$, obtained from the Fourier transform of the probability distribution in Fig.~\ref{fig:single_x}, is plotted in Fig.~\ref{fig:single_k}. The plot shows a clear peak at $q\approx 2k_*$, consistent with the interference argument.
When the wave packet approaches the boundary, the peak initially appears at slightly larger momenta; as the reflection develops, it shifts toward $2k_*$ and increases in magnitude. The early appearance of the peak at larger $q$ can be attributed to interference involving the high-momentum components of the wave packet, which propagate with a larger group velocity. Their contribution, however, is weaker, since they originate from the small high-$k$ tail of the wave packet.

\section{Many-body Hamiltonian}
\label{sec:MB}

In this Section, we show how the protocols for state preparation and detection that we introduced in the single-particle model can be generalized to a many-body Hamiltonian. We will consider the model of a one-dimensional array of Rydberg atoms, where each atom of the chain can be in its internal ground state $\ket{g}$ or in a Rydberg state $\ket{r}$. The model is described by the Hamiltonian

\begin{equation}
\label{eq:Ryd}
    H= \sum_{j=1}^L \left(\frac{\Omega}{2}\sigma_j^x -\Delta\, n_j \right)+\sum_{1\le j<k\le L} \frac{C}{|j-k|^6}n_j n_k.
\end{equation}

Here we defined $\sigma_j^x=(\ket{r}\bra{g}+\ket{g}\bra{r})_j$ and $n_j=\ket{r}\bra{r}_j$ is the occupation of the Rydberg state in the $j$-th atom. The Rydberg-Rydberg interaction is of the form $V_{jk}=C/|j-k|^6$.

The model (\ref{eq:Ryd}) supports several phases, including a disordered phase and various ordered phases with spontaneous breaking of translational symmetry. Here we work in the disordered phase and make the following choice of parameters for the numerical simulations \footnote{In our plots, the time is measured in units of a Rabi cycle, i.e. $2\pi/\Omega$}: $\Delta=-0.5 \,\Omega,\, C=4\, \Omega$. This point in the disordered phase can be adiabatically connected to the limit $\Delta\rightarrow-\infty$ where the ground state is the empty state $\ket{ggg\dots}$ and quasiparticles correspond to Rydberg excitations. The approximate identification of quasiparticles with Rydberg excitations will be useful to make connections with the single-particle model in Sec.~\ref{sec:SP}.

\subsection{State preparation and detection}

To prepare a propagating wave packet, we introduce an auxiliary atom (labeled $j=0$) positioned at the edge of the chain, at a distance $d$ from site $j=1$. We choose $d$ to be larger than the spacing $a$ between the other atoms, such that the auxiliary atom is only weakly coupled with the rest of the system. This setup mirrors the role of the $j=0$ site in the single-particle model (Fig.~\ref{fig:scheme}). We control the state of the auxiliary atom by modulating its time-dependent detuning $\Delta_0(t)$, leading to the single-site Hamiltonian
\begin{equation}
    H_0(t)= \frac{\Omega}{2}\sigma_0^x -\Delta_0(t)\, n_0= \frac{1}{2}\Big(\Omega \sigma_0^x -\Delta_0(t)\sigma_0^z\Big)+\text{constant}.
\end{equation}

This local control over $\Delta_0(t)$ allows us to engineer the emission of a quasiparticle from the edge, analogously to the single-particle case. The interaction between the auxiliary atom and the chain is described by
\begin{equation}
    H_{\mathrm{int}}=\sum_{j=1}^L \frac{C}{|\kappa +j|^6}n_0 n_j,
\end{equation}
where the dimensionless parameter $\kappa=(d/a)-1$ effectively tunes the coupling strength between the auxiliary atom and the chain.
For large $\kappa$, the auxiliary atom is nearly decoupled from the chain and behaves as an isolated two-level system with an energy gap $V_0=\sqrt{\Omega^2+\Delta_0^2}$. 

To prepare the wave packet, we initialize the auxiliary qubit in its (approximate) excited state, while the chain remains in its ground state.
The detuning $\Delta_0 = \Delta_*$ is selected so that $V_0$ matches the energy of the desired quasiparticle excitation.
As the system evolves, the weak coupling between the auxiliary site and the chain enables a gradual energy transfer from the qubit to the chain,
exciting a propagating mode with energy $V_0$.

\begin{figure*}[t]
    \centering
    \begin{minipage}[t]{0.32\textwidth}
    \vspace{0pt}
    \includegraphics[width=0.98\linewidth]{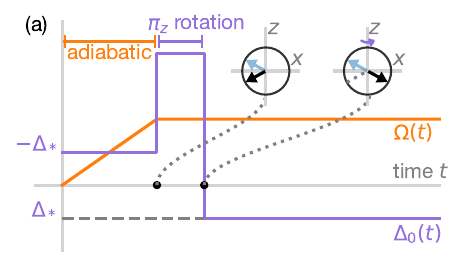}
    \includegraphics[width=\linewidth]{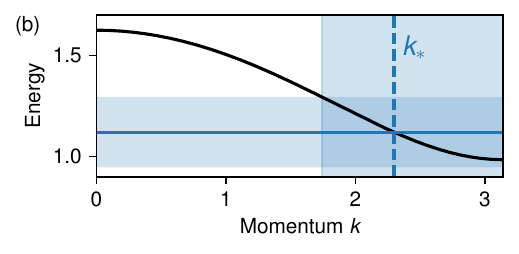}
    \includegraphics[width=\linewidth]{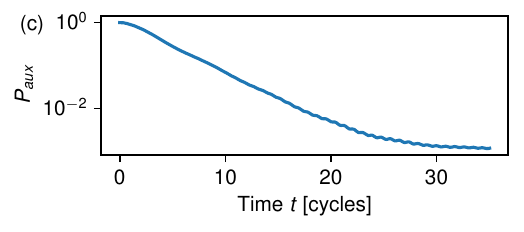}
    \end{minipage}
    \begin{minipage}[t]{0.67\textwidth}
    \vspace{0pt}
    \includegraphics[width=\linewidth]{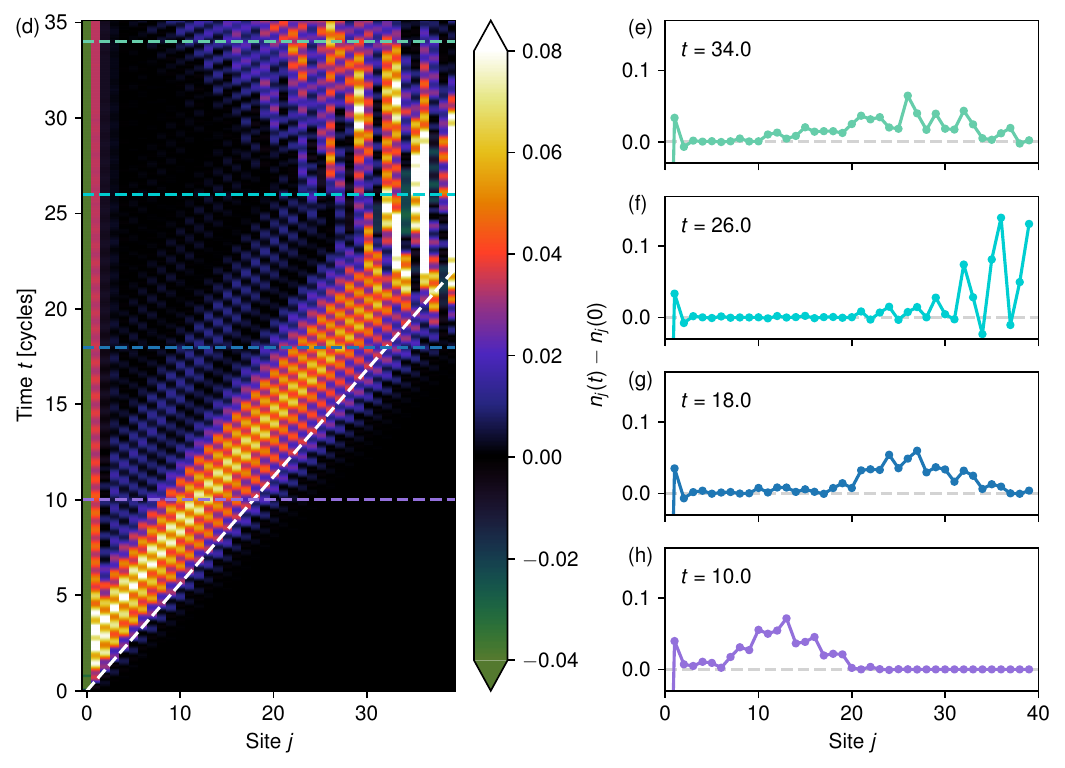}
    \end{minipage}
    \caption{Wave packet preparation in a Rydberg model. (a) Time dependence of the parameters $\Omega$ and $\Delta_0$ of the Rydberg atom Hamiltonian. This procedure prepares the auxiliary qubit in its excited state. (b) Dispersion relation of the excitations in the Rydberg atom chain. As in the single-particle model, a wave packet with given momentum $k_*$ and momentum width is prepared by tuning the average energy and energy variance. (c) Population of the excited state of the auxiliary qubit. The auxiliary qubit gradually decays, and its energy is transferred as a wave packet propagating in the chain.  (d) Variation of the local Rydberg occupation $n_j(t)-n_j(0)$, showing propagation of the wave packet with the predicted group velocity (white dashed line). (e-h)  Variation of the local Rydberg occupation at selected fixed times, indicated in (d) by horizontal dashed lines. The wave packet reaches the opposite end of the chain and undergoes reflection, producing an interference pattern. The bond dimension used in the MPS simulation is $\chi=200$. }
    \label{fig:seq}
\end{figure*}

As in the single-particle case, the dynamics can be understood in terms of the conserved energy and energy variance.
The initial state carries energy $V_0$ (relative to the ground state) and has a narrow energy variance $(\delta E)^2$ determined by the weak interaction $H_{\mathrm{int}}$.
Consequently, only system eigenmodes with energy within $V_0 \pm \delta E$ are significantly populated, leading to the formation of a localized 
wave packet with a well-defined momentum peak at $k_*$ and a small momentum variance.

An important caveat in this protocol is that the single-particle modes must be well separated from the continuum of multi-particle excitations. In other words, the mass gap should be sufficiently large to ensure that only a single quasiparticle is excited during the process. If this condition is not satisfied, the energy injected into the system may instead populate multi-particle states. In Section \ref{sec:inelastic}, we will discuss how to prepare a wave packet even when the mass gap is small and the single-particle band is not well separated from the rest of the spectrum.

\subsubsection{Time modulation of the Hamiltonian parameters}
We now describe how to prepare the qubit in its excited state by modulating the (global) Rabi frequency $\Omega(t)$ and the local detuning on the auxiliary spin $\Delta_0(t)$. The Rabi frequency is varied only globally, and the time-dependent modulation of the local detuning that we propose is compatible with current experimental capabilities in Rydberg atom array platforms \cite{chen2023continuous,manovitz2025quantum,deOlivera25,Wang2025,Wang2025individual}.

The system is first initialized in the global ground state with $\Delta_0 = -\Delta_*$, using an adiabatic ramp of the Rabi frequency from $0$ to $\Omega$, starting from the fully polarized state $\ket{ggg\ldots}$. Since we operate in the disordered phase, this ramp does not encounter any phase transition, allowing the ground state to be prepared reliably.

Next, we apply a $\pi$-rotation around the $\sigma^z$ axis to the auxiliary site by evolving the system for a short time $\delta t$ with a large detuning $\Delta_0= \pi / \delta t$. The evolution of the qubit's state on the Bloch sphere is illustrated in Fig.~\ref{fig:seq}a. This pulse flips the sign of the qubit’s $x$-polarization, effectively preparing the ground state of the Hamiltonian $(-\Omega \sigma_0^x + \Delta_* \sigma_0^z)/2$. This state is, by construction, the excited state of the target Hamiltonian $H_0=(\Omega \sigma_0^x - \Delta_* \sigma_0^z)/2$.

\subsubsection{Numerical results}
We now verify, using matrix-product state (MPS) numerical simulations \footnote{Using the TenPy library, we employ, in particular, the two-site density matrix renormalization group (DMRG) method to prepare the initial ground state and the two-site time-dependent variational principle (TDVP) for the time evolution. In all the simulations reported in this paper, the scattering involves only few particles, and the entanglement growth is bounded, such that relatively small bond dimensions can be used, even for long-time evolutions. Where not explicitly stated, the bond dimension used in the simulations was $\chi=200$.}, that the protocol we propose is preparing the desired wave packet, and that the momentum can be efficiently reconstructed with our detection method. 

In our simulations we choose $\Delta_*=-0.5\, \Omega$, $\kappa=0.5\, \Omega$, and $\delta t = 0.1\Omega^{-1}=0.1(2\pi)^{-1}$ cycles. We define $t=0$ as the time at which the $\pi_z$ rotation is completed.
The energy of the auxiliary qubit is $V_0 = \sqrt{\Omega^2 + \Delta_*^2} \approx 1.12\, \Omega $,
which corresponds to a target momentum $k_*\approx 2.3$ (see dispersion relation in Fig.~\ref{fig:seq}b, obtained as discussed in Appendix \ref{app:disp}). 
We choose a relatively small value of $\kappa$, which results in a large energy variance and, consequently, in a sizeable momentum spread of the emitted wave packet. 
This choice is instrumental in generating a wave packet that extends over only $\mathcal O(10)$ sites, suited for a proof-of-principle experimental implementation on small systems that are currently within reach. 
We will demonstrate this protocol for narrower momentum spreads in Sec.~\ref{sec:inelastic}.

\begin{figure}
    \centering
    \includegraphics[width=\linewidth]{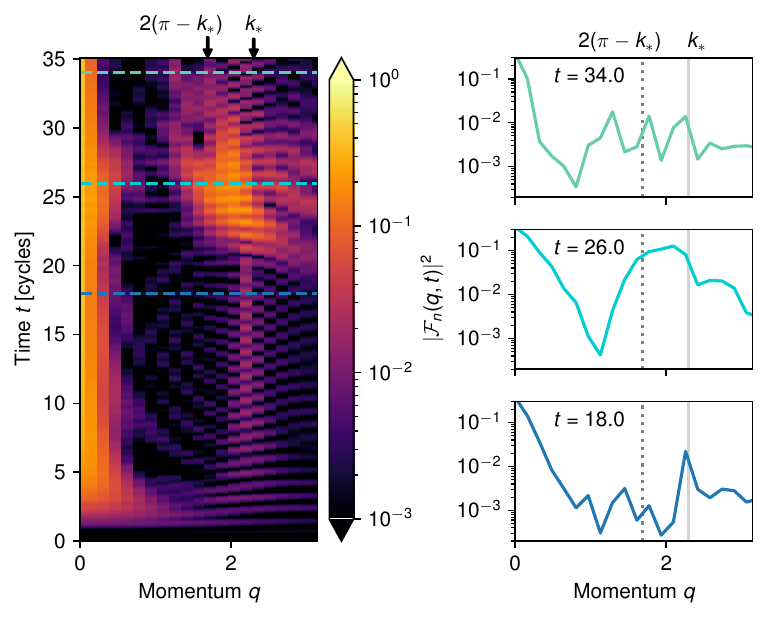}
    \caption{State detection in a Rydberg model. Left: absolute value squared of $\mathcal F_n(q,t)$, defined in Eq.~(\ref{eq:Fn}). Right: The same function $|\mathcal F_n(q,t)|^2$ at selected fixed times $t$, indicated in the left panel by horizontal dashed lines. The function develops a peak at $q=2k_*$ (and, equivalently, at $q=2\pi-2k_*$) when the wave packet is reflected at the boundary, originating from the interference between the incoming and reflected wave packets. A smaller peak is present at $q=k_*$ and appears well before the wave packet has reached the boundary. This smaller peak originates from a non-vanishing overlap with the ground state, as explained in the text.}
    \label{fig:rydk}
\end{figure}

As shown in Fig.~\ref{fig:seq}c, the auxiliary qubit, initially prepared in its excited state, gradually decays to the ground state. 
The probability $P_{\mathrm{aux}}$ of finding it in the excited state decays exponentially in time.
As the auxiliary qubit decays, it releases an excitation into the system (Fig.~\ref{fig:seq}d--h), which propagates from the left edge of the chain. 
A simple way to observe this propagation experimentally is to monitor the variation of a local observable, for instance the local Rydberg occupation $n_j$, between time $t$ and time $0$, when the system is locally in the ground state. 
As indicated by the white dashed line, the resulting wave packet propagates with the expected group velocity $v(k_*)$. 
Upon reaching the right boundary, the wave packet is reflected, leading to pronounced spatial oscillations. 
As in the single-particle case, these oscillations provide a useful probe of the wave packet momentum. 
In Fig.~\ref{fig:rydk} we therefore plot the absolute value squared of $\mathcal F_n(q,t)$, defined as the Fourier transform, with respect to the spatial coordinate, of $n_j(t)-n_j(0)$ (we again omit the site $j=0$):
\begin{equation}
\label{eq:Fn}
    \mathcal F_n(q,t)=\sum_{j=1}^{L-1}[n_j(t)-n_j(0)]e^{iqj}.
\end{equation}

\begin{figure*}
    \centering
    \includegraphics[width=0.57\linewidth]{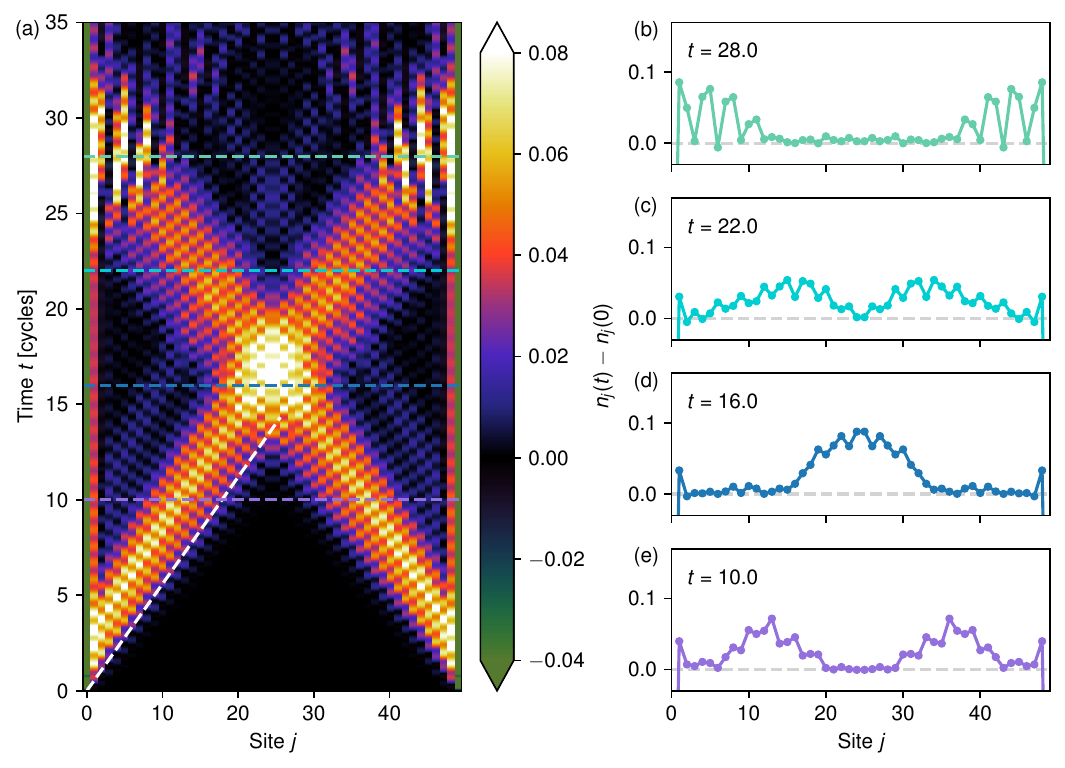}
    \includegraphics[width=0.42\linewidth]{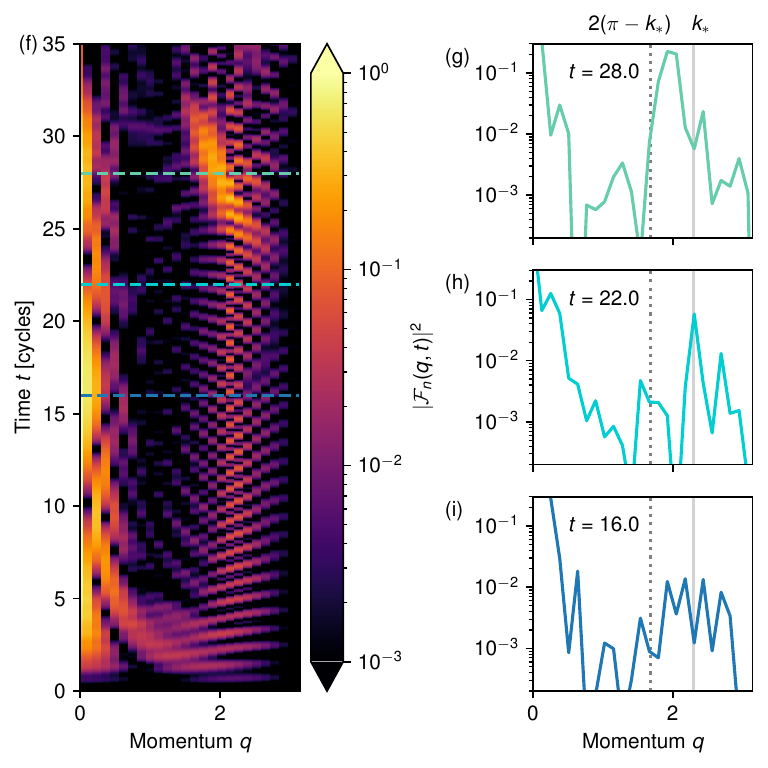}
    \caption{Two-particle scattering in a Rydberg model. (a) Variation of the local Rydberg occupation $n_j(t)-n_j(0)$ as a function of $(j,t)$, and (b-e) as a function of $j$ at selected fixed times. The two wave packets propagate from the two ends of the chain, scatter elastically among themselves, and are then scattered elastically at the boundaries. (f) Absolute value squared of $\mathcal F_n(q,t)$ as a function of $(q,t)$, and (g-i) as a function of $q$ for selected fixed times. A peak at $q\approx k_*$ (whose origin is explained in the text) is observed throughout the evolution, while the peak at $q\approx 2k_*$ (here, equivalently, $q\approx 2\pi-2k_*$) appears when the wave packets are scattered at the boundaries, resulting in interference patterns.}
    \label{fig:RydSc}
\end{figure*}

Similarly to the single-particle case, we observe that a peak develops when the wave packet is reflected at the boundary (at time $t\approx 26$). 
Because the momentum uncertainty is relatively large, this peak is not very sharp in momentum space and extends over a broad region around $q=2k_*\approx 4.6$. 
Note that in Fig.~\ref{fig:rydk} we plot only momenta in the interval $[0,\pi]$; as a result, the peak appears at the equivalent momentum $q=2\pi-2k_*$. 
We also observe a similar temporal trend from higher-velocity components to lower-velocity ones, as in the single-particle case. 
Here, however, the faster components correspond to smaller momenta, as can be inferred from the dispersion relation in Fig.~\ref{fig:seq}b. 
Consequently, the peak around $q=2(\pi-k_*)$ moves from larger to smaller values of $q$ as a function of time.

In addition to this main feature, we identify a distinct signal at $q\approx k_*$, which is already visible before the wave packet reaches the boundary and persists after its reflection (see, for example, the profiles at $t=18$ and $t=34$).
This peak is significantly smaller in amplitude than the one at $2(\pi-k_*)$ observed at $t=26$, but is observed systematically over a wide range of parameters $\Delta_*$ and $\kappa$. 
This feature can be understood as follows. 
Suppose that the preparation of the wave packet is not perfect, so that the time-evolved state retains a small overlap with the ground state of the system. 
We may then write, schematically,
\begin{equation}
    \ket{\psi(t)}\approx c_{g}\ket{g}+\sum_k c_ke^{-iE_kt}\ket{e_k},
\end{equation}
where $\ket{g}$ denotes the ground state and $\ket{e_k}$ is the eigenstate corresponding to a single-particle excitation with momentum $k$ and energy $E_k$. 
For the purpose of this argument, we treat the system as translationally invariant with periodic boundary conditions, which is justified since we are interested in the propagation of the wave packet in the bulk, far from the boundaries. 
We also assume that the population of higher-energy eigenstates involving two or more particles can be neglected.

Using translational invariance, the local operator $n_j$ can be related to the operator at a reference site $j=0$ via the translation operator $T$ as
\begin{equation}
    T^{-j}n_j T^j=n_0.
\end{equation}
Together with the relations $T\ket{g}=\ket{g}$ and $T\ket{e_k}=e^{ik}\ket{e_k}$, this leads to
\begin{align}
    \label{eq:nj}\braket{n_j(t)}=&|c_g|^2\braket{g|n_0|g}\nonumber\\
    &+\sum_{k,k'}c_kc_{k'}^*e^{-i(E_k-E_{k'})t+i(k-k')j}\braket{e_{k'}|n_0|e_k}\nonumber\\
    &+\left(\sum_{k}c_k c_g^*e^{-iE_kt}e^{ikj} \braket{g|n_0|e_k}+\mathrm{H.c.}\right).
\end{align}
Since the coefficients $c_k$ are sizeable only within a narrow window around $k_*$, the first two lines contribute to $\mathcal F_n(q,t)$ only at small values of $q$. 
The third line, instead, gives rise to a peak at $q\approx k_*$, which oscillates in time with a frequency set by $E_{k_*}$. 
This argument fully accounts for the features observed in Fig.~\ref{fig:rydk}, including both the low-$q$ signal and the oscillating peak at $q\approx k_*$. 

Although this latter peak originates from an undesired effect—--namely the imperfect preparation of the wave packet and its residual overlap with the ground state—--it can nevertheless be exploited as an alternative probe of the momentum. 
Finally, we note that this discussion is not specific to the choice of the operator $n_j$, but is quite general and applies to generic local observables.

\subsection{Scattering}
Having established methods for both the preparation of wave packets and the detection of their momenta, we now turn to the simulation of a scattering process. 
An auxiliary qubit is placed at each end of the chain and prepared in its excited state, following the same protocol used for the preparation of a single wave packet. 
The results for a chain of length $L=50$ are shown in Fig.~\ref{fig:RydSc}, using the same set of parameters as in the previous subsection. 
The two wave packets propagate inward from opposite ends of the chain, collide near the center, and are subsequently reflected at the boundaries (Fig.~\ref{fig:RydSc}a--e).

During the collision, conservation of energy and momentum restricts the dynamics to elastic scattering. 
As a result, two outgoing wave packets emerge from the scattering region with momenta $\pm k_*$, identical to those of the incoming wave packets. 
The outgoing momenta are detected by analyzing the Fourier transform of $n_j(t)-n_j(0)$, as described in the previous subsection (Fig.~\ref{fig:RydSc}f-i). 
When the wave packets are reflected at the boundaries, a peak develops near $q=2k_*$ (and, equivalently, at $q=2\pi-2k_*$), in agreement with the expectation for elastic scattering.

\begin{figure}
    \centering
    \includegraphics[width=\linewidth]{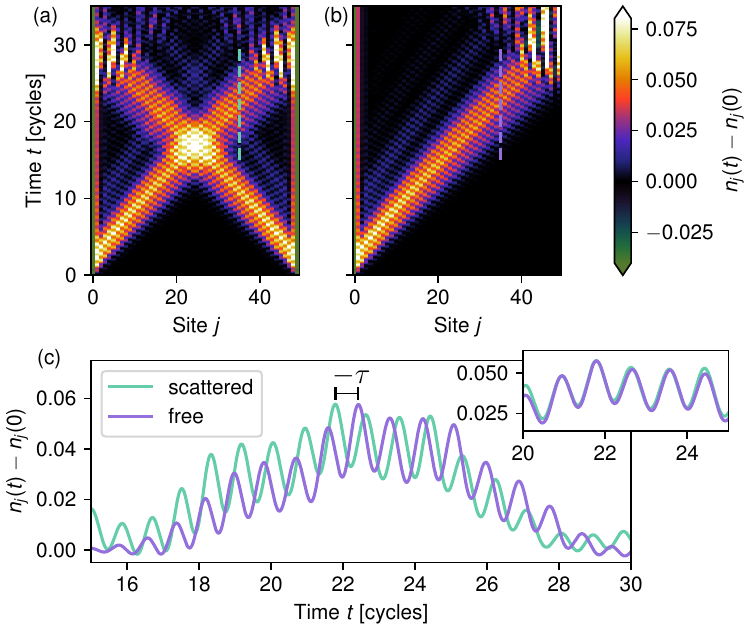}
    \caption{
    The time delay is extracted by comparing the propagation of (a) a wave packet that undergoes a collision, with (b) the propagation of a free wave packet. Panel~(c) displays the quantity $n_j(t)-n_j(0)$ at site $j=35$ for the two cases. The scattered wave packet arrives earlier than the free one, corresponding to a negative time delay $\tau \approx -0.1$ cycles. As shown in the inset, shifting the free-propagation signal by a time $\tau$ yields an excellent overlap between the two curves.}
    \label{fig:delay}
\end{figure}

For elastic scattering of this type, the scattering matrix reduces to a single parameter, namely a phase shift. 
This phase shift can be extracted by measuring the time delay accumulated during the collision, for different values of the momentum $k_*$. 
As shown in Fig.~\ref{fig:delay}, the time delay $\tau$ is obtained by comparing the free propagation of a single wave packet with the propagation of a wave packet that undergoes a collision with another wave packet.

\section{Probing inelastic scattering at high energies}
\label{sec:inelastic}
In the previous section, we demonstrated our methods for state preparation and detection, and showed how to extract the time delay associated with elastic scattering.
A key requirement for applying this protocol was the presence of a sufficiently isolated mode, so that an energy filter (implemented by fixing the average energy and its variance) selectively targets a single-particle excitation with well-defined momentum. However, many physically interesting scattering regimes do not satisfy this condition. In particular, one often aims to probe scattering at relatively high energies, approaching a relativistic regime in which the bandwidth greatly exceeds the mass gap and strongly inelastic processes can occur.

In this section, we show how to address this limitation by employing adiabatic ramps, and demonstrate that this approach enables the study of inelastic scattering at high energies.
To this end, we consider a model different from the Rydberg-atom chain discussed above, and instead focus on a system that supports multiple bands, or ``species'', of single-particle excitations: the mixed-field quantum Ising chain. 
This model enables inelastic scattering between two-particle states with discrete allowed momenta, providing a clean setting in which such processes can be resolved. Furthermore, the mixed-field quantum Ising chain can be realized in several experimental platforms, including ultracold atoms in optical lattices \cite{Simon2011-gs}, Rydberg atom arrays \cite{Labuhn2016-jw,deLeseleuc2018} and, most notably, trapped-ion systems \cite{Monroe2021}, where multiple quasiparticle excitations have already been observed \cite{Tan2021-eg}. The protocol proposed here is directly compatible with modern analog trapped-ion architectures featuring local control \cite{de2024,luo2025}. Scattering in the same model has also been investigated on digital platforms based on superconducting qubits \cite{farrell2025digital}, providing a natural benchmark for comparison with analog implementations.

We note that, for a different choice of parameters, the Rydberg-atom Hamiltonian considered in the previous section can also support inelastic scattering processes in which two incoming particles scatter into three outgoing particles. 
However, in that case the allowed final-state momenta form a continuum, which significantly complicates their detection. 
For this reason, we defer a detailed study of such processes to future work.

In this section, we also introduce a slightly different preparation scheme, in which the single-site terms are tuned in a site-dependent manner while the interaction terms remain uniform across the chain, without the need for weak-interaction links. 
This variation may be advantageous for a range of experimental platforms.

\subsection{Ramp protocol and inelastic scattering at the boundary}
\begin{figure*}
    \centering
    \includegraphics[width=\linewidth]{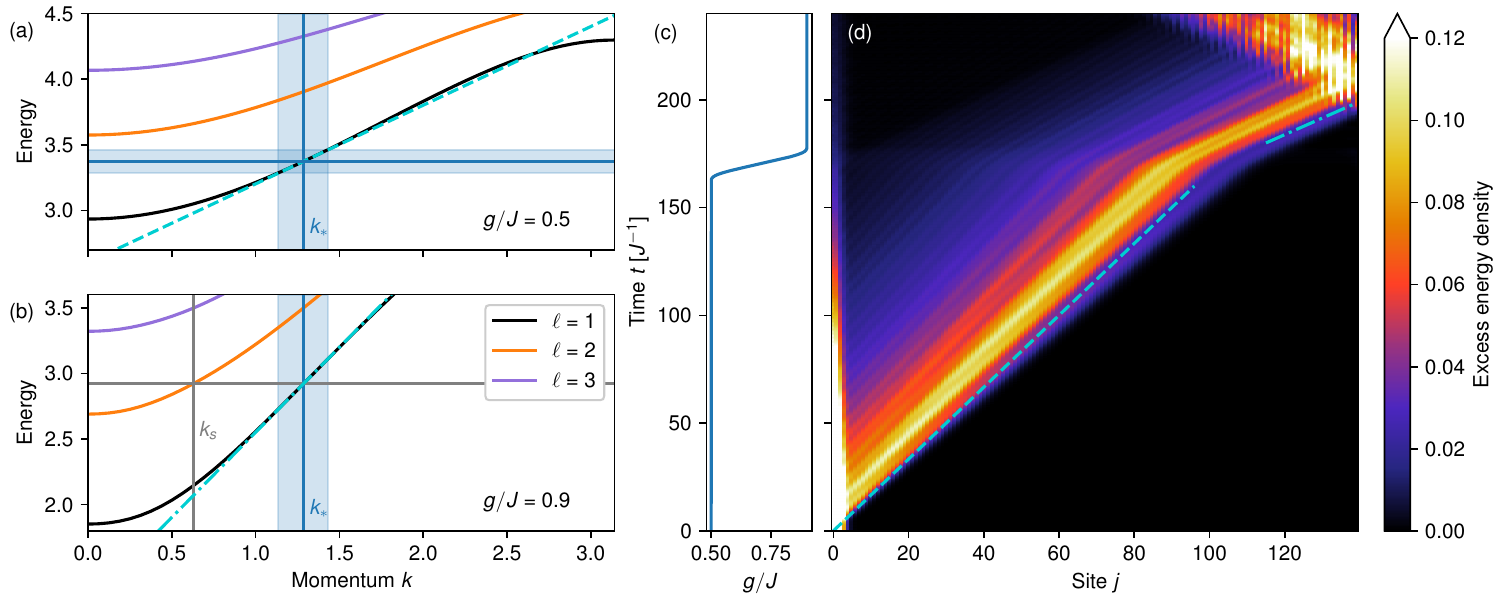}
    \caption{Preparation of a high-energy (above inelastic threshold) wave packet in the Ising chain using  an adiabatic ramp. (a) Dispersion relation of the model for $g/J=0.5$ and $h/J=0.1$. A wave packet with momentum $k_*$ and quantum number $\ell=1$ can be prepared with our method, as it is well separated in energy. (b) Dispersion relation of the model for $g/J=0.9$ and $h/J=0.1$. A wave packet with momentum $k_*$ and quantum number $\ell=1$ is not well separated in energy. To prepare it, we prepare the wave packet with $g/J=0.5$, then tune the transverse field in time to adiabatically reach the desired regime $g/J=0.9$. (c) Profile of the adiabatic ramp of $g/J$. (d) Excess energy density as a function of site $j$ and time $t$. Before the ramp, the wave packet propagates with the expected group velocity for $g/J=0.5$ (dashed line). After the ramp, the group velocity is the one expected for $g/J=0.9$ (dotted-dashed line).   }
    \label{fig:ramp}
\end{figure*}

To demonstrate the ramp protocol, we consider here the mixed-field quantum Ising chain, defined by the Hamiltonian
\begin{equation}
\label{eq:ising}
    H=-J\sum_{j=1}^{L-1}\sigma_j^z\sigma_{j+1}^z+\sum_{j=1}^{L}(g\sigma_j^x+h\sigma_{j}^z).
\end{equation}
For $J>|g|$ and nonzero $h$ (which we here set to $h=0.1\, J$), the low-energy spectrum of this model consists of a tower of single-particle excitations. These excitations are known as {\it mesons}, as they correspond to confined states of two domain walls (which are the elementary excitations of the model for $h=0$) \cite{McCoyWu1978,DELFINO1996469,Kormos2017-ar}. The bands corresponding to different meson species are well separated for small $g$, and increasingly overlap as $g$ approaches $J$ from below (Fig.~\ref{fig:ramp}a-b).

If the energy ranges of two bands $\ell$ and $\ell'$ overlap, inelastic scattering processes can occur in which two $\ell$-mesons scatter into two $\ell'$-mesons while conserving the total energy and total momentum $k_{tot}=0$. For example, for a specific choice of parameters, shown in Fig.~\ref{fig:ramp}b, a $1$-meson with momentum $k_*$ has the same energy as a $2$-meson with momentum $k_s$. 
As a consequence, an incoming state consisting of two $1$-mesons with momenta $\pm k_*$ can scatter inelastically into a pair of $2$-mesons with momenta $\pm k_s$. Another possible inelastic channel, which may be present even if the two bands do not overlap in energy, is the conversion of two $1$-mesons into a final state consisting of a $1$-meson and a $2$-meson.  These various inelastic processes are the phenomena we here aim to investigate.

Because the protocol relies on energy conservation, a scheme such as the one used above would generically prepare a superposition of states at the target energy (marked by a grey horizontal line in Fig.~\ref{fig:ramp}b) belonging to different bands ($\ell=1$ and $\ell=2$) and carrying different momenta ($k_*$ and $k_s$). 
To achieve a faithful preparation of an excitation within a single band and with a well-defined momentum, we therefore employ an adiabatic ramp.
Specifically, we first prepare the wave packet at a smaller value of $g$, where the $1$-meson with momentum $k_*$ is well separated in energy from other bands (Fig.~\ref{fig:ramp}a). 
Once the wave packet has been completely emitted from the first site, we adiabatically ramp $g$ to its final value (Fig.~\ref{fig:ramp}c). 
Since $g$ is ramped uniformly throughout the system, momentum is conserved during the evolution, provided the wave packet remains far from the boundaries. 
Moreover, if the ramp is sufficiently slow, adiabaticity ensures that transitions between different bands are suppressed. 
As a result, we obtain the desired wave packet: an excitation confined to a single band with a narrow momentum distribution.

We demonstrate this protocol for the Ising chain defined in Eq.~(\ref{eq:ising}) by coupling the system to an auxiliary site, following the approach introduced in the previous sections. 
The full Hamiltonian used for state preparation reads
\begin{equation}
\label{eq:Isingprep}
    H_{\mathrm{prep}} = g_0 \sigma_0^x + h_0 \sigma_0^z - J \sigma_0^z \sigma_1^z + H .
\end{equation}
The Ising coupling between the auxiliary site and the first site of the chain is taken to be equal to $J$, matching the couplings within the bulk of the chain, while the on-site fields $g_0$ and $h_0$ acting on the auxiliary site are distinct and can be tuned independently.  A model of this type, with local time-dependent control of the transverse and longitudinal fields, has been demonstrated, for example, in recent trapped ion experiments \cite{de2024,luo2025}.

In this setup, the auxiliary qubit is not weakly coupled to the system. 
Nevertheless, the state-preparation protocol can still be applied, provided that one can prepare a state with the desired energy and a sufficiently small energy variance. 
One possible approach is to first prepare the ground state of the system with $g_0 = 0$ and a large negative $h_0$, which initializes the auxiliary qubit in the $\uparrow$ state. 
Subsequently, one performs a sudden quench of the on-site fields, setting $g_0$ to a small but nonzero value and $h_0$ to a positive value: 
after this quench, the energy variance is $g_0^2$, while the energy can be controlled by tuning the value of $h_0$.

In our numerical simulation, this first step of the wave packet preparation is carried out at $g/J = 0.5$. 
The parameters $g_0=0.09$ and $h_0=0.95$ are chosen so as to match the energy of a $1$-meson with momentum $k_*\approx 1.28$, while ensuring a small momentum spread of the prepared wave packet (Fig.~\ref{fig:ramp}a). 

 This state preparation protocol is demonstrated in Fig.~\ref{fig:ramp}d, where we plot the excess energy density after the quench. The latter is defined as the expectation value of the local energy-density operator
\begin{equation}
    \mathcal E_j = -J\sigma_j^z \sigma_{j+1}^z+\frac{g}{2}(\sigma_j^x +  \sigma_{j+1}^x)+\frac{h}{2}(\sigma_j^z +  \sigma_{j+1}^z),
\end{equation}
from which we subtract the expectation value of the same operator evaluated in the ground state.
As shown in Fig.~\ref{fig:ramp}d, the wave packet propagates with the expected group velocity.

\begin{figure}
    \centering
    \includegraphics[width=\linewidth]{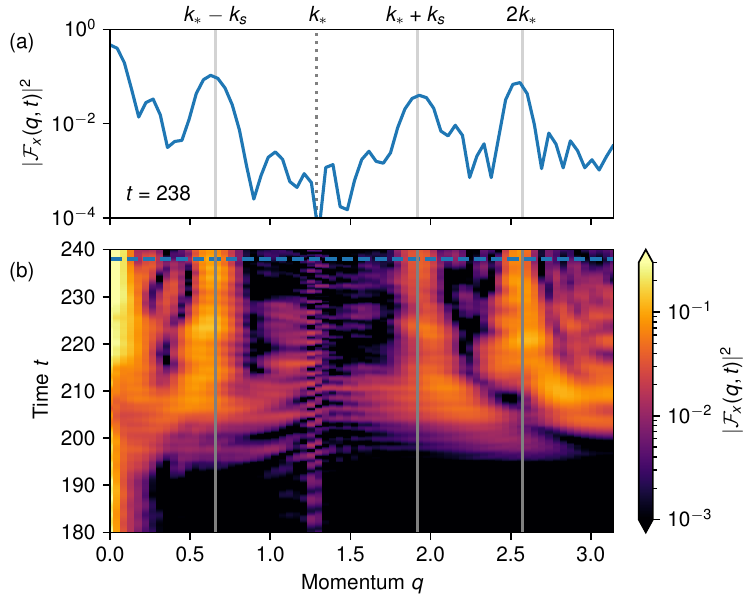}
    \caption{Inelastic scattering at the boundary. (a) Fourier transform of the local $\sigma_j^x(t)$, after subtracting the ground state expectation value. After the collision with the boundary, distinct peaks signal the presence of elastic scattering ($q=2k_*$) and inelastic processes to a state with $\ell=2$ and momentum $k_s$ ($q=k_*\pm k_s$). (b) Fourier transform of $\sigma_j^x(t)$ (same as in (a)) for a  range of times after the end of the ramp.}
    \label{fig:inelasticwall}
\end{figure}
Once the wave packet has been fully emitted from site $j=0$, we slowly ramp the transverse field $g$ from $0.5$ to $0.9$ (Fig.~\ref{fig:ramp}c), thereby entering a regime in which inelastic scattering processes are allowed. 
As a consequence of the ramp, the group velocity of the wave packet changes, matching the group velocity at the target momentum in the final Hamiltonian. 
We then let the wave packet scatter off the boundary to probe the inelastic process converting a 1-meson into a 2-meson. Importantly, no symmetry-based selection rules forbid such transitions. To characterize the resulting dynamics, we analyze the Fourier transform of a local observable. In this case, we focus on the operator $\sigma_j^x$, as we expect it to have sizeable matrix elements between different energy eigenstates.
Such off-diagonal matrix elements play an important role in this detection scheme, as already implied by Eq.~(\ref{eq:nj}).

In this signal, shown in Fig.~\ref{fig:inelasticwall}, we observe three distinct peaks: one located at $q = 2k_*$, as expected for elastic scattering, and two peaks at $q = k_* \pm k_s$, demonstrating that in the collision with the boundary the $1$-meson with momentum $+k_*$ was partially scattered into a $2$-meson with momentum $-k_s$. Similar peaks at these values of $q$ are observed for other observables, such as $\mathcal E_j$ or $\sigma_j^z$. However, in these cases the signal is noticeably less clean than for $\sigma_j^x$.

\subsection{Inelastic two-particle scattering}
Having demonstrated the state-preparation protocol based on adiabatic ramps, and after testing the inelastic scattering from a collision with a boundary, we now analyze the case of a two-particle inelastic scattering process.

\begin{figure}
    \centering
    \includegraphics[width=\linewidth]{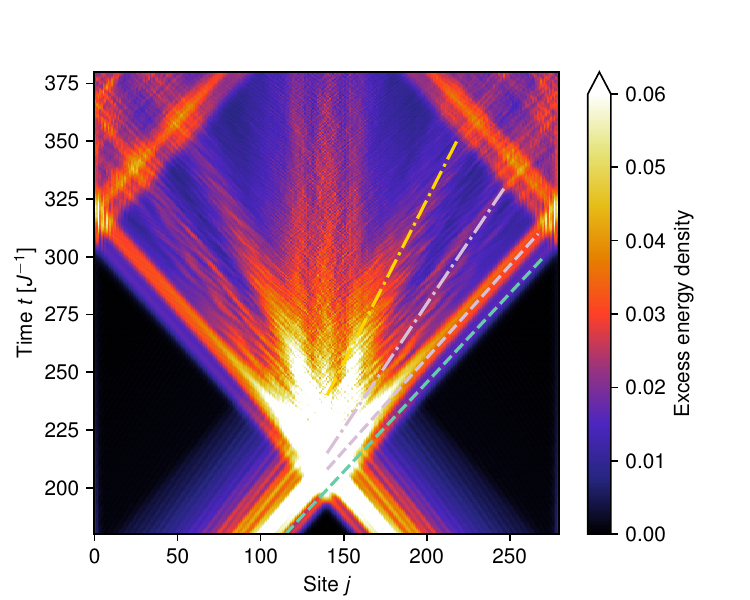}
    \caption{Inelastic two-particle collision. Two wave packets, prepared as in Fig.~\ref{fig:ramp} collide, resulting in a final state with multiple available channels: (i) a state with two $\ell=1$ mesons corresponding to elastic scattering (blue dashed line), (ii) a state with a $\ell=1$ (dashed pink line) and a $\ell=2$ (dotted-dashed pink line) mesons, (iii) a state with two $\ell=2$ mesons (dotted-dashed yellow line) (iv) states with three $\ell=1$ mesons (approximately vertical traces). Another possible channel, consisting of an $\ell=1$ and an $\ell=3$ meson (not shown), cannot be resolved reliably because its group velocity is too similar to those of the other channels. The bond dimension used in the MPS simulation is $\chi=350$.}
    \label{fig:inelastic_scatt}
\end{figure}

We employ the same ramp-based state preparation scheme as in the previous section, but we apply it at both ends of the chain to generate two incoming wave packets. The ramp brings the system into a regime where several inelastic channels are kinematically allowed (see Appendix \ref{app:disp} for a detailed analysis of the kinematics). The outcome of the collision is shown in Fig.~\ref{fig:inelastic_scatt}. In addition to the elastic channel with two $1$-mesons in the final state, we observe several outgoing traces consistent with inelastic processes: (i) a final state with a $1$-meson and a $2$-meson, (ii) a final state with two $2$-mesons, (iii) a continuum of states corresponding to three $1$-mesons. The contributions of channels (i) and (ii) can be distinguished from the elastic channel by their characteristic group velocities, which can be predicted from a simple kinematic analysis (see Appendix \ref{app:disp}).  We note, however, that the $(3+1)$ channel is also kinematically allowed. For the system sizes considered here, its contribution cannot be resolved because its group velocity is too similar to those of the other inelastic channels, and we therefore cannot exclude its presence. The kinematically allowed three-$1$-meson states involve mesons with very small momenta and, consequently, very small group velocities. As a result, these excitations produce nearly vertical traces in the spacetime plot in Fig.~\ref{fig:inelastic_scatt}.

\section{Controlled shaping of the wave packet}
\label{sec:slide}

While having a sharp momentum distribution, the wave packets produced with our method are not Gaussian. They tend to have a rather irregular shape and long tails. This is a drawback that adversely affects the sensitivity of detection. A wave packet with long tails corresponds to smaller local amplitudes in the relevant observables, like the excess energy density. Moreover, since in our ramp scheme we need to wait until the wave packet has been emitted completely before applying the ramp, we may have to wait for a very long time if the wave packet has a long trailing tail. In this section we show how to address this limitation and improve the shape of the wave packet.

To achieve this goal, we take inspiration, once again, from the simple single-particle model presented in Sec.~\ref{sec:SP}. Even in that case, our protocol produced wave packets with long tails and irregular shapes with wiggles, rather than Gaussian wave packets. A protocol that has previously been shown to produce nearly Gaussian wave packets is the so-called ``quantum slide''. The idea is to modulate the parameters of the model in a region near the edge of the chain: while in Sec.~\ref{sec:SP} the hopping amplitude and local potential have a sudden jump between $j=0$ and $j=1$, in a slide the parameters vary more smoothly in a region near the edge. A careful choice of the parameters along the slide is guaranteed to produce an approximate Gaussian wave packet (see App.~\ref{app:slide} and Refs.~\cite{Wang2020Integrated,wang2022quantum}).

\begin{figure}
    \centering
    \includegraphics[width=\linewidth]{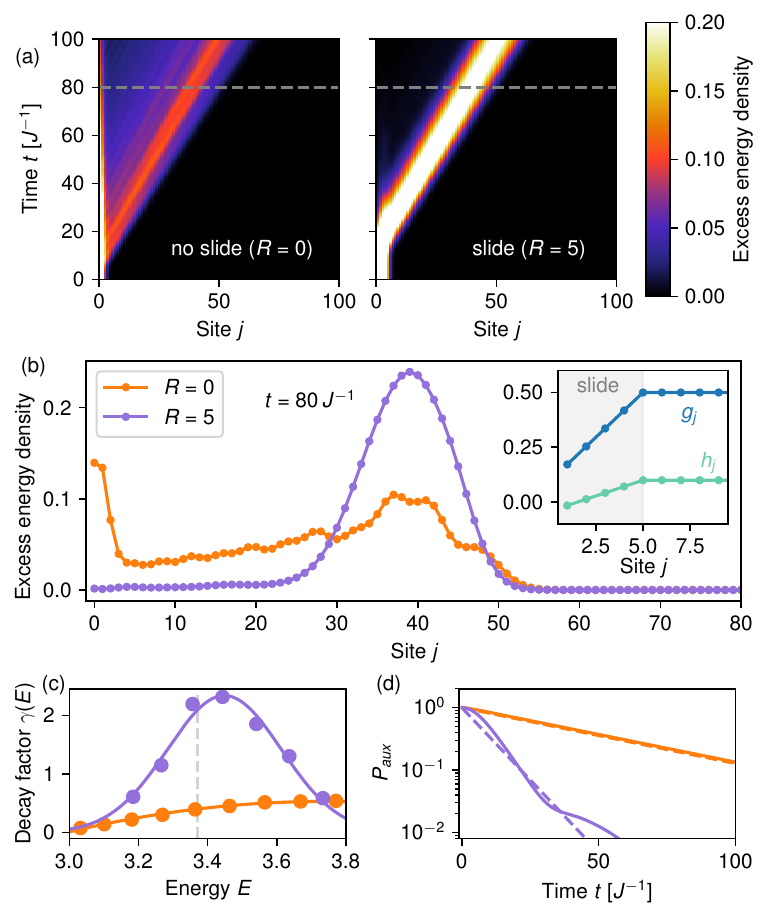}
    \caption{ Shaping of the wave packet. (a) Preparation of a wave packet in the Ising chain using the standard protocol (left) and the slide protocol (right). (b) Profile of the wave packet at time $t=80 J^{-1}$ using the two methods. The slide method produces a much more regular wave packet, with less spreading in real space and an enhanced amplitude at its center. The inset shows the site-dependent parameters employed for the slide of size $R=5$. (c) Decay factor $\gamma(E)$, computed using exact diagonalization (dots) and fitted (solid lines, see App.~\ref{app:decay}). In the slide protocol (purple), the decay factor $\gamma(E_0)$ is enhanced at the energy $E_0$ of the target wave packet (indicated by a vertical dashed line), resulting in a faster decay of the auxiliary degree of freedom and, consequently, a narrower wave packet. (d) Population of the excited state of the auxiliary qubit for the two protocols. The exponential decays with rates $\Gamma=2\pi\sigma^2\gamma(E)$ are shown as dashed lines for the two protocols (with and without a slide) for the same energy variance $\sigma^2$.  }
    \label{fig:shaping}
\end{figure}

We here would like to employ a similar scheme to control the shape of the wave packets in the many-body case, as exemplified by the mixed-field quantum Ising chain. In contrast with the single-particle case, where the shape of the slide can be justified analytically, in the many-body case we do not have a similarly rigorous approach. We choose therefore a simple linear slide of size $R$, where the (site-dependent) transverse and longitudinal fields take the form \footnote{Explicitly, the Hamiltonian is 
$H=-J\sum_{j=1}^{L-1}\sigma_j^z\sigma_{j+1}^z+\sum_{j=1}^{L}(g_j\sigma_j^x+h_j\sigma_{j}^z)$.
} (see inset in Fig.~\ref{fig:shaping}b)
\begin{equation}
    g_j =\begin{cases}
    g_0+\frac{j}{R}(g-g_0) & \text{for $1\le j\le R$,}\\
    g & \text{for $j>R$,}
    \end{cases}
\end{equation}
\begin{equation}
    h_j =\begin{cases}
    h_0-1 +\frac{j}{R}(h-h_0+1) & \text{for $1\le j\le R$,}\\
    h & \text{for $j>R$.}
    \end{cases}
\end{equation}
Here $g$ and $h$ are chosen to match the values in the interior of the system, while $h_0$ and $g_0$ are the (post-quench) values of the longitudinal field and transverse field on the auxiliary site in Eq.~(\ref{eq:Isingprep}) \footnote{It may seem natural to choose a linear interpolation from $h_0$ to $h$ for the longitudinal field. However, one must take into account that the auxiliary site, unlike the sites in the chain, has only a single neighbor rather than two. Ensuring a smooth evolution of the excitation energy along the interpolation therefore requires accounting for this asymmetry, modeling such effect as an additional longitudinal field. To obtain a smooth potential for the meson, we then design a ramp that linearly interpolates between $h_0-1$ and $h$. We find empirically that the choice of interpolation reported here performs well for our chosen value of $h_0$, although a different choice may be required for other values. }. 

The results of the wave packet preparation using the previous protocol (without the slide) and the modified protocol with a slide of size $R=5$ are compared in Fig.~\ref{fig:shaping}a-b. In the slide protocol, the wave packet is significantly narrower, with a larger local amplitude and a more regular shape. As shown in Fig.~\ref{fig:shaping}b, in the slide protocol the wave packet is fully emitted from the first site at a time when, in the simple protocol, it has not yet been emitted. This represents a major practical advantage, as it substantially reduces the waiting time before applying the adiabatic ramp.

\subsection{Analysis of the quantum slide method}

To gain a theoretical understanding of these numerical observations, we analyze the time required to eject the wave packet from the auxiliary site. This timescale, multiplied by the group velocity, sets the spatial extent of the wave packet. It cannot be made arbitrarily short, since the minimal spatial extent is constrained by the inverse of the momentum uncertainty. Our goal is therefore to minimize the decay time of the auxiliary site, subject to a fixed energy variance (which determines the momentum uncertainty). This optimization leads to a wave packet that is more localized in real space without increasing its momentum spread, thereby yielding a profile closer to a Gaussian wave packet.

We now show how to compute the decay rate. As before, the system is initialized in the ground state with a large negative $h_0$: The ground state $\ket{\psi_0}$ of the combined system and auxiliary site has the latter in the $\uparrow$ state, while the system is in the ground state of $H - J \sigma_1^z$, i.e., the ground state subject to the boundary condition imposed by the auxiliary site. We denote this state by $\ket{GS_\uparrow}$:
\begin{equation}
    (H-J\sigma_1^z)\ket{GS_\uparrow}=\epsilon_\uparrow^{(GS)}.
\end{equation}
The combined system is thus in the initial state
\begin{equation}
    \ket{\psi_0}=\ket{\uparrow}\otimes \ket{GS_\uparrow},
\end{equation}
and evolves after the local quench under $H_{\text{prep}}=H'+V$, where
\begin{equation}
    H'=H+h_0\sigma_0^z-J\sigma_0^z\sigma_1^z, 
\end{equation}
and 
\begin{equation}
    V=g_0\sigma_0^x.
\end{equation}

The initial state $\ket{\psi_0}$ is an eigenstate of $H'$ with energy $E_0=h_0+\epsilon_\uparrow^{(GS)}$ (although not its ground state), and $V$ can be treated as a weak perturbation. The spectrum of $H'$ separates into two sectors, depending on whether the auxiliary site is in the $\uparrow$ or $\downarrow$ state. In the post-quench Hamiltonian, the parameters are chosen such that the ground state lies in the $\downarrow$ sector. For ease of notation, we set the zero of the energy as the ground state energy of $H'$, i.e., $-h_0+\epsilon_\downarrow^{(GS)}=0$ (note that this choice needs to be accounted for in the definition of $\epsilon_\uparrow^{(GS)}$). Here $-h_0$ is the energy of the auxiliary site in the $\ket{\downarrow}$ state, $\epsilon_\downarrow^{(GS)}$ is the ground state energy of $H+J\sigma_1^z$ (i.e., the Hamiltonian of the system subject to the boundary condition corresponding to having the auxiliary site in the $\downarrow$ state).

As before, a key requirement of our protocol is that the energy $E_0$ of $\ket{\psi_0}$ matches the energy of the single-particle excitation at the desired momentum.
The decay rate of the auxiliary excitation into a bulk excitation can then be estimated using Fermi's golden rule:

\begin{equation}
\label{eq:Gamma}
    \Gamma = 2\pi |\bra{E_0}V\ket{\psi_0}|^2\rho(E_0),
\end{equation}
where $\ket{E}$ are eigenstates of $H'$ with energy $E$, and $\rho(E)$ is the density of states.

On the other hand, the energy variance of the initial state is
\begin{equation}
    \label{eq:sigma}\sigma^2(H_\text{prep})=\bra{\psi_0}V^2\ket{\psi_0}=\int dE\, |\braket{E|V|\psi_0}|^2 \rho(E),
\end{equation}
where we have used that $\braket{\psi_0|V|\psi_0}=0$.
Comparing Eqs.~(\ref{eq:Gamma}) and (\ref{eq:sigma}), we see that enhancing the decay rate $\Gamma$ while keeping the energy variance fixed requires $V$ to couple $\ket{\psi_0}$ preferentially to states at the target energy. In the standard language of a small system (here, the auxiliary qubit) coupled to a large bath (the chain), this amounts to requiring that the \emph{spectral density} of the system-bath coupling be peaked at the energy of the target excitation.
Although this argument was presented in the context of the Ising model, the same considerations apply more generally to this class of protocols.

We now examine the decay rate explicitly in this model. The relevant matrix elements are those between $\ket{\psi_0}$ and states in the opposite sector, $\ket{E} = \ket{\downarrow} \otimes \ket{\epsilon_\downarrow}$, where $\ket{\epsilon_\downarrow}$ is an eigenstate of $H + J\sigma_1^z$  with eigenvalue $\epsilon_\downarrow(E)=E+h_0$ (such that $H'\ket{E}=E\ket{E}$):
\begin{equation}
    \bra{E}V\ket{\psi_0}= g_0\braket{\epsilon_\downarrow|GS_{\uparrow}}.
\end{equation}
The corresponding decay rate is therefore
\begin{equation}
\label{eq:rate}
    \Gamma=2\pi\sigma^2 \gamma(E_0),
\end{equation}
where $\sigma^2=\braket{\psi_0|V^2|\psi_0}=|g_0|^2$ is the energy variance after the local quench,
and
\begin{equation}
   \gamma(E_0) =|\braket{\epsilon_{\downarrow,0} | GS_\uparrow}|^2 \, \rho_\downarrow(\epsilon_{\downarrow,0}). 
\end{equation}
Here  $\rho_\downarrow$ is the density of states of the Hamiltonian $H+J\sigma_1^z$, and $\epsilon_{\downarrow,0}=\epsilon_{\downarrow}(E_0)=E_0+h_0$. Note that $\gamma(E_0)$ does not depend on $g_0$, and therefore the decay factor $\gamma(E_0)$ can be tuned independently from the energy variance. 

The slide protocol modifies the factor $ \gamma(E)$,  
effectively enhancing the spectral density at the target energy $E_0$, as shown in Fig.~\ref{fig:shaping}c (see App.~\ref{app:decay} for details on the calculations). As a result, for a fixed energy variance $\sigma^2 = |g_0|^2$, the decay rate of the auxiliary qubit can be significantly increased, as illustrated in Fig.~\ref{fig:shaping}d.

We now comment on the generality of this result and on how to design quantum slides more broadly. The key intuition is that the parameters should vary smoothly in space, rather than exhibiting the discontinuities present in the absence of a slide. When the variation is sufficiently smooth, a local density approximation can be employed to analyze how the dispersion relation depends on position. 
Within this framework, the slide should be designed so that the energy after the local quench is never in a gapped region of the (local) spectrum. This ensures that an excitation can propagate resonantly across the slide. A detailed investigation of this approach and of the optimal slide design is left for future work.

\subsection{Scattering}
\begin{figure}
    \centering
    \includegraphics[width=\linewidth]{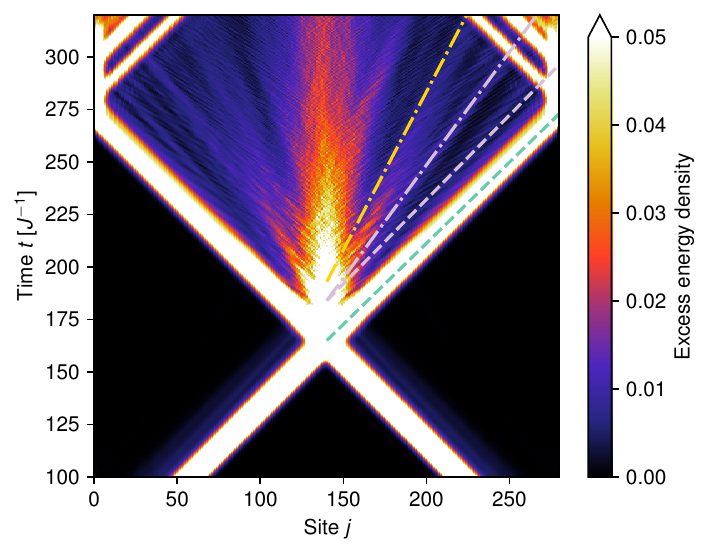}
    \caption{Two-particle scattering following wave packet preparation by the quantum slide method and the adiabatic ramp. The simulation is performed in the same regime as Fig.~\ref{fig:inelastic_scatt}. The elastic and inelastic scattering channels are visible, although the inelastic channels are less pronounced than in Fig.~\ref{fig:inelastic_scatt}. The bond dimension used in the MPS simulation is $\chi=350$.}
    \label{fig:slide-ramp-scatter}
\end{figure}

We now apply the quantum slide method for wave packet shaping, as demonstrated in Fig.~\ref{fig:shaping}, followed by the adiabatic ramp, and simulate the two-particle scattering in the same regime as in Fig.~\ref{fig:inelastic_scatt}. The results are shown in Fig.~\ref{fig:slide-ramp-scatter}. The scattering channels remain visible in this case; however, the inelastic channels are less prominent. The origin of this suppression is not yet understood and is left for future investigation.

\section{Higher dimensions}
\label{sec:2D}
Our protocol can be straightforwardly extended to higher dimensions. For instance, auxiliary sites can be placed at opposite corners of a square lattice. As in the one-dimensional case, we use conservation of energy and energy variance to constrain the momentum of the propagating wave packet.

\begin{figure}
    \centering
    \includegraphics[width=\linewidth]{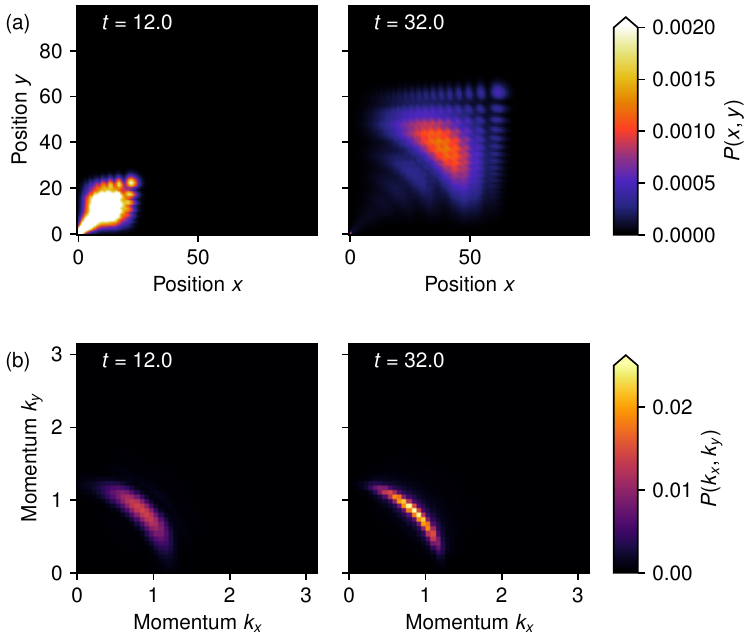}
    \caption{
    Wave packet preparation in the two-dimensional single-particle model. Parameters: $w=1, w'=0.4, V_0=-2.6$. 
(a) Probability distribution of the particle position at times $t=12.0$ and $t=32.0$. The particle is initially localized at a corner of the lattice, with energy $V_0 = E(\vec{k}_*)$. The corner site is weakly coupled to the rest of the system via small hopping amplitudes. As time evolves, the wave packet propagates from the initial corner toward the opposite corner of the lattice. 
(b) Probability distribution of the particle momentum at times $t=12.0$ and $t=32.0$. The momentum distribution becomes peaked at values of $\vec{k}$ satisfying $E(\vec{k}) = V_0$. The resulting wave packet is therefore characterized by a well-defined magnitude $|\vec{k}|$, while remaining spread over the angular direction in the $(k_x, k_y)$ plane. }
    \label{fig:2D}
\end{figure}

A key difference from the one-dimensional case is that energy conservation no longer selects a single momentum $\vec{k}$, but rather a manifold of momenta. In systems with emergent rotational symmetry, this constant energy contour reduces approximately to a fixed-$|\vec{k}|$ arc at low energy. This is illustrated in Fig.~\ref{fig:2D} for the two-dimensional tight-binding model with hopping amplitude $w$ along both the $x$ and $y$ directions.

In the initial state, the particle is localized at a corner site. The hopping amplitudes from this site to its neighbors, denoted $w'$, can be tuned to control the momentum spread of the resulting wave packet. The onsite energy of the corner site determines the energy of the state, and hence the target value of $|\vec{k}|$.

As shown in Fig.~\ref{fig:2D}a, the wave packet propagates from the corner toward the center of the system. During this evolution, its momentum distribution becomes peaked along an arc with fixed $|\vec{k}| \approx 1.27$ and arbitrary angle (Fig.~\ref{fig:2D}b). These momentum values correspond to those satisfying $E(\vec{k}) \approx V_0$. As shown in App.~\ref{app:2Ddet}, the momentum components can be measured using a method analogous to that employed in the one-dimensional case, based on reflection from a boundary.

\begin{figure}
    \centering
    \includegraphics[width=\linewidth]{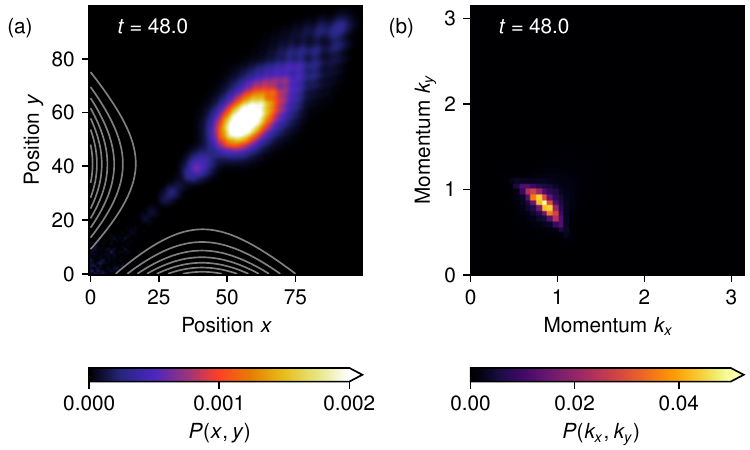}
    \caption{ Wave packet preparation using a focusing potential near the particle-injection corner (site $x=0, y=0$). The other parameters are the same as in Fig.~\ref{fig:2D}. (a) Probability distribution of the particle position at time $t=48.0$. The grey lines indicate equipotential contours corresponding to potential values $4, 8, 12, \dots$. (b) Probability distribution of the particle momentum at time $t=48.0$. The focusing potential reduces the uncertainty in the emission angle, resulting in a more localized momentum distribution. }
    \label{fig:funnel}
\end{figure}

This protocol could be optimized to achieve a better control of the momentum distribution and target a specific angle. Possible strategies include modifying the lattice geometry, by attaching a one-dimensional system that acts as a waveguide, or by introducing a funnel-like structure at the corner. Alternatively, a site-dependent potential could be used to bias the particle toward a given direction. A demonstration of the latter scheme is shown in Fig.~\ref{fig:funnel}a, where the focusing potential is illustrated via its equipotential lines. The potential near the corner is chosen to be flat along a diagonal region and to increase away from it, thereby energetically favoring propagation in the diagonal direction. This funnel-like profile gradually opens up, so that the emitted wave packet eventually evolves in a homogeneous two-dimensional region. The resulting wave packet is less dispersed in the transverse direction and exhibits a more sharply defined propagation angle in momentum space (Fig.~\ref{fig:funnel}b), leading to improved momentum resolution.

\section{Conclusions}

We propose and numerically validate a protocol for the preparation and detection of particle collisions. The protocol is compatible with currently available quantum simulation platforms and can be implemented on both analog simulators and digital simulators via Trotterization.

A key consideration is the resource cost associated with the implementation. In our scheme, the time required to generate a wave packet is fundamentally limited by the desired momentum resolution. Specifically, achieving a small momentum uncertainty $\delta k$ requires a wave packet of spatial width $W \gtrsim \delta k^{-1}$. Since the packet has to be emitted from the boundary, the corresponding preparation time scales as $\Delta t \sim W/v \gtrsim (v\,\delta k)^{-1}$, where $v$ denotes the group velocity at the target momentum. In general, the width $W$ is larger than the optimal value $\sim \delta k^{-1}$, but we have shown that the use of a quantum slide enables the preparation of significantly narrower wave packets, approaching the optimal width.

Another contribution to the preparation cost is the adiabatic ramp, which is used to reach the regime of inelastic scattering, where the mass gap is much smaller than the bandwidth. Accessing inelastic channels with $n$ particles in the final state requires a bandwidth-to-mass-gap ratio of order $n$, making small mass gaps necessary for probing highly inelastic processes. However, as the mass gap decreases, maintaining adiabaticity becomes increasingly challenging. The adiabatic condition requires the rate of change of the control parameter to scale no faster than the square of the instantaneous energy gap. Consequently, preparing states in the small-gap regime demands progressively slower ramps, making the adiabatic evolution a potentially significant contribution to the overall state-preparation time. (Note, however, that this limitation does not arise specifically from the wave packet preparation itself, but rather from the more general problem of ground-state preparation.) Moreover, this strategy is only applicable provided the target regime is adiabatically connected to a parameter regime with an isolated single-particle band. In particular, the interpolation path must avoid quantum phase transitions or other gap-closing points that would invalidate adiabatic state preparation.

Another important resource is the system size. While elastic scattering processes can be investigated in relatively small systems, the observation of inelastic scattering is considerably more demanding. To unambiguously identify the different scattering channels, the outgoing particles must propagate sufficiently far apart after the scattering event so that the final states can be resolved. This limitation is not specific to our protocol; rather, it reflects a fundamental requirement of scattering theory, namely that asymptotic particle states must be well separated in order to define and characterize scattering outcomes.

These considerations suggest that generating wave packets at the two ends of a large system may not always be the most efficient strategy. Since the packets must travel a substantial distance before colliding, the protocol requires a correspondingly long evolution time. A potentially more efficient approach is to create the wave packets closer to the center of the chain by introducing temporary, fictitious boundaries. Once the desired wave packets have been prepared, these boundaries can be removed adiabatically, allowing the outgoing particles to evolve in the full system after the collision. Such a scheme could substantially reduce the overall state-preparation time.

Additional reductions in the required preparation time may be achieved through further optimization of the protocol. In particular, one could improve the design of the adiabatic ramp used to generate the wave packets or refine the implementation of the quantum slide to approach the optimal packet shape more efficiently. Exploring such optimizations is an interesting direction for future work and could further enhance the practicality of scattering experiments on quantum simulators.

Beyond the study of asymptotic states and the scattering matrix, investigating the transient dynamics immediately after a collision is another interesting direction. In particular, the emergence of hydrodynamic behavior as a localized lump of energy spreads and cools raises important questions in the high-energy regime, already before the lump fragments into well-separated particles. Such physics can be accessed at relatively short times after the collision and represents a promising avenue for near-term investigation on quantum simulation platforms.

Finally, testing our protocols in a two-dimensional many-body system is another interesting direction for future work. While tensor-network simulations in higher dimensions are considerably more computationally demanding than in one dimension, recent studies have demonstrated the simulation of wave packet dynamics on lattices as large as $24\times 24$ \cite{pavevsic2026scattering}. Such system sizes may already be sufficient to verify the preparation of wave packets that are relatively narrow in momentum space.

\begin{acknowledgments}
We acknowledge useful discussions with Zohreh Davoudi, Manuel Endres, Roland Farrell, and Alessio Lerose. This work was supported by RIT (Research IT, Trinity College Dublin). FMS and JP acknowledge support provided by the U.S.\ Department of Energy Office of Science, Office of Advanced Scientific Computing Research, (DE-SC0025572); DOE QuantISED program through the theory consortium ``Intersections of QIS and Theoretical Particle Physics'' at Fermilab; DOE National Quantum Information Science Research Centers, Quantum Systems Accelerator; and by the Institute for Quantum Information and Matter, an NSF Physics Frontiers Center (PHY-2317110). FMS acknowledges support from Amazon Web Services, AWS Quantum Program. This research was supported in part by grant NSF PHY-2309135 to the Kavli Institute for Theoretical Physics (KITP). SB acknowledges support from Caltech's Summer Undergraduate Research Fellowship (SURF) program. Calculations were performed using the TeNPy Library (version 1.0.0) \cite{tenpy2024}.
\end{acknowledgments}

\bibliography{bib}

\begin{thebibliography}{71}%
\makeatletter
\providecommand \@ifxundefined [1]{%
 \@ifx{#1\undefined}
}%
\providecommand \@ifnum [1]{%
 \ifnum #1\expandafter \@firstoftwo
 \else \expandafter \@secondoftwo
 \fi
}%
\providecommand \@ifx [1]{%
 \ifx #1\expandafter \@firstoftwo
 \else \expandafter \@secondoftwo
 \fi
}%
\providecommand \natexlab [1]{#1}%
\providecommand \enquote  [1]{``#1''}%
\providecommand \bibnamefont  [1]{#1}%
\providecommand \bibfnamefont [1]{#1}%
\providecommand \citenamefont [1]{#1}%
\providecommand \href@noop [0]{\@secondoftwo}%
\providecommand \href [0]{\begingroup \@sanitize@url \@href}%
\providecommand \@href[1]{\@@startlink{#1}\@@href}%
\providecommand \@@href[1]{\endgroup#1\@@endlink}%
\providecommand \@sanitize@url [0]{\catcode `\\12\catcode `\$12\catcode `\&12\catcode `\#12\catcode `\^12\catcode `\_12\catcode `\%12\relax}%
\providecommand \@@startlink[1]{}%
\providecommand \@@endlink[0]{}%
\providecommand \url  [0]{\begingroup\@sanitize@url \@url }%
\providecommand \@url [1]{\endgroup\@href {#1}{\urlprefix }}%
\providecommand \urlprefix  [0]{URL }%
\providecommand \Eprint [0]{\href }%
\providecommand \doibase [0]{https://doi.org/}%
\providecommand \selectlanguage [0]{\@gobble}%
\providecommand \bibinfo  [0]{\@secondoftwo}%
\providecommand \bibfield  [0]{\@secondoftwo}%
\providecommand \translation [1]{[#1]}%
\providecommand \BibitemOpen [0]{}%
\providecommand \bibitemStop [0]{}%
\providecommand \bibitemNoStop [0]{.\EOS\space}%
\providecommand \EOS [0]{\spacefactor3000\relax}%
\providecommand \BibitemShut  [1]{\csname bibitem#1\endcsname}%
\let\auto@bib@innerbib\@empty
\bibitem [{\citenamefont {Daley}\ \emph {et~al.}(2022)\citenamefont {Daley}, \citenamefont {Bloch}, \citenamefont {Kokail}, \citenamefont {Flannigan}, \citenamefont {Pearson}, \citenamefont {Troyer},\ and\ \citenamefont {Zoller}}]{Daley_2022}%
  \BibitemOpen
  \bibfield  {author} {\bibinfo {author} {\bibfnamefont {A.~J.}\ \bibnamefont {Daley}}, \bibinfo {author} {\bibfnamefont {I.}~\bibnamefont {Bloch}}, \bibinfo {author} {\bibfnamefont {C.}~\bibnamefont {Kokail}}, \bibinfo {author} {\bibfnamefont {S.}~\bibnamefont {Flannigan}}, \bibinfo {author} {\bibfnamefont {N.}~\bibnamefont {Pearson}}, \bibinfo {author} {\bibfnamefont {M.}~\bibnamefont {Troyer}},\ and\ \bibinfo {author} {\bibfnamefont {P.}~\bibnamefont {Zoller}},\ }\bibfield  {title} {\bibinfo {title} {Practical quantum advantage in quantum simulation},\ }\href {https://www.nature.com/articles/s41586-022-04940-6#citeas} {\bibfield  {journal} {\bibinfo  {journal} {Nature News}\ } (\bibinfo {year} {2022})}\BibitemShut {NoStop}%
\bibitem [{\citenamefont {Eisert}\ and\ \citenamefont {Preskill}(2025)}]{eisert2025mind}%
  \BibitemOpen
  \bibfield  {author} {\bibinfo {author} {\bibfnamefont {J.}~\bibnamefont {Eisert}}\ and\ \bibinfo {author} {\bibfnamefont {J.}~\bibnamefont {Preskill}},\ }\bibfield  {title} {\bibinfo {title} {Mind the gaps: The fraught road to quantum advantage},\ }\href {https://doi.org/10.48550/arXiv.2510.19928} {\bibfield  {journal} {\bibinfo  {journal} {arXiv preprint arXiv:2510.19928}\ } (\bibinfo {year} {2025})}\BibitemShut {NoStop}%
\bibitem [{\citenamefont {Jordan}\ \emph {et~al.}(2014)\citenamefont {Jordan}, \citenamefont {Lee},\ and\ \citenamefont {Preskill}}]{jordan2011quantum}%
  \BibitemOpen
  \bibfield  {author} {\bibinfo {author} {\bibfnamefont {S.~P.}\ \bibnamefont {Jordan}}, \bibinfo {author} {\bibfnamefont {K.~S.~M.}\ \bibnamefont {Lee}},\ and\ \bibinfo {author} {\bibfnamefont {J.}~\bibnamefont {Preskill}},\ }\bibfield  {title} {\bibinfo {title} {Quantum computation of scattering in scalar quantum field theories},\ }\href@noop {} {\bibfield  {journal} {\bibinfo  {journal} {Quantum Info. Comput.}\ }\textbf {\bibinfo {volume} {14}},\ \bibinfo {pages} {1014–1080} (\bibinfo {year} {2014})}\BibitemShut {NoStop}%
\bibitem [{\citenamefont {Jordan}\ \emph {et~al.}(2012)\citenamefont {Jordan}, \citenamefont {Lee},\ and\ \citenamefont {Preskill}}]{Jordan2012}%
  \BibitemOpen
  \bibfield  {author} {\bibinfo {author} {\bibfnamefont {S.~P.}\ \bibnamefont {Jordan}}, \bibinfo {author} {\bibfnamefont {K.~S.~M.}\ \bibnamefont {Lee}},\ and\ \bibinfo {author} {\bibfnamefont {J.}~\bibnamefont {Preskill}},\ }\bibfield  {title} {\bibinfo {title} {Quantum algorithms for quantum field theories},\ }\href {https://doi.org/10.1126/science.1217069} {\bibfield  {journal} {\bibinfo  {journal} {Science}\ }\textbf {\bibinfo {volume} {336}},\ \bibinfo {pages} {1130} (\bibinfo {year} {2012})}\BibitemShut {NoStop}%
\bibitem [{\citenamefont {Preskill}(2018)}]{preskill2018simulating}%
  \BibitemOpen
  \bibfield  {author} {\bibinfo {author} {\bibfnamefont {J.}~\bibnamefont {Preskill}},\ }\bibfield  {title} {\bibinfo {title} {Simulating quantum field theory with a quantum computer},\ }\href@noop {} {\bibfield  {journal} {\bibinfo  {journal} {arXiv preprint arXiv:1811.10085}\ } (\bibinfo {year} {2018})}\BibitemShut {NoStop}%
\bibitem [{\citenamefont {Bauer}\ \emph {et~al.}(2023)\citenamefont {Bauer}, \citenamefont {Davoudi}, \citenamefont {Balantekin}, \citenamefont {Bhattacharya}, \citenamefont {Carena}, \citenamefont {de~Jong}, \citenamefont {Draper}, \citenamefont {El-Khadra}, \citenamefont {Gemelke}, \citenamefont {Hanada}, \citenamefont {Kharzeev}, \citenamefont {Lamm}, \citenamefont {Li}, \citenamefont {Liu}, \citenamefont {Lukin}, \citenamefont {Meurice}, \citenamefont {Monroe}, \citenamefont {Nachman}, \citenamefont {Pagano}, \citenamefont {Preskill}, \citenamefont {Rinaldi}, \citenamefont {Roggero}, \citenamefont {Santiago}, \citenamefont {Savage}, \citenamefont {Siddiqi}, \citenamefont {Siopsis}, \citenamefont {Van~Zanten}, \citenamefont {Wiebe}, \citenamefont {Yamauchi}, \citenamefont {Yeter-Aydeniz},\ and\ \citenamefont {Zorzetti}}]{Bauer2023}%
  \BibitemOpen
  \bibfield  {author} {\bibinfo {author} {\bibfnamefont {C.~W.}\ \bibnamefont {Bauer}}, \bibinfo {author} {\bibfnamefont {Z.}~\bibnamefont {Davoudi}}, \bibinfo {author} {\bibfnamefont {A.~B.}\ \bibnamefont {Balantekin}}, \bibinfo {author} {\bibfnamefont {T.}~\bibnamefont {Bhattacharya}}, \bibinfo {author} {\bibfnamefont {M.}~\bibnamefont {Carena}}, \bibinfo {author} {\bibfnamefont {W.~A.}\ \bibnamefont {de~Jong}}, \bibinfo {author} {\bibfnamefont {P.}~\bibnamefont {Draper}}, \bibinfo {author} {\bibfnamefont {A.}~\bibnamefont {El-Khadra}}, \bibinfo {author} {\bibfnamefont {N.}~\bibnamefont {Gemelke}}, \bibinfo {author} {\bibfnamefont {M.}~\bibnamefont {Hanada}}, \bibinfo {author} {\bibfnamefont {D.}~\bibnamefont {Kharzeev}}, \bibinfo {author} {\bibfnamefont {H.}~\bibnamefont {Lamm}}, \bibinfo {author} {\bibfnamefont {Y.-Y.}\ \bibnamefont {Li}}, \bibinfo {author} {\bibfnamefont {J.}~\bibnamefont {Liu}}, \bibinfo {author} {\bibfnamefont {M.}~\bibnamefont {Lukin}}, \bibinfo {author} {\bibfnamefont {Y.}~\bibnamefont {Meurice}}, \bibinfo {author} {\bibfnamefont {C.}~\bibnamefont {Monroe}}, \bibinfo {author} {\bibfnamefont {B.}~\bibnamefont {Nachman}}, \bibinfo {author} {\bibfnamefont {G.}~\bibnamefont {Pagano}}, \bibinfo {author} {\bibfnamefont {J.}~\bibnamefont {Preskill}}, \bibinfo {author} {\bibfnamefont {E.}~\bibnamefont {Rinaldi}}, \bibinfo {author} {\bibfnamefont {A.}~\bibnamefont {Roggero}}, \bibinfo {author} {\bibfnamefont {D.~I.}\ \bibnamefont {Santiago}}, \bibinfo {author} {\bibfnamefont {M.~J.}\ \bibnamefont {Savage}}, \bibinfo {author} {\bibfnamefont {I.}~\bibnamefont {Siddiqi}}, \bibinfo {author} {\bibfnamefont {G.}~\bibnamefont {Siopsis}}, \bibinfo {author} {\bibfnamefont {D.}~\bibnamefont {Van~Zanten}}, \bibinfo {author} {\bibfnamefont {N.}~\bibnamefont {Wiebe}}, \bibinfo {author} {\bibfnamefont {Y.}~\bibnamefont {Yamauchi}}, \bibinfo {author} {\bibfnamefont {K.}~\bibnamefont {Yeter-Aydeniz}},\ and\ \bibinfo {author} {\bibfnamefont {S.}~\bibnamefont {Zorzetti}},\ }\bibfield  {title} {\bibinfo {title} {Quantum simulation for high-energy physics},\ }\href {https://doi.org/10.1103/PRXQuantum.4.027001} {\bibfield  {journal} {\bibinfo  {journal} {PRX Quantum}\ }\textbf {\bibinfo {volume} {4}},\ \bibinfo {pages} {027001} (\bibinfo {year} {2023})}\BibitemShut {NoStop}%
\bibitem [{\citenamefont {Wang}\ \emph {et~al.}(2024)\citenamefont {Wang}, \citenamefont {Du}, \citenamefont {Zuo},\ and\ \citenamefont {Vary}}]{Wang2024}%
  \BibitemOpen
  \bibfield  {author} {\bibinfo {author} {\bibfnamefont {P.}~\bibnamefont {Wang}}, \bibinfo {author} {\bibfnamefont {W.}~\bibnamefont {Du}}, \bibinfo {author} {\bibfnamefont {W.}~\bibnamefont {Zuo}},\ and\ \bibinfo {author} {\bibfnamefont {J.~P.}\ \bibnamefont {Vary}},\ }\bibfield  {title} {\bibinfo {title} {Nuclear scattering via quantum computing},\ }\href {https://doi.org/10.1103/PhysRevC.109.064623} {\bibfield  {journal} {\bibinfo  {journal} {Phys. Rev. C}\ }\textbf {\bibinfo {volume} {109}},\ \bibinfo {pages} {064623} (\bibinfo {year} {2024})}\BibitemShut {NoStop}%
\bibitem [{\citenamefont {Di~Meglio}\ \emph {et~al.}(2024)\citenamefont {Di~Meglio}, \citenamefont {Jansen}, \citenamefont {Tavernelli}, \citenamefont {Alexandrou}, \citenamefont {Arunachalam}, \citenamefont {Bauer}, \citenamefont {Borras}, \citenamefont {Carrazza}, \citenamefont {Crippa}, \citenamefont {Croft}, \citenamefont {de~Putter}, \citenamefont {Delgado}, \citenamefont {Dunjko}, \citenamefont {Egger}, \citenamefont {Fern\'andez-Combarro}, \citenamefont {Fuchs}, \citenamefont {Funcke}, \citenamefont {Gonz\'alez-Cuadra}, \citenamefont {Grossi}, \citenamefont {Halimeh}, \citenamefont {Holmes}, \citenamefont {K\"uhn}, \citenamefont {Lacroix}, \citenamefont {Lewis}, \citenamefont {Lucchesi}, \citenamefont {Martinez}, \citenamefont {Meloni}, \citenamefont {Mezzacapo}, \citenamefont {Montangero}, \citenamefont {Nagano}, \citenamefont {Pascuzzi}, \citenamefont {Radescu}, \citenamefont {Ortega}, \citenamefont {Roggero}, \citenamefont {Schuhmacher}, \citenamefont {Seixas}, \citenamefont {Silvi}, \citenamefont {Spentzouris}, \citenamefont {Tacchino}, \citenamefont {Temme}, \citenamefont {Terashi}, \citenamefont {Tura}, \citenamefont {T\"uys\"uz}, \citenamefont {Vallecorsa}, \citenamefont {Wiese}, \citenamefont {Yoo},\ and\ \citenamefont {Zhang}}]{DiMeglio2024}%
  \BibitemOpen
  \bibfield  {author} {\bibinfo {author} {\bibfnamefont {A.}~\bibnamefont {Di~Meglio}}, \bibinfo {author} {\bibfnamefont {K.}~\bibnamefont {Jansen}}, \bibinfo {author} {\bibfnamefont {I.}~\bibnamefont {Tavernelli}}, \bibinfo {author} {\bibfnamefont {C.}~\bibnamefont {Alexandrou}}, \bibinfo {author} {\bibfnamefont {S.}~\bibnamefont {Arunachalam}}, \bibinfo {author} {\bibfnamefont {C.~W.}\ \bibnamefont {Bauer}}, \bibinfo {author} {\bibfnamefont {K.}~\bibnamefont {Borras}}, \bibinfo {author} {\bibfnamefont {S.}~\bibnamefont {Carrazza}}, \bibinfo {author} {\bibfnamefont {A.}~\bibnamefont {Crippa}}, \bibinfo {author} {\bibfnamefont {V.}~\bibnamefont {Croft}}, \bibinfo {author} {\bibfnamefont {R.}~\bibnamefont {de~Putter}}, \bibinfo {author} {\bibfnamefont {A.}~\bibnamefont {Delgado}}, \bibinfo {author} {\bibfnamefont {V.}~\bibnamefont {Dunjko}}, \bibinfo {author} {\bibfnamefont {D.~J.}\ \bibnamefont {Egger}}, \bibinfo {author} {\bibfnamefont {E.}~\bibnamefont {Fern\'andez-Combarro}}, \bibinfo {author} {\bibfnamefont {E.}~\bibnamefont {Fuchs}}, \bibinfo {author} {\bibfnamefont {L.}~\bibnamefont {Funcke}}, \bibinfo {author} {\bibfnamefont {D.}~\bibnamefont {Gonz\'alez-Cuadra}}, \bibinfo {author} {\bibfnamefont {M.}~\bibnamefont {Grossi}}, \bibinfo {author} {\bibfnamefont {J.~C.}\ \bibnamefont {Halimeh}}, \bibinfo {author} {\bibfnamefont {Z.}~\bibnamefont {Holmes}}, \bibinfo {author} {\bibfnamefont {S.}~\bibnamefont {K\"uhn}}, \bibinfo {author} {\bibfnamefont {D.}~\bibnamefont {Lacroix}}, \bibinfo {author} {\bibfnamefont {R.}~\bibnamefont {Lewis}}, \bibinfo {author} {\bibfnamefont {D.}~\bibnamefont {Lucchesi}}, \bibinfo {author} {\bibfnamefont {M.~L.}\ \bibnamefont {Martinez}}, \bibinfo {author} {\bibfnamefont {F.}~\bibnamefont {Meloni}}, \bibinfo {author} {\bibfnamefont {A.}~\bibnamefont {Mezzacapo}}, \bibinfo {author} {\bibfnamefont {S.}~\bibnamefont {Montangero}}, \bibinfo {author} {\bibfnamefont {L.}~\bibnamefont {Nagano}}, \bibinfo {author} {\bibfnamefont {V.~R.}\ \bibnamefont {Pascuzzi}}, \bibinfo {author} {\bibfnamefont {V.}~\bibnamefont {Radescu}}, \bibinfo {author} {\bibfnamefont {E.~R.}\ \bibnamefont {Ortega}}, \bibinfo {author} {\bibfnamefont {A.}~\bibnamefont {Roggero}}, \bibinfo {author} {\bibfnamefont {J.}~\bibnamefont {Schuhmacher}}, \bibinfo {author} {\bibfnamefont {J.}~\bibnamefont {Seixas}}, \bibinfo {author} {\bibfnamefont {P.}~\bibnamefont {Silvi}}, \bibinfo {author} {\bibfnamefont {P.}~\bibnamefont {Spentzouris}}, \bibinfo {author} {\bibfnamefont {F.}~\bibnamefont {Tacchino}}, \bibinfo {author} {\bibfnamefont {K.}~\bibnamefont {Temme}}, \bibinfo {author} {\bibfnamefont {K.}~\bibnamefont {Terashi}}, \bibinfo {author} {\bibfnamefont {J.}~\bibnamefont {Tura}}, \bibinfo {author} {\bibfnamefont {C.}~\bibnamefont {T\"uys\"uz}}, \bibinfo {author} {\bibfnamefont {S.}~\bibnamefont {Vallecorsa}}, \bibinfo {author} {\bibfnamefont {U.-J.}\ \bibnamefont {Wiese}}, \bibinfo {author} {\bibfnamefont {S.}~\bibnamefont {Yoo}},\ and\ \bibinfo {author} {\bibfnamefont {J.}~\bibnamefont {Zhang}},\ }\bibfield  {title} {\bibinfo {title} {Quantum computing for high-energy physics: State of the art and challenges},\ }\href {https://doi.org/10.1103/PRXQuantum.5.037001} {\bibfield  {journal} {\bibinfo  {journal} {PRX Quantum}\ }\textbf {\bibinfo {volume} {5}},\ \bibinfo {pages} {037001} (\bibinfo {year} {2024})}\BibitemShut {NoStop}%
\bibitem [{\citenamefont {Bauer}(2025)}]{bauer2025efficient}%
  \BibitemOpen
  \bibfield  {author} {\bibinfo {author} {\bibfnamefont {C.~W.}\ \bibnamefont {Bauer}},\ }\bibfield  {title} {\bibinfo {title} {Efficient use of quantum computers for collider physics},\ }\href {https://doi.org/10.1007/JHEP11(2025)108} {\bibfield  {journal} {\bibinfo  {journal} {Journal of High Energy Physics}\ }\textbf {\bibinfo {volume} {2025}},\ \bibinfo {pages} {108} (\bibinfo {year} {2025})}\BibitemShut {NoStop}%
\bibitem [{\citenamefont {Burbano}\ \emph {et~al.}(2026)\citenamefont {Burbano}, \citenamefont {Carrillo}, \citenamefont {Urek}, \citenamefont {Ciavarella},\ and\ \citenamefont {Brice\~no}}]{Burbano2026}%
  \BibitemOpen
  \bibfield  {author} {\bibinfo {author} {\bibfnamefont {I.~M.}\ \bibnamefont {Burbano}}, \bibinfo {author} {\bibfnamefont {M.~A.}\ \bibnamefont {Carrillo}}, \bibinfo {author} {\bibfnamefont {R.}~\bibnamefont {Urek}}, \bibinfo {author} {\bibfnamefont {A.~N.}\ \bibnamefont {Ciavarella}},\ and\ \bibinfo {author} {\bibfnamefont {R.~A.}\ \bibnamefont {Brice\~no}},\ }\bibfield  {title} {\bibinfo {title} {Real-time estimators for scattering observables: A full account of finite-volume errors for quantum simulation},\ }\href {https://doi.org/10.1103/dc17-4zjy} {\bibfield  {journal} {\bibinfo  {journal} {Phys. Rev. D}\ }\textbf {\bibinfo {volume} {113}},\ \bibinfo {pages} {L071502} (\bibinfo {year} {2026})}\BibitemShut {NoStop}%
\bibitem [{\citenamefont {Hardy}\ \emph {et~al.}(2026)\citenamefont {Hardy}, \citenamefont {Mukhopadhyay}, \citenamefont {Alam}, \citenamefont {Konik}, \citenamefont {Hormozi}, \citenamefont {Rieffel}, \citenamefont {Hadfield}, \citenamefont {Barata}, \citenamefont {Venugopalan}, \citenamefont {Kharzeev},\ and\ \citenamefont {Wiebe}}]{Hardy2026}%
  \BibitemOpen
  \bibfield  {author} {\bibinfo {author} {\bibfnamefont {A.}~\bibnamefont {Hardy}}, \bibinfo {author} {\bibfnamefont {P.}~\bibnamefont {Mukhopadhyay}}, \bibinfo {author} {\bibfnamefont {M.~S.}\ \bibnamefont {Alam}}, \bibinfo {author} {\bibfnamefont {R.}~\bibnamefont {Konik}}, \bibinfo {author} {\bibfnamefont {L.}~\bibnamefont {Hormozi}}, \bibinfo {author} {\bibfnamefont {E.}~\bibnamefont {Rieffel}}, \bibinfo {author} {\bibfnamefont {S.}~\bibnamefont {Hadfield}}, \bibinfo {author} {\bibfnamefont {J.~a.}\ \bibnamefont {Barata}}, \bibinfo {author} {\bibfnamefont {R.}~\bibnamefont {Venugopalan}}, \bibinfo {author} {\bibfnamefont {D.~E.}\ \bibnamefont {Kharzeev}},\ and\ \bibinfo {author} {\bibfnamefont {N.}~\bibnamefont {Wiebe}},\ }\bibfield  {title} {\bibinfo {title} {Scattering processes from quantum simulation algorithms for scalar field theories},\ }\href {https://doi.org/10.1103/3krb-wwfx} {\bibfield  {journal} {\bibinfo  {journal} {PRX Quantum}\ }\textbf {\bibinfo {volume} {7}},\ \bibinfo {pages} {010343} (\bibinfo {year} {2026})}\BibitemShut {NoStop}%
\bibitem [{\citenamefont {{Barata, Joao}}(2026)}]{Barata2026}%
  \BibitemOpen
  \bibfield  {author} {\bibinfo {author} {\bibnamefont {{Barata, Joao}}},\ }\bibfield  {title} {\bibinfo {title} {Quantum computing for heavy-ion physics: Near-term status and future prospects},\ }\href {https://doi.org/10.1051/epjconf/202636401020} {\bibfield  {journal} {\bibinfo  {journal} {EPJ Web Conf.}\ }\textbf {\bibinfo {volume} {364}},\ \bibinfo {pages} {01020} (\bibinfo {year} {2026})}\BibitemShut {NoStop}%
\bibitem [{\citenamefont {Berges}\ \emph {et~al.}(2021)\citenamefont {Berges}, \citenamefont {Heller}, \citenamefont {Mazeliauskas},\ and\ \citenamefont {Venugopalan}}]{BergesQCD}%
  \BibitemOpen
  \bibfield  {author} {\bibinfo {author} {\bibfnamefont {J.}~\bibnamefont {Berges}}, \bibinfo {author} {\bibfnamefont {M.~P.}\ \bibnamefont {Heller}}, \bibinfo {author} {\bibfnamefont {A.}~\bibnamefont {Mazeliauskas}},\ and\ \bibinfo {author} {\bibfnamefont {R.}~\bibnamefont {Venugopalan}},\ }\bibfield  {title} {\bibinfo {title} {{QCD thermalization: Ab initio approaches and interdisciplinary connections}},\ }\href {https://doi.org/10.1103/RevModPhys.93.035003} {\bibfield  {journal} {\bibinfo  {journal} {Rev. Mod. Phys.}\ }\textbf {\bibinfo {volume} {93}},\ \bibinfo {pages} {035003} (\bibinfo {year} {2021})}\BibitemShut {NoStop}%
\bibitem [{\citenamefont {Busza}\ \emph {et~al.}(2018)\citenamefont {Busza}, \citenamefont {Rajagopal},\ and\ \citenamefont {van~der Schee}}]{Busza2018}%
  \BibitemOpen
  \bibfield  {author} {\bibinfo {author} {\bibfnamefont {W.}~\bibnamefont {Busza}}, \bibinfo {author} {\bibfnamefont {K.}~\bibnamefont {Rajagopal}},\ and\ \bibinfo {author} {\bibfnamefont {W.}~\bibnamefont {van~der Schee}},\ }\bibfield  {title} {\bibinfo {title} {Heavy ion collisions: The big picture and the big questions},\ }\href {https://doi.org/https://doi.org/10.1146/annurev-nucl-101917-020852} {\bibfield  {journal} {\bibinfo  {journal} {Annual Review of Nuclear and Particle Science}\ }\textbf {\bibinfo {volume} {68}},\ \bibinfo {pages} {339} (\bibinfo {year} {2018})}\BibitemShut {NoStop}%
\bibitem [{\citenamefont {Milsted}\ \emph {et~al.}(2022)\citenamefont {Milsted}, \citenamefont {Liu}, \citenamefont {Preskill},\ and\ \citenamefont {Vidal}}]{milsted2022collisions}%
  \BibitemOpen
  \bibfield  {author} {\bibinfo {author} {\bibfnamefont {A.}~\bibnamefont {Milsted}}, \bibinfo {author} {\bibfnamefont {J.}~\bibnamefont {Liu}}, \bibinfo {author} {\bibfnamefont {J.}~\bibnamefont {Preskill}},\ and\ \bibinfo {author} {\bibfnamefont {G.}~\bibnamefont {Vidal}},\ }\bibfield  {title} {\bibinfo {title} {Collisions of false-vacuum bubble walls in a quantum spin chain},\ }\href {https://doi.org/10.1103/PRXQuantum.3.020316} {\bibfield  {journal} {\bibinfo  {journal} {PRX Quantum}\ }\textbf {\bibinfo {volume} {3}},\ \bibinfo {pages} {020316} (\bibinfo {year} {2022})}\BibitemShut {NoStop}%
\bibitem [{\citenamefont {Vanderstraeten}\ \emph {et~al.}(2014)\citenamefont {Vanderstraeten}, \citenamefont {Haegeman}, \citenamefont {Osborne},\ and\ \citenamefont {Verstraete}}]{Vanderstraeten2014}%
  \BibitemOpen
  \bibfield  {author} {\bibinfo {author} {\bibfnamefont {L.}~\bibnamefont {Vanderstraeten}}, \bibinfo {author} {\bibfnamefont {J.}~\bibnamefont {Haegeman}}, \bibinfo {author} {\bibfnamefont {T.~J.}\ \bibnamefont {Osborne}},\ and\ \bibinfo {author} {\bibfnamefont {F.}~\bibnamefont {Verstraete}},\ }\bibfield  {title} {\bibinfo {title} {$s$ matrix from matrix product states},\ }\href {https://doi.org/10.1103/PhysRevLett.112.257202} {\bibfield  {journal} {\bibinfo  {journal} {Phys. Rev. Lett.}\ }\textbf {\bibinfo {volume} {112}},\ \bibinfo {pages} {257202} (\bibinfo {year} {2014})}\BibitemShut {NoStop}%
\bibitem [{\citenamefont {Van~Damme}\ \emph {et~al.}(2021)\citenamefont {Van~Damme}, \citenamefont {Vanderstraeten}, \citenamefont {De~Nardis}, \citenamefont {Haegeman},\ and\ \citenamefont {Verstraete}}]{VanDamme2021}%
  \BibitemOpen
  \bibfield  {author} {\bibinfo {author} {\bibfnamefont {M.}~\bibnamefont {Van~Damme}}, \bibinfo {author} {\bibfnamefont {L.}~\bibnamefont {Vanderstraeten}}, \bibinfo {author} {\bibfnamefont {J.}~\bibnamefont {De~Nardis}}, \bibinfo {author} {\bibfnamefont {J.}~\bibnamefont {Haegeman}},\ and\ \bibinfo {author} {\bibfnamefont {F.}~\bibnamefont {Verstraete}},\ }\bibfield  {title} {\bibinfo {title} {Real-time scattering of interacting quasiparticles in quantum spin chains},\ }\href {https://doi.org/10.1103/PhysRevResearch.3.013078} {\bibfield  {journal} {\bibinfo  {journal} {Phys. Rev. Res.}\ }\textbf {\bibinfo {volume} {3}},\ \bibinfo {pages} {013078} (\bibinfo {year} {2021})}\BibitemShut {NoStop}%
\bibitem [{\citenamefont {Rigobello}\ \emph {et~al.}(2021)\citenamefont {Rigobello}, \citenamefont {Notarnicola}, \citenamefont {Magnifico},\ and\ \citenamefont {Montangero}}]{Rigobello2021}%
  \BibitemOpen
  \bibfield  {author} {\bibinfo {author} {\bibfnamefont {M.}~\bibnamefont {Rigobello}}, \bibinfo {author} {\bibfnamefont {S.}~\bibnamefont {Notarnicola}}, \bibinfo {author} {\bibfnamefont {G.}~\bibnamefont {Magnifico}},\ and\ \bibinfo {author} {\bibfnamefont {S.}~\bibnamefont {Montangero}},\ }\bibfield  {title} {\bibinfo {title} {Entanglement generation in $(1+1)\mathrm{D}$ qed scattering processes},\ }\href {https://doi.org/10.1103/PhysRevD.104.114501} {\bibfield  {journal} {\bibinfo  {journal} {Phys. Rev. D}\ }\textbf {\bibinfo {volume} {104}},\ \bibinfo {pages} {114501} (\bibinfo {year} {2021})}\BibitemShut {NoStop}%
\bibitem [{\citenamefont {Karpov}\ \emph {et~al.}(2022)\citenamefont {Karpov}, \citenamefont {Zhu}, \citenamefont {Heller},\ and\ \citenamefont {Heyl}}]{Karpov2022}%
  \BibitemOpen
  \bibfield  {author} {\bibinfo {author} {\bibfnamefont {P.~I.}\ \bibnamefont {Karpov}}, \bibinfo {author} {\bibfnamefont {G.-Y.}\ \bibnamefont {Zhu}}, \bibinfo {author} {\bibfnamefont {M.~P.}\ \bibnamefont {Heller}},\ and\ \bibinfo {author} {\bibfnamefont {M.}~\bibnamefont {Heyl}},\ }\bibfield  {title} {\bibinfo {title} {Spatiotemporal dynamics of particle collisions in quantum spin chains},\ }\href {https://doi.org/10.1103/PhysRevResearch.4.L032001} {\bibfield  {journal} {\bibinfo  {journal} {Phys. Rev. Res.}\ }\textbf {\bibinfo {volume} {4}},\ \bibinfo {pages} {L032001} (\bibinfo {year} {2022})}\BibitemShut {NoStop}%
\bibitem [{\citenamefont {Belyansky}\ \emph {et~al.}(2024)\citenamefont {Belyansky}, \citenamefont {Whitsitt}, \citenamefont {Mueller}, \citenamefont {Fahimniya}, \citenamefont {Bennewitz}, \citenamefont {Davoudi},\ and\ \citenamefont {Gorshkov}}]{Belyansky2024}%
  \BibitemOpen
  \bibfield  {author} {\bibinfo {author} {\bibfnamefont {R.}~\bibnamefont {Belyansky}}, \bibinfo {author} {\bibfnamefont {S.}~\bibnamefont {Whitsitt}}, \bibinfo {author} {\bibfnamefont {N.}~\bibnamefont {Mueller}}, \bibinfo {author} {\bibfnamefont {A.}~\bibnamefont {Fahimniya}}, \bibinfo {author} {\bibfnamefont {E.~R.}\ \bibnamefont {Bennewitz}}, \bibinfo {author} {\bibfnamefont {Z.}~\bibnamefont {Davoudi}},\ and\ \bibinfo {author} {\bibfnamefont {A.~V.}\ \bibnamefont {Gorshkov}},\ }\bibfield  {title} {\bibinfo {title} {{High-Energy Collision of Quarks and Mesons in the Schwinger Model: From Tensor Networks to Circuit QED}},\ }\href {https://doi.org/10.1103/PhysRevLett.132.091903} {\bibfield  {journal} {\bibinfo  {journal} {Phys. Rev. Lett.}\ }\textbf {\bibinfo {volume} {132}},\ \bibinfo {pages} {091903} (\bibinfo {year} {2024})}\BibitemShut {NoStop}%
\bibitem [{\citenamefont {Jha}\ \emph {et~al.}(2025)\citenamefont {Jha}, \citenamefont {Milsted}, \citenamefont {Neuenfeld}, \citenamefont {Preskill},\ and\ \citenamefont {Vieira}}]{Jha2025}%
  \BibitemOpen
  \bibfield  {author} {\bibinfo {author} {\bibfnamefont {R.~G.}\ \bibnamefont {Jha}}, \bibinfo {author} {\bibfnamefont {A.}~\bibnamefont {Milsted}}, \bibinfo {author} {\bibfnamefont {D.}~\bibnamefont {Neuenfeld}}, \bibinfo {author} {\bibfnamefont {J.}~\bibnamefont {Preskill}},\ and\ \bibinfo {author} {\bibfnamefont {P.}~\bibnamefont {Vieira}},\ }\bibfield  {title} {\bibinfo {title} {Real-time scattering in ising field theory using matrix product states},\ }\href {https://doi.org/10.1103/9dxz-k5wb} {\bibfield  {journal} {\bibinfo  {journal} {Phys. Rev. Res.}\ }\textbf {\bibinfo {volume} {7}},\ \bibinfo {pages} {023266} (\bibinfo {year} {2025})}\BibitemShut {NoStop}%
\bibitem [{\citenamefont {Papaefstathiou}\ \emph {et~al.}(2025)\citenamefont {Papaefstathiou}, \citenamefont {Knolle},\ and\ \citenamefont {Ba\~nuls}}]{Papaefstathiou2025}%
  \BibitemOpen
  \bibfield  {author} {\bibinfo {author} {\bibfnamefont {I.}~\bibnamefont {Papaefstathiou}}, \bibinfo {author} {\bibfnamefont {J.}~\bibnamefont {Knolle}},\ and\ \bibinfo {author} {\bibfnamefont {M.~C.}\ \bibnamefont {Ba\~nuls}},\ }\bibfield  {title} {\bibinfo {title} {Real-time scattering in the lattice schwinger model},\ }\href {https://doi.org/10.1103/PhysRevD.111.014504} {\bibfield  {journal} {\bibinfo  {journal} {Phys. Rev. D}\ }\textbf {\bibinfo {volume} {111}},\ \bibinfo {pages} {014504} (\bibinfo {year} {2025})}\BibitemShut {NoStop}%
\bibitem [{\citenamefont {Pave{\v{s}}i{\'c}}\ \emph {et~al.}(2026)\citenamefont {Pave{\v{s}}i{\'c}}, \citenamefont {Di~Liberto},\ and\ \citenamefont {Montangero}}]{pavevsic2026scattering}%
  \BibitemOpen
  \bibfield  {author} {\bibinfo {author} {\bibfnamefont {L.}~\bibnamefont {Pave{\v{s}}i{\'c}}}, \bibinfo {author} {\bibfnamefont {M.}~\bibnamefont {Di~Liberto}},\ and\ \bibinfo {author} {\bibfnamefont {S.}~\bibnamefont {Montangero}},\ }\bibfield  {title} {\bibinfo {title} {Scattering and induced false vacuum decay in the two-dimensional quantum ising model},\ }\href {https://doi.org/10.48550/arXiv.2509.02702} {\bibfield  {journal} {\bibinfo  {journal} {Nature Communications}\ } (\bibinfo {year} {2026})}\BibitemShut {NoStop}%
\bibitem [{\citenamefont {Surace}\ and\ \citenamefont {Lerose}(2021)}]{surace2021scattering}%
  \BibitemOpen
  \bibfield  {author} {\bibinfo {author} {\bibfnamefont {F.~M.}\ \bibnamefont {Surace}}\ and\ \bibinfo {author} {\bibfnamefont {A.}~\bibnamefont {Lerose}},\ }\bibfield  {title} {\bibinfo {title} {Scattering of mesons in quantum simulators},\ }\href {https://doi.org/10.1088/1367-2630/abfc40} {\bibfield  {journal} {\bibinfo  {journal} {New Journal of Physics}\ }\textbf {\bibinfo {volume} {23}},\ \bibinfo {pages} {062001} (\bibinfo {year} {2021})}\BibitemShut {NoStop}%
\bibitem [{\citenamefont {Barata}\ \emph {et~al.}(2021)\citenamefont {Barata}, \citenamefont {Mueller}, \citenamefont {Tarasov},\ and\ \citenamefont {Venugopalan}}]{Barata2021}%
  \BibitemOpen
  \bibfield  {author} {\bibinfo {author} {\bibfnamefont {J.}~\bibnamefont {Barata}}, \bibinfo {author} {\bibfnamefont {N.}~\bibnamefont {Mueller}}, \bibinfo {author} {\bibfnamefont {A.}~\bibnamefont {Tarasov}},\ and\ \bibinfo {author} {\bibfnamefont {R.}~\bibnamefont {Venugopalan}},\ }\bibfield  {title} {\bibinfo {title} {Single-particle digitization strategy for quantum computation of a ${\ensuremath{\phi}}^{4}$ scalar field theory},\ }\href {https://doi.org/10.1103/PhysRevA.103.042410} {\bibfield  {journal} {\bibinfo  {journal} {Phys. Rev. A}\ }\textbf {\bibinfo {volume} {103}},\ \bibinfo {pages} {042410} (\bibinfo {year} {2021})}\BibitemShut {NoStop}%
\bibitem [{\citenamefont {Turco}\ \emph {et~al.}(2024)\citenamefont {Turco}, \citenamefont {Quinta}, \citenamefont {Seixas},\ and\ \citenamefont {Omar}}]{Turco2024}%
  \BibitemOpen
  \bibfield  {author} {\bibinfo {author} {\bibfnamefont {M.}~\bibnamefont {Turco}}, \bibinfo {author} {\bibfnamefont {G.}~\bibnamefont {Quinta}}, \bibinfo {author} {\bibfnamefont {J.}~\bibnamefont {Seixas}},\ and\ \bibinfo {author} {\bibfnamefont {Y.}~\bibnamefont {Omar}},\ }\bibfield  {title} {\bibinfo {title} {Quantum simulation of bound state scattering},\ }\href {https://doi.org/10.1103/PRXQuantum.5.020311} {\bibfield  {journal} {\bibinfo  {journal} {PRX Quantum}\ }\textbf {\bibinfo {volume} {5}},\ \bibinfo {pages} {020311} (\bibinfo {year} {2024})}\BibitemShut {NoStop}%
\bibitem [{\citenamefont {Farrell}\ \emph {et~al.}(2024)\citenamefont {Farrell}, \citenamefont {Illa}, \citenamefont {Ciavarella},\ and\ \citenamefont {Savage}}]{Farrell2024}%
  \BibitemOpen
  \bibfield  {author} {\bibinfo {author} {\bibfnamefont {R.~C.}\ \bibnamefont {Farrell}}, \bibinfo {author} {\bibfnamefont {M.}~\bibnamefont {Illa}}, \bibinfo {author} {\bibfnamefont {A.~N.}\ \bibnamefont {Ciavarella}},\ and\ \bibinfo {author} {\bibfnamefont {M.~J.}\ \bibnamefont {Savage}},\ }\bibfield  {title} {\bibinfo {title} {Quantum simulations of hadron dynamics in the schwinger model using 112 qubits},\ }\href {https://doi.org/10.1103/PhysRevD.109.114510} {\bibfield  {journal} {\bibinfo  {journal} {Phys. Rev. D}\ }\textbf {\bibinfo {volume} {109}},\ \bibinfo {pages} {114510} (\bibinfo {year} {2024})}\BibitemShut {NoStop}%
\bibitem [{\citenamefont {Su}\ \emph {et~al.}(2024)\citenamefont {Su}, \citenamefont {Osborne},\ and\ \citenamefont {Halimeh}}]{Su2024}%
  \BibitemOpen
  \bibfield  {author} {\bibinfo {author} {\bibfnamefont {G.-X.}\ \bibnamefont {Su}}, \bibinfo {author} {\bibfnamefont {J.~J.}\ \bibnamefont {Osborne}},\ and\ \bibinfo {author} {\bibfnamefont {J.~C.}\ \bibnamefont {Halimeh}},\ }\bibfield  {title} {\bibinfo {title} {Cold-atom particle collider},\ }\href {https://doi.org/10.1103/PRXQuantum.5.040310} {\bibfield  {journal} {\bibinfo  {journal} {PRX Quantum}\ }\textbf {\bibinfo {volume} {5}},\ \bibinfo {pages} {040310} (\bibinfo {year} {2024})}\BibitemShut {NoStop}%
\bibitem [{\citenamefont {Bennewitz}\ \emph {et~al.}(2025)\citenamefont {Bennewitz}, \citenamefont {Ware}, \citenamefont {Schuckert}, \citenamefont {Lerose}, \citenamefont {Surace}, \citenamefont {Belyansky}, \citenamefont {Morong}, \citenamefont {Luo}, \citenamefont {De}, \citenamefont {Collins}, \citenamefont {Katz}, \citenamefont {Monroe}, \citenamefont {Davoudi},\ and\ \citenamefont {Gorshkov}}]{Bennewitz2025simulatingmeson}%
  \BibitemOpen
  \bibfield  {author} {\bibinfo {author} {\bibfnamefont {E.~R.}\ \bibnamefont {Bennewitz}}, \bibinfo {author} {\bibfnamefont {B.}~\bibnamefont {Ware}}, \bibinfo {author} {\bibfnamefont {A.}~\bibnamefont {Schuckert}}, \bibinfo {author} {\bibfnamefont {A.}~\bibnamefont {Lerose}}, \bibinfo {author} {\bibfnamefont {F.~M.}\ \bibnamefont {Surace}}, \bibinfo {author} {\bibfnamefont {R.}~\bibnamefont {Belyansky}}, \bibinfo {author} {\bibfnamefont {W.}~\bibnamefont {Morong}}, \bibinfo {author} {\bibfnamefont {D.}~\bibnamefont {Luo}}, \bibinfo {author} {\bibfnamefont {A.}~\bibnamefont {De}}, \bibinfo {author} {\bibfnamefont {K.~S.}\ \bibnamefont {Collins}}, \bibinfo {author} {\bibfnamefont {O.}~\bibnamefont {Katz}}, \bibinfo {author} {\bibfnamefont {C.}~\bibnamefont {Monroe}}, \bibinfo {author} {\bibfnamefont {Z.}~\bibnamefont {Davoudi}},\ and\ \bibinfo {author} {\bibfnamefont {A.~V.}\ \bibnamefont {Gorshkov}},\ }\bibfield  {title} {\bibinfo {title} {Simulating {M}eson {S}cattering on {S}pin {Q}uantum {S}imulators},\ }\href {https://doi.org/10.22331/q-2025-06-17-1773} {\bibfield  {journal} {\bibinfo  {journal} {{Quantum}}\ }\textbf {\bibinfo {volume} {9}},\ \bibinfo {pages} {1773} (\bibinfo {year} {2025})}\BibitemShut {NoStop}%
\bibitem [{\citenamefont {Turco}\ \emph {et~al.}(2025)\citenamefont {Turco}, \citenamefont {Quinta}, \citenamefont {Seixas},\ and\ \citenamefont {Omar}}]{Turco2025}%
  \BibitemOpen
  \bibfield  {author} {\bibinfo {author} {\bibfnamefont {M.}~\bibnamefont {Turco}}, \bibinfo {author} {\bibfnamefont {G.}~\bibnamefont {Quinta}}, \bibinfo {author} {\bibfnamefont {J.}~\bibnamefont {Seixas}},\ and\ \bibinfo {author} {\bibfnamefont {Y.}~\bibnamefont {Omar}},\ }\bibfield  {title} {\bibinfo {title} {Creation of wave packets for quantum chromodynamics on quantum computers},\ }\href {https://doi.org/10.1103/2nb4-171l} {\bibfield  {journal} {\bibinfo  {journal} {Phys. Rev. D}\ }\textbf {\bibinfo {volume} {112}},\ \bibinfo {pages} {034506} (\bibinfo {year} {2025})}\BibitemShut {NoStop}%
\bibitem [{\citenamefont {Joshi}\ \emph {et~al.}(2025)\citenamefont {Joshi}, \citenamefont {Louw}, \citenamefont {Meth}, \citenamefont {Osborne}, \citenamefont {Mato}, \citenamefont {Su}, \citenamefont {Ringbauer},\ and\ \citenamefont {Halimeh}}]{joshi2025probing}%
  \BibitemOpen
  \bibfield  {author} {\bibinfo {author} {\bibfnamefont {R.}~\bibnamefont {Joshi}}, \bibinfo {author} {\bibfnamefont {J.~C.}\ \bibnamefont {Louw}}, \bibinfo {author} {\bibfnamefont {M.}~\bibnamefont {Meth}}, \bibinfo {author} {\bibfnamefont {J.~J.}\ \bibnamefont {Osborne}}, \bibinfo {author} {\bibfnamefont {K.}~\bibnamefont {Mato}}, \bibinfo {author} {\bibfnamefont {G.-X.}\ \bibnamefont {Su}}, \bibinfo {author} {\bibfnamefont {M.}~\bibnamefont {Ringbauer}},\ and\ \bibinfo {author} {\bibfnamefont {J.~C.}\ \bibnamefont {Halimeh}},\ }\bibfield  {title} {\bibinfo {title} {Probing hadron scattering in lattice gauge theories on qudit quantum computers},\ }\href {https://doi.org/10.48550/arXiv.2507.12614} {\bibfield  {journal} {\bibinfo  {journal} {arXiv preprint arXiv:2507.12614}\ } (\bibinfo {year} {2025})}\BibitemShut {NoStop}%
\bibitem [{\citenamefont {Ingoldby}\ \emph {et~al.}(2025)\citenamefont {Ingoldby}, \citenamefont {Spannowsky}, \citenamefont {Sypchenko}, \citenamefont {Williams},\ and\ \citenamefont {Wingate}}]{ingoldby2025real}%
  \BibitemOpen
  \bibfield  {author} {\bibinfo {author} {\bibfnamefont {J.}~\bibnamefont {Ingoldby}}, \bibinfo {author} {\bibfnamefont {M.}~\bibnamefont {Spannowsky}}, \bibinfo {author} {\bibfnamefont {T.}~\bibnamefont {Sypchenko}}, \bibinfo {author} {\bibfnamefont {S.}~\bibnamefont {Williams}},\ and\ \bibinfo {author} {\bibfnamefont {M.}~\bibnamefont {Wingate}},\ }\bibfield  {title} {\bibinfo {title} {Real-time scattering on quantum computers via hamiltonian truncation},\ }\href {https://doi.org/10.48550/arXiv.2505.03878} {\bibfield  {journal} {\bibinfo  {journal} {arXiv preprint arXiv:2505.03878}\ } (\bibinfo {year} {2025})}\BibitemShut {NoStop}%
\bibitem [{\citenamefont {Abel}\ \emph {et~al.}(2025)\citenamefont {Abel}, \citenamefont {Spannowsky},\ and\ \citenamefont {Williams}}]{Abel2025}%
  \BibitemOpen
  \bibfield  {author} {\bibinfo {author} {\bibfnamefont {S.}~\bibnamefont {Abel}}, \bibinfo {author} {\bibfnamefont {M.}~\bibnamefont {Spannowsky}},\ and\ \bibinfo {author} {\bibfnamefont {S.}~\bibnamefont {Williams}},\ }\bibfield  {title} {\bibinfo {title} {Real-time scattering processes with continuous-variable quantum computers},\ }\href {https://doi.org/10.1103/q36d-w649} {\bibfield  {journal} {\bibinfo  {journal} {Phys. Rev. A}\ }\textbf {\bibinfo {volume} {112}},\ \bibinfo {pages} {012614} (\bibinfo {year} {2025})}\BibitemShut {NoStop}%
\bibitem [{\citenamefont {Lee}\ and\ \citenamefont {Farrell}(2026)}]{lee2026studying}%
  \BibitemOpen
  \bibfield  {author} {\bibinfo {author} {\bibfnamefont {M.}~\bibnamefont {Lee}}\ and\ \bibinfo {author} {\bibfnamefont {R.~C.}\ \bibnamefont {Farrell}},\ }\bibfield  {title} {\bibinfo {title} {Studying energy-resolved transport with wavepacket dynamics on quantum computers},\ }\href {https://arxiv.org/abs/2601.16180} {\bibfield  {journal} {\bibinfo  {journal} {arXiv preprint arXiv:2601.16180}\ } (\bibinfo {year} {2026})}\BibitemShut {NoStop}%
\bibitem [{\citenamefont {Morgavi}\ \emph {et~al.}(2026)\citenamefont {Morgavi}, \citenamefont {Majcen}, \citenamefont {Rigobello}, \citenamefont {Montangero},\ and\ \citenamefont {Silvi}}]{morgavi2026preparation}%
  \BibitemOpen
  \bibfield  {author} {\bibinfo {author} {\bibfnamefont {M.}~\bibnamefont {Morgavi}}, \bibinfo {author} {\bibfnamefont {P.}~\bibnamefont {Majcen}}, \bibinfo {author} {\bibfnamefont {M.}~\bibnamefont {Rigobello}}, \bibinfo {author} {\bibfnamefont {S.}~\bibnamefont {Montangero}},\ and\ \bibinfo {author} {\bibfnamefont {P.}~\bibnamefont {Silvi}},\ }\bibfield  {title} {\bibinfo {title} {Preparation and detection of quasiparticles for quantum simulations of scattering},\ }\href {https://doi.org/10.48550/arXiv.2604.16210} {\bibfield  {journal} {\bibinfo  {journal} {arXiv preprint arXiv:2604.16210}\ } (\bibinfo {year} {2026})}\BibitemShut {NoStop}%
\bibitem [{\citenamefont {Zemlevskiy}(2026)}]{zemlevskiy2026exclusive}%
  \BibitemOpen
  \bibfield  {author} {\bibinfo {author} {\bibfnamefont {N.~A.}\ \bibnamefont {Zemlevskiy}},\ }\bibfield  {title} {\bibinfo {title} {Exclusive scattering channels from entanglement structure in real-time simulations},\ }\href {https://doi.org/10.48550/arXiv.2603.15621} {\bibfield  {journal} {\bibinfo  {journal} {arXiv preprint arXiv:2603.15621}\ } (\bibinfo {year} {2026})}\BibitemShut {NoStop}%
\bibitem [{\citenamefont {Davoudi}\ \emph {et~al.}(2024)\citenamefont {Davoudi}, \citenamefont {Hsieh},\ and\ \citenamefont {Kadam}}]{davoudi2024scattering}%
  \BibitemOpen
  \bibfield  {author} {\bibinfo {author} {\bibfnamefont {Z.}~\bibnamefont {Davoudi}}, \bibinfo {author} {\bibfnamefont {C.-C.}\ \bibnamefont {Hsieh}},\ and\ \bibinfo {author} {\bibfnamefont {S.~V.}\ \bibnamefont {Kadam}},\ }\bibfield  {title} {\bibinfo {title} {Scattering wave packets of hadrons in gauge theories: {P}reparation on a quantum computer},\ }\href {https://doi.org/10.22331/q-2024-11-11-1520} {\bibfield  {journal} {\bibinfo  {journal} {{Quantum}}\ }\textbf {\bibinfo {volume} {8}},\ \bibinfo {pages} {1520} (\bibinfo {year} {2024})}\BibitemShut {NoStop}%
\bibitem [{\citenamefont {Chai}\ \emph {et~al.}(2025)\citenamefont {Chai}, \citenamefont {Crippa}, \citenamefont {Jansen}, \citenamefont {K{\"{u}}hn}, \citenamefont {Pascuzzi}, \citenamefont {Tacchino},\ and\ \citenamefont {Tavernelli}}]{Chai2025fermionicwavepacket}%
  \BibitemOpen
  \bibfield  {author} {\bibinfo {author} {\bibfnamefont {Y.}~\bibnamefont {Chai}}, \bibinfo {author} {\bibfnamefont {A.}~\bibnamefont {Crippa}}, \bibinfo {author} {\bibfnamefont {K.}~\bibnamefont {Jansen}}, \bibinfo {author} {\bibfnamefont {S.}~\bibnamefont {K{\"{u}}hn}}, \bibinfo {author} {\bibfnamefont {V.~R.}\ \bibnamefont {Pascuzzi}}, \bibinfo {author} {\bibfnamefont {F.}~\bibnamefont {Tacchino}},\ and\ \bibinfo {author} {\bibfnamefont {I.}~\bibnamefont {Tavernelli}},\ }\bibfield  {title} {\bibinfo {title} {Fermionic wave packet scattering: a quantum computing approach},\ }\href {https://doi.org/10.22331/q-2025-02-19-1638} {\bibfield  {journal} {\bibinfo  {journal} {{Quantum}}\ }\textbf {\bibinfo {volume} {9}},\ \bibinfo {pages} {1638} (\bibinfo {year} {2025})}\BibitemShut {NoStop}%
\bibitem [{\citenamefont {Zemlevskiy}(2025)}]{Zemlevskiy2025}%
  \BibitemOpen
  \bibfield  {author} {\bibinfo {author} {\bibfnamefont {N.~A.}\ \bibnamefont {Zemlevskiy}},\ }\bibfield  {title} {\bibinfo {title} {Scalable quantum simulations of scattering in scalar field theory on 120 qubits},\ }\href {https://doi.org/10.1103/qr72-51v1} {\bibfield  {journal} {\bibinfo  {journal} {Phys. Rev. D}\ }\textbf {\bibinfo {volume} {112}},\ \bibinfo {pages} {034502} (\bibinfo {year} {2025})}\BibitemShut {NoStop}%
\bibitem [{\citenamefont {Farrell}\ \emph {et~al.}(2025)\citenamefont {Farrell}, \citenamefont {Zemlevskiy}, \citenamefont {Illa},\ and\ \citenamefont {Preskill}}]{farrell2025digital}%
  \BibitemOpen
  \bibfield  {author} {\bibinfo {author} {\bibfnamefont {R.~C.}\ \bibnamefont {Farrell}}, \bibinfo {author} {\bibfnamefont {N.~A.}\ \bibnamefont {Zemlevskiy}}, \bibinfo {author} {\bibfnamefont {M.}~\bibnamefont {Illa}},\ and\ \bibinfo {author} {\bibfnamefont {J.}~\bibnamefont {Preskill}},\ }\bibfield  {title} {\bibinfo {title} {{Digital quantum simulations of scattering in quantum field theories using W states}},\ }\href {https://doi.org/10.48550/arXiv.2505.03111} {\bibfield  {journal} {\bibinfo  {journal} {arXiv preprint arXiv:2505.03111}\ } (\bibinfo {year} {2025})}\BibitemShut {NoStop}%
\bibitem [{\citenamefont {Schuhmacher}\ \emph {et~al.}(2025)\citenamefont {Schuhmacher}, \citenamefont {Su}, \citenamefont {Osborne}, \citenamefont {Gandon}, \citenamefont {Halimeh},\ and\ \citenamefont {Tavernelli}}]{schuhmacher2025observation}%
  \BibitemOpen
  \bibfield  {author} {\bibinfo {author} {\bibfnamefont {J.}~\bibnamefont {Schuhmacher}}, \bibinfo {author} {\bibfnamefont {G.-X.}\ \bibnamefont {Su}}, \bibinfo {author} {\bibfnamefont {J.~J.}\ \bibnamefont {Osborne}}, \bibinfo {author} {\bibfnamefont {A.}~\bibnamefont {Gandon}}, \bibinfo {author} {\bibfnamefont {J.~C.}\ \bibnamefont {Halimeh}},\ and\ \bibinfo {author} {\bibfnamefont {I.}~\bibnamefont {Tavernelli}},\ }\bibfield  {title} {\bibinfo {title} {Observation of hadron scattering in a lattice gauge theory on a quantum computer},\ }\href {https://doi.org/10.48550/arXiv.2505.20387} {\bibfield  {journal} {\bibinfo  {journal} {arXiv preprint arXiv:2505.20387}\ } (\bibinfo {year} {2025})}\BibitemShut {NoStop}%
\bibitem [{\citenamefont {Roy}\ \emph {et~al.}(2017)\citenamefont {Roy}, \citenamefont {Wilson},\ and\ \citenamefont {Firstenberg}}]{Roy2017}%
  \BibitemOpen
  \bibfield  {author} {\bibinfo {author} {\bibfnamefont {D.}~\bibnamefont {Roy}}, \bibinfo {author} {\bibfnamefont {C.~M.}\ \bibnamefont {Wilson}},\ and\ \bibinfo {author} {\bibfnamefont {O.}~\bibnamefont {Firstenberg}},\ }\bibfield  {title} {\bibinfo {title} {Colloquium: Strongly interacting photons in one-dimensional continuum},\ }\href {https://doi.org/10.1103/RevModPhys.89.021001} {\bibfield  {journal} {\bibinfo  {journal} {Rev. Mod. Phys.}\ }\textbf {\bibinfo {volume} {89}},\ \bibinfo {pages} {021001} (\bibinfo {year} {2017})}\BibitemShut {NoStop}%
\bibitem [{Note1()}]{Note1}%
  \BibitemOpen
  \bibinfo {note} {In our plots, the time is measured in units of a Rabi cycle, i.e. $2\pi /\Omega $}\BibitemShut {NoStop}%
\bibitem [{\citenamefont {Chen}\ \emph {et~al.}(2023)\citenamefont {Chen}, \citenamefont {Bornet}, \citenamefont {Bintz}, \citenamefont {Emperauger}, \citenamefont {Leclerc}, \citenamefont {Liu}, \citenamefont {Scholl}, \citenamefont {Barredo}, \citenamefont {Hauschild}, \citenamefont {Chatterjee} \emph {et~al.}}]{chen2023continuous}%
  \BibitemOpen
  \bibfield  {author} {\bibinfo {author} {\bibfnamefont {C.}~\bibnamefont {Chen}}, \bibinfo {author} {\bibfnamefont {G.}~\bibnamefont {Bornet}}, \bibinfo {author} {\bibfnamefont {M.}~\bibnamefont {Bintz}}, \bibinfo {author} {\bibfnamefont {G.}~\bibnamefont {Emperauger}}, \bibinfo {author} {\bibfnamefont {L.}~\bibnamefont {Leclerc}}, \bibinfo {author} {\bibfnamefont {V.~S.}\ \bibnamefont {Liu}}, \bibinfo {author} {\bibfnamefont {P.}~\bibnamefont {Scholl}}, \bibinfo {author} {\bibfnamefont {D.}~\bibnamefont {Barredo}}, \bibinfo {author} {\bibfnamefont {J.}~\bibnamefont {Hauschild}}, \bibinfo {author} {\bibfnamefont {S.}~\bibnamefont {Chatterjee}}, \emph {et~al.},\ }\bibfield  {title} {\bibinfo {title} {Continuous symmetry breaking in a two-dimensional rydberg array},\ }\href {https://doi.org/https://doi.org/10.1038/s41586-023-05859-2} {\bibfield  {journal} {\bibinfo  {journal} {Nature}\ }\textbf {\bibinfo {volume} {616}},\ \bibinfo {pages} {691} (\bibinfo {year} {2023})}\BibitemShut {NoStop}%
\bibitem [{\citenamefont {Manovitz}\ \emph {et~al.}(2025)\citenamefont {Manovitz}, \citenamefont {Li}, \citenamefont {Ebadi}, \citenamefont {Samajdar}, \citenamefont {Geim}, \citenamefont {Evered}, \citenamefont {Bluvstein}, \citenamefont {Zhou}, \citenamefont {Koyluoglu}, \citenamefont {Feldmeier} \emph {et~al.}}]{manovitz2025quantum}%
  \BibitemOpen
  \bibfield  {author} {\bibinfo {author} {\bibfnamefont {T.}~\bibnamefont {Manovitz}}, \bibinfo {author} {\bibfnamefont {S.~H.}\ \bibnamefont {Li}}, \bibinfo {author} {\bibfnamefont {S.}~\bibnamefont {Ebadi}}, \bibinfo {author} {\bibfnamefont {R.}~\bibnamefont {Samajdar}}, \bibinfo {author} {\bibfnamefont {A.~A.}\ \bibnamefont {Geim}}, \bibinfo {author} {\bibfnamefont {S.~J.}\ \bibnamefont {Evered}}, \bibinfo {author} {\bibfnamefont {D.}~\bibnamefont {Bluvstein}}, \bibinfo {author} {\bibfnamefont {H.}~\bibnamefont {Zhou}}, \bibinfo {author} {\bibfnamefont {N.~U.}\ \bibnamefont {Koyluoglu}}, \bibinfo {author} {\bibfnamefont {J.}~\bibnamefont {Feldmeier}}, \emph {et~al.},\ }\bibfield  {title} {\bibinfo {title} {Quantum coarsening and collective dynamics on a programmable simulator},\ }\href {https://doi.org/https://doi.org/10.1038/s41586-024-08353-5} {\bibfield  {journal} {\bibinfo  {journal} {Nature}\ }\textbf {\bibinfo {volume} {638}},\ \bibinfo {pages} {86} (\bibinfo {year} {2025})}\BibitemShut {NoStop}%
\bibitem [{\citenamefont {de~Oliveira}\ \emph {et~al.}(2025)\citenamefont {de~Oliveira}, \citenamefont {Diamond-Hitchcock}, \citenamefont {Walker}, \citenamefont {Wells-Pestell}, \citenamefont {Pelegr\'{\i}}, \citenamefont {Picken}, \citenamefont {Malcolm}, \citenamefont {Daley}, \citenamefont {Bass},\ and\ \citenamefont {Pritchard}}]{deOlivera25}%
  \BibitemOpen
  \bibfield  {author} {\bibinfo {author} {\bibfnamefont {A.~G.}\ \bibnamefont {de~Oliveira}}, \bibinfo {author} {\bibfnamefont {E.}~\bibnamefont {Diamond-Hitchcock}}, \bibinfo {author} {\bibfnamefont {D.~M.}\ \bibnamefont {Walker}}, \bibinfo {author} {\bibfnamefont {M.~T.}\ \bibnamefont {Wells-Pestell}}, \bibinfo {author} {\bibfnamefont {G.}~\bibnamefont {Pelegr\'{\i}}}, \bibinfo {author} {\bibfnamefont {C.~J.}\ \bibnamefont {Picken}}, \bibinfo {author} {\bibfnamefont {G.~P.~A.}\ \bibnamefont {Malcolm}}, \bibinfo {author} {\bibfnamefont {A.~J.}\ \bibnamefont {Daley}}, \bibinfo {author} {\bibfnamefont {J.}~\bibnamefont {Bass}},\ and\ \bibinfo {author} {\bibfnamefont {J.~D.}\ \bibnamefont {Pritchard}},\ }\bibfield  {title} {\bibinfo {title} {Demonstration of weighted-graph optimization on a rydberg-atom array using local light shifts},\ }\href {https://doi.org/10.1103/PRXQuantum.6.010301} {\bibfield  {journal} {\bibinfo  {journal} {PRX Quantum}\ }\textbf {\bibinfo {volume} {6}},\ \bibinfo {pages} {010301} (\bibinfo {year} {2025})}\BibitemShut {NoStop}%
\bibitem [{\citenamefont {Wang}\ \emph {et~al.}(2025{\natexlab{a}})\citenamefont {Wang}, \citenamefont {Tang}, \citenamefont {Du},\ and\ \citenamefont {Zhang}}]{Wang2025}%
  \BibitemOpen
  \bibfield  {author} {\bibinfo {author} {\bibfnamefont {J.-J.}\ \bibnamefont {Wang}}, \bibinfo {author} {\bibfnamefont {L.-Z.}\ \bibnamefont {Tang}}, \bibinfo {author} {\bibfnamefont {Y.-X.}\ \bibnamefont {Du}},\ and\ \bibinfo {author} {\bibfnamefont {D.-W.}\ \bibnamefont {Zhang}},\ }\bibfield  {title} {\bibinfo {title} {Discrete time crystals enhanced by stark potentials in rydberg atom arrays},\ }\href {https://doi.org/https://doi.org/10.1016/j.physleta.2025.130896} {\bibfield  {journal} {\bibinfo  {journal} {Physics Letters A}\ }\textbf {\bibinfo {volume} {558}},\ \bibinfo {pages} {130896} (\bibinfo {year} {2025}{\natexlab{a}})}\BibitemShut {NoStop}%
\bibitem [{\citenamefont {Wang}\ \emph {et~al.}(2025{\natexlab{b}})\citenamefont {Wang}, \citenamefont {Xu}, \citenamefont {Li}, \citenamefont {Vuleti\ifmmode~\acute{c}\else \'{c}\fi{}},\ and\ \citenamefont {Cappellaro}}]{Wang2025individual}%
  \BibitemOpen
  \bibfield  {author} {\bibinfo {author} {\bibfnamefont {G.}~\bibnamefont {Wang}}, \bibinfo {author} {\bibfnamefont {W.}~\bibnamefont {Xu}}, \bibinfo {author} {\bibfnamefont {C.}~\bibnamefont {Li}}, \bibinfo {author} {\bibfnamefont {V.}~\bibnamefont {Vuleti\ifmmode~\acute{c}\else \'{c}\fi{}}},\ and\ \bibinfo {author} {\bibfnamefont {P.}~\bibnamefont {Cappellaro}},\ }\bibfield  {title} {\bibinfo {title} {Individual-atom control in an array through phase modulation},\ }\href {https://doi.org/10.1103/PhysRevApplied.23.024072} {\bibfield  {journal} {\bibinfo  {journal} {Phys. Rev. Appl.}\ }\textbf {\bibinfo {volume} {23}},\ \bibinfo {pages} {024072} (\bibinfo {year} {2025}{\natexlab{b}})}\BibitemShut {NoStop}%
\bibitem [{Note2()}]{Note2}%
  \BibitemOpen
  \bibinfo {note} {Using the TenPy library, we employ, in particular, the two-site density matrix renormalization group (DMRG) method to prepare the initial ground state and the two-site time-dependent variational principle (TDVP) for the time evolution. In all the simulations reported in this paper, the scattering involves only few particles, and the entanglement growth is bounded, such that relatively small bond dimensions can be used, even for long-time evolutions. Where not explicitly stated, the bond dimension used in the simulations was $\chi =200$.}\BibitemShut {Stop}%
\bibitem [{\citenamefont {Simon}\ \emph {et~al.}(2011)\citenamefont {Simon}, \citenamefont {Bakr}, \citenamefont {Ma}, \citenamefont {Tai}, \citenamefont {Preiss},\ and\ \citenamefont {Greiner}}]{Simon2011-gs}%
  \BibitemOpen
  \bibfield  {author} {\bibinfo {author} {\bibfnamefont {J.}~\bibnamefont {Simon}}, \bibinfo {author} {\bibfnamefont {W.~S.}\ \bibnamefont {Bakr}}, \bibinfo {author} {\bibfnamefont {R.}~\bibnamefont {Ma}}, \bibinfo {author} {\bibfnamefont {M.~E.}\ \bibnamefont {Tai}}, \bibinfo {author} {\bibfnamefont {P.~M.}\ \bibnamefont {Preiss}},\ and\ \bibinfo {author} {\bibfnamefont {M.}~\bibnamefont {Greiner}},\ }\bibfield  {title} {\bibinfo {title} {Quantum simulation of antiferromagnetic spin chains in an optical lattice},\ }\href {https://doi.org/10.1038/nature09994} {\bibfield  {journal} {\bibinfo  {journal} {Nature}\ }\textbf {\bibinfo {volume} {472}},\ \bibinfo {pages} {307} (\bibinfo {year} {2011})}\BibitemShut {NoStop}%
\bibitem [{\citenamefont {Labuhn}\ \emph {et~al.}(2016)\citenamefont {Labuhn}, \citenamefont {Barredo}, \citenamefont {Ravets}, \citenamefont {de~L{\'e}s{\'e}leuc}, \citenamefont {Macr{\`\i}}, \citenamefont {Lahaye},\ and\ \citenamefont {Browaeys}}]{Labuhn2016-jw}%
  \BibitemOpen
  \bibfield  {author} {\bibinfo {author} {\bibfnamefont {H.}~\bibnamefont {Labuhn}}, \bibinfo {author} {\bibfnamefont {D.}~\bibnamefont {Barredo}}, \bibinfo {author} {\bibfnamefont {S.}~\bibnamefont {Ravets}}, \bibinfo {author} {\bibfnamefont {S.}~\bibnamefont {de~L{\'e}s{\'e}leuc}}, \bibinfo {author} {\bibfnamefont {T.}~\bibnamefont {Macr{\`\i}}}, \bibinfo {author} {\bibfnamefont {T.}~\bibnamefont {Lahaye}},\ and\ \bibinfo {author} {\bibfnamefont {A.}~\bibnamefont {Browaeys}},\ }\bibfield  {title} {\bibinfo {title} {Tunable two-dimensional arrays of single rydberg atoms for realizing quantum ising models},\ }\href {https://doi.org/10.1038/nature18274} {\bibfield  {journal} {\bibinfo  {journal} {Nature}\ }\textbf {\bibinfo {volume} {534}},\ \bibinfo {pages} {667} (\bibinfo {year} {2016})}\BibitemShut {NoStop}%
\bibitem [{\citenamefont {de~L\'es\'eleuc}\ \emph {et~al.}(2018)\citenamefont {de~L\'es\'eleuc}, \citenamefont {Weber}, \citenamefont {Lienhard}, \citenamefont {Barredo}, \citenamefont {B\"uchler}, \citenamefont {Lahaye},\ and\ \citenamefont {Browaeys}}]{deLeseleuc2018}%
  \BibitemOpen
  \bibfield  {author} {\bibinfo {author} {\bibfnamefont {S.}~\bibnamefont {de~L\'es\'eleuc}}, \bibinfo {author} {\bibfnamefont {S.}~\bibnamefont {Weber}}, \bibinfo {author} {\bibfnamefont {V.}~\bibnamefont {Lienhard}}, \bibinfo {author} {\bibfnamefont {D.}~\bibnamefont {Barredo}}, \bibinfo {author} {\bibfnamefont {H.~P.}\ \bibnamefont {B\"uchler}}, \bibinfo {author} {\bibfnamefont {T.}~\bibnamefont {Lahaye}},\ and\ \bibinfo {author} {\bibfnamefont {A.}~\bibnamefont {Browaeys}},\ }\bibfield  {title} {\bibinfo {title} {Accurate mapping of multilevel rydberg atoms on interacting spin-$1/2$ particles for the quantum simulation of ising models},\ }\href {https://doi.org/10.1103/PhysRevLett.120.113602} {\bibfield  {journal} {\bibinfo  {journal} {Phys. Rev. Lett.}\ }\textbf {\bibinfo {volume} {120}},\ \bibinfo {pages} {113602} (\bibinfo {year} {2018})}\BibitemShut {NoStop}%
\bibitem [{\citenamefont {Monroe}\ \emph {et~al.}(2021)\citenamefont {Monroe}, \citenamefont {Campbell}, \citenamefont {Duan}, \citenamefont {Gong}, \citenamefont {Gorshkov}, \citenamefont {Hess}, \citenamefont {Islam}, \citenamefont {Kim}, \citenamefont {Linke}, \citenamefont {Pagano}, \citenamefont {Richerme}, \citenamefont {Senko},\ and\ \citenamefont {Yao}}]{Monroe2021}%
  \BibitemOpen
  \bibfield  {author} {\bibinfo {author} {\bibfnamefont {C.}~\bibnamefont {Monroe}}, \bibinfo {author} {\bibfnamefont {W.~C.}\ \bibnamefont {Campbell}}, \bibinfo {author} {\bibfnamefont {L.-M.}\ \bibnamefont {Duan}}, \bibinfo {author} {\bibfnamefont {Z.-X.}\ \bibnamefont {Gong}}, \bibinfo {author} {\bibfnamefont {A.~V.}\ \bibnamefont {Gorshkov}}, \bibinfo {author} {\bibfnamefont {P.~W.}\ \bibnamefont {Hess}}, \bibinfo {author} {\bibfnamefont {R.}~\bibnamefont {Islam}}, \bibinfo {author} {\bibfnamefont {K.}~\bibnamefont {Kim}}, \bibinfo {author} {\bibfnamefont {N.~M.}\ \bibnamefont {Linke}}, \bibinfo {author} {\bibfnamefont {G.}~\bibnamefont {Pagano}}, \bibinfo {author} {\bibfnamefont {P.}~\bibnamefont {Richerme}}, \bibinfo {author} {\bibfnamefont {C.}~\bibnamefont {Senko}},\ and\ \bibinfo {author} {\bibfnamefont {N.~Y.}\ \bibnamefont {Yao}},\ }\bibfield  {title} {\bibinfo {title} {Programmable quantum simulations of spin systems with trapped ions},\ }\href {https://doi.org/10.1103/RevModPhys.93.025001} {\bibfield  {journal} {\bibinfo  {journal} {Rev. Mod. Phys.}\ }\textbf {\bibinfo {volume} {93}},\ \bibinfo {pages} {025001} (\bibinfo {year} {2021})}\BibitemShut {NoStop}%
\bibitem [{\citenamefont {Tan}\ \emph {et~al.}(2021)\citenamefont {Tan}, \citenamefont {Becker}, \citenamefont {Liu}, \citenamefont {Pagano}, \citenamefont {Collins}, \citenamefont {De}, \citenamefont {Feng}, \citenamefont {Kaplan}, \citenamefont {Kyprianidis}, \citenamefont {Lundgren}, \citenamefont {Morong}, \citenamefont {Whitsitt}, \citenamefont {Gorshkov},\ and\ \citenamefont {Monroe}}]{Tan2021-eg}%
  \BibitemOpen
  \bibfield  {author} {\bibinfo {author} {\bibfnamefont {W.~L.}\ \bibnamefont {Tan}}, \bibinfo {author} {\bibfnamefont {P.}~\bibnamefont {Becker}}, \bibinfo {author} {\bibfnamefont {F.}~\bibnamefont {Liu}}, \bibinfo {author} {\bibfnamefont {G.}~\bibnamefont {Pagano}}, \bibinfo {author} {\bibfnamefont {K.~S.}\ \bibnamefont {Collins}}, \bibinfo {author} {\bibfnamefont {A.}~\bibnamefont {De}}, \bibinfo {author} {\bibfnamefont {L.}~\bibnamefont {Feng}}, \bibinfo {author} {\bibfnamefont {H.~B.}\ \bibnamefont {Kaplan}}, \bibinfo {author} {\bibfnamefont {A.}~\bibnamefont {Kyprianidis}}, \bibinfo {author} {\bibfnamefont {R.}~\bibnamefont {Lundgren}}, \bibinfo {author} {\bibfnamefont {W.}~\bibnamefont {Morong}}, \bibinfo {author} {\bibfnamefont {S.}~\bibnamefont {Whitsitt}}, \bibinfo {author} {\bibfnamefont {A.~V.}\ \bibnamefont {Gorshkov}},\ and\ \bibinfo {author} {\bibfnamefont {C.}~\bibnamefont {Monroe}},\ }\bibfield  {title} {\bibinfo {title} {Domain-wall confinement and dynamics in a quantum simulator},\ }\href {https://doi.org/10.1038/s41567-021-01194-3} {\bibfield  {journal} {\bibinfo  {journal} {Nat. Phys.}\ }\textbf {\bibinfo {volume} {17}},\ \bibinfo {pages} {742} (\bibinfo {year} {2021})}\BibitemShut {NoStop}%
\bibitem [{\citenamefont {De}\ \emph {et~al.}(2024)\citenamefont {De}, \citenamefont {Lerose}, \citenamefont {Luo}, \citenamefont {Surace}, \citenamefont {Schuckert}, \citenamefont {Bennewitz}, \citenamefont {Ware}, \citenamefont {Morong}, \citenamefont {Collins}, \citenamefont {Davoudi} \emph {et~al.}}]{de2024}%
  \BibitemOpen
  \bibfield  {author} {\bibinfo {author} {\bibfnamefont {A.}~\bibnamefont {De}}, \bibinfo {author} {\bibfnamefont {A.}~\bibnamefont {Lerose}}, \bibinfo {author} {\bibfnamefont {D.}~\bibnamefont {Luo}}, \bibinfo {author} {\bibfnamefont {F.~M.}\ \bibnamefont {Surace}}, \bibinfo {author} {\bibfnamefont {A.}~\bibnamefont {Schuckert}}, \bibinfo {author} {\bibfnamefont {E.~R.}\ \bibnamefont {Bennewitz}}, \bibinfo {author} {\bibfnamefont {B.}~\bibnamefont {Ware}}, \bibinfo {author} {\bibfnamefont {W.}~\bibnamefont {Morong}}, \bibinfo {author} {\bibfnamefont {K.~S.}\ \bibnamefont {Collins}}, \bibinfo {author} {\bibfnamefont {Z.}~\bibnamefont {Davoudi}}, \emph {et~al.},\ }\bibfield  {title} {\bibinfo {title} {Observation of string-breaking dynamics in a quantum simulator},\ }\href {https://arxiv.org/abs/2410.13815} {\bibfield  {journal} {\bibinfo  {journal} {arXiv preprint arXiv:2410.13815}\ } (\bibinfo {year} {2024})}\BibitemShut {NoStop}%
\bibitem [{\citenamefont {Luo}\ \emph {et~al.}(2025)\citenamefont {Luo}, \citenamefont {Surace}, \citenamefont {De}, \citenamefont {Lerose}, \citenamefont {Bennewitz}, \citenamefont {Ware}, \citenamefont {Schuckert}, \citenamefont {Davoudi}, \citenamefont {Gorshkov}, \citenamefont {Katz} \emph {et~al.}}]{luo2025}%
  \BibitemOpen
  \bibfield  {author} {\bibinfo {author} {\bibfnamefont {D.}~\bibnamefont {Luo}}, \bibinfo {author} {\bibfnamefont {F.~M.}\ \bibnamefont {Surace}}, \bibinfo {author} {\bibfnamefont {A.}~\bibnamefont {De}}, \bibinfo {author} {\bibfnamefont {A.}~\bibnamefont {Lerose}}, \bibinfo {author} {\bibfnamefont {E.~R.}\ \bibnamefont {Bennewitz}}, \bibinfo {author} {\bibfnamefont {B.}~\bibnamefont {Ware}}, \bibinfo {author} {\bibfnamefont {A.}~\bibnamefont {Schuckert}}, \bibinfo {author} {\bibfnamefont {Z.}~\bibnamefont {Davoudi}}, \bibinfo {author} {\bibfnamefont {A.~V.}\ \bibnamefont {Gorshkov}}, \bibinfo {author} {\bibfnamefont {O.}~\bibnamefont {Katz}}, \emph {et~al.},\ }\bibfield  {title} {\bibinfo {title} {Quantum simulation of bubble nucleation across a quantum phase transition},\ }\href {https://arxiv.org/abs/2505.09607} {\bibfield  {journal} {\bibinfo  {journal} {arXiv preprint arXiv:2505.09607}\ } (\bibinfo {year} {2025})}\BibitemShut {NoStop}%
\bibitem [{\citenamefont {McCoy}\ and\ \citenamefont {Wu}(1978)}]{McCoyWu1978}%
  \BibitemOpen
  \bibfield  {author} {\bibinfo {author} {\bibfnamefont {B.~M.}\ \bibnamefont {McCoy}}\ and\ \bibinfo {author} {\bibfnamefont {T.~T.}\ \bibnamefont {Wu}},\ }\bibfield  {title} {\bibinfo {title} {Two-dimensional ising field theory in a magnetic field: Breakup of the cut in the two-point function},\ }\href {https://doi.org/10.1103/PhysRevD.18.1259} {\bibfield  {journal} {\bibinfo  {journal} {Phys. Rev. D}\ }\textbf {\bibinfo {volume} {18}},\ \bibinfo {pages} {1259} (\bibinfo {year} {1978})}\BibitemShut {NoStop}%
\bibitem [{\citenamefont {Delfino}\ \emph {et~al.}(1996)\citenamefont {Delfino}, \citenamefont {Mussardo},\ and\ \citenamefont {Simonetti}}]{DELFINO1996469}%
  \BibitemOpen
  \bibfield  {author} {\bibinfo {author} {\bibfnamefont {G.}~\bibnamefont {Delfino}}, \bibinfo {author} {\bibfnamefont {G.}~\bibnamefont {Mussardo}},\ and\ \bibinfo {author} {\bibfnamefont {P.}~\bibnamefont {Simonetti}},\ }\bibfield  {title} {\bibinfo {title} {Non-integrable quantum field theories as perturbations of certain integrable models},\ }\href {https://doi.org/https://doi.org/10.1016/0550-3213(96)00265-9} {\bibfield  {journal} {\bibinfo  {journal} {Nuclear Physics B}\ }\textbf {\bibinfo {volume} {473}},\ \bibinfo {pages} {469} (\bibinfo {year} {1996})}\BibitemShut {NoStop}%
\bibitem [{\citenamefont {Kormos}\ \emph {et~al.}(2017)\citenamefont {Kormos}, \citenamefont {Collura}, \citenamefont {Tak{\'a}cs},\ and\ \citenamefont {Calabrese}}]{Kormos2017-ar}%
  \BibitemOpen
  \bibfield  {author} {\bibinfo {author} {\bibfnamefont {M.}~\bibnamefont {Kormos}}, \bibinfo {author} {\bibfnamefont {M.}~\bibnamefont {Collura}}, \bibinfo {author} {\bibfnamefont {G.}~\bibnamefont {Tak{\'a}cs}},\ and\ \bibinfo {author} {\bibfnamefont {P.}~\bibnamefont {Calabrese}},\ }\bibfield  {title} {\bibinfo {title} {Real-time confinement following a quantum quench to a non-integrable model},\ }\href {https://doi.org/10.1038/nphys3934} {\bibfield  {journal} {\bibinfo  {journal} {Nat. Phys.}\ }\textbf {\bibinfo {volume} {13}},\ \bibinfo {pages} {246} (\bibinfo {year} {2017})}\BibitemShut {NoStop}%
\bibitem [{\citenamefont {Wang}\ \emph {et~al.}(2020)\citenamefont {Wang}, \citenamefont {Cui}, \citenamefont {Lu}, \citenamefont {Zhang}, \citenamefont {Gao}, \citenamefont {Chang}, \citenamefont {Yung},\ and\ \citenamefont {Jin}}]{Wang2020Integrated}%
  \BibitemOpen
  \bibfield  {author} {\bibinfo {author} {\bibfnamefont {Y.}~\bibnamefont {Wang}}, \bibinfo {author} {\bibfnamefont {Z.-W.}\ \bibnamefont {Cui}}, \bibinfo {author} {\bibfnamefont {Y.-H.}\ \bibnamefont {Lu}}, \bibinfo {author} {\bibfnamefont {X.-M.}\ \bibnamefont {Zhang}}, \bibinfo {author} {\bibfnamefont {J.}~\bibnamefont {Gao}}, \bibinfo {author} {\bibfnamefont {Y.-J.}\ \bibnamefont {Chang}}, \bibinfo {author} {\bibfnamefont {M.-H.}\ \bibnamefont {Yung}},\ and\ \bibinfo {author} {\bibfnamefont {X.-M.}\ \bibnamefont {Jin}},\ }\bibfield  {title} {\bibinfo {title} {Integrated quantum-walk structure and nand tree on a photonic chip},\ }\href {https://doi.org/10.1103/PhysRevLett.125.160502} {\bibfield  {journal} {\bibinfo  {journal} {Phys. Rev. Lett.}\ }\textbf {\bibinfo {volume} {125}},\ \bibinfo {pages} {160502} (\bibinfo {year} {2020})}\BibitemShut {NoStop}%
\bibitem [{\citenamefont {Wang}\ \emph {et~al.}(2022)\citenamefont {Wang}, \citenamefont {Cheng}, \citenamefont {Cui},\ and\ \citenamefont {Yung}}]{wang2022quantum}%
  \BibitemOpen
  \bibfield  {author} {\bibinfo {author} {\bibfnamefont {F.}~\bibnamefont {Wang}}, \bibinfo {author} {\bibfnamefont {B.}~\bibnamefont {Cheng}}, \bibinfo {author} {\bibfnamefont {Z.-W.}\ \bibnamefont {Cui}},\ and\ \bibinfo {author} {\bibfnamefont {M.-H.}\ \bibnamefont {Yung}},\ }\bibfield  {title} {\bibinfo {title} {Quantum computing by quantum walk on quantum slide},\ }\href {https://arxiv.org/abs/2211.08659} {\bibfield  {journal} {\bibinfo  {journal} {arXiv preprint arXiv:2211.08659}\ } (\bibinfo {year} {2022})}\BibitemShut {NoStop}%
\bibitem [{Note3()}]{Note3}%
  \BibitemOpen
  \bibinfo {note} {Explicitly, the Hamiltonian is $H=-J\DOTSB \sum@ \slimits@ _{j=1}^{L-1}\sigma _j^z\sigma _{j+1}^z+\DOTSB \sum@ \slimits@ _{j=1}^{L}(g_j\sigma _j^x+h_j\sigma _{j}^z)$.}\BibitemShut {Stop}%
\bibitem [{Note4()}]{Note4}%
  \BibitemOpen
  \bibinfo {note} {It may seem natural to choose a linear interpolation from $h_0$ to $h$ for the longitudinal field. However, one must take into account that the auxiliary site, unlike the sites in the chain, has only a single neighbor rather than two. Ensuring a smooth evolution of the excitation energy along the interpolation therefore requires accounting for this asymmetry, modeling such effect as an additional longitudinal field. To obtain a smooth potential for the meson, we then design a ramp that linearly interpolates between $h_0-1$ and $h$. We find empirically that the choice of interpolation reported here performs well for our chosen value of $h_0$, although a different choice may be required for other values.}\BibitemShut {Stop}%
\bibitem [{\citenamefont {Hauschild}\ \emph {et~al.}(2024)\citenamefont {Hauschild}, \citenamefont {Unfried}, \citenamefont {Anand}, \citenamefont {Andrews}, \citenamefont {Bintz}, \citenamefont {Borla}, \citenamefont {Divic}, \citenamefont {Drescher}, \citenamefont {Geiger}, \citenamefont {Hefel}, \citenamefont {Hémery}, \citenamefont {Kadow}, \citenamefont {Kemp}, \citenamefont {Kirchner}, \citenamefont {Liu}, \citenamefont {Möller}, \citenamefont {Parker}, \citenamefont {Rader}, \citenamefont {Romen}, \citenamefont {Scalet}, \citenamefont {Schoonderwoerd}, \citenamefont {Schulz}, \citenamefont {Soejima}, \citenamefont {Thoma}, \citenamefont {Wu}, \citenamefont {Zechmann}, \citenamefont {Zweng}, \citenamefont {Mong}, \citenamefont {Zaletel},\ and\ \citenamefont {Pollmann}}]{tenpy2024}%
  \BibitemOpen
  \bibfield  {author} {\bibinfo {author} {\bibfnamefont {J.}~\bibnamefont {Hauschild}}, \bibinfo {author} {\bibfnamefont {J.}~\bibnamefont {Unfried}}, \bibinfo {author} {\bibfnamefont {S.}~\bibnamefont {Anand}}, \bibinfo {author} {\bibfnamefont {B.}~\bibnamefont {Andrews}}, \bibinfo {author} {\bibfnamefont {M.}~\bibnamefont {Bintz}}, \bibinfo {author} {\bibfnamefont {U.}~\bibnamefont {Borla}}, \bibinfo {author} {\bibfnamefont {S.}~\bibnamefont {Divic}}, \bibinfo {author} {\bibfnamefont {M.}~\bibnamefont {Drescher}}, \bibinfo {author} {\bibfnamefont {J.}~\bibnamefont {Geiger}}, \bibinfo {author} {\bibfnamefont {M.}~\bibnamefont {Hefel}}, \bibinfo {author} {\bibfnamefont {K.}~\bibnamefont {Hémery}}, \bibinfo {author} {\bibfnamefont {W.}~\bibnamefont {Kadow}}, \bibinfo {author} {\bibfnamefont {J.}~\bibnamefont {Kemp}}, \bibinfo {author} {\bibfnamefont {N.}~\bibnamefont {Kirchner}}, \bibinfo {author} {\bibfnamefont {V.~S.}\ \bibnamefont {Liu}}, \bibinfo {author} {\bibfnamefont {G.}~\bibnamefont {Möller}}, \bibinfo {author} {\bibfnamefont {D.}~\bibnamefont {Parker}}, \bibinfo {author} {\bibfnamefont {M.}~\bibnamefont {Rader}}, \bibinfo {author} {\bibfnamefont {A.}~\bibnamefont {Romen}}, \bibinfo {author} {\bibfnamefont {S.}~\bibnamefont {Scalet}}, \bibinfo {author} {\bibfnamefont {L.}~\bibnamefont {Schoonderwoerd}}, \bibinfo {author} {\bibfnamefont {M.}~\bibnamefont {Schulz}}, \bibinfo {author} {\bibfnamefont {T.}~\bibnamefont {Soejima}}, \bibinfo {author} {\bibfnamefont {P.}~\bibnamefont {Thoma}}, \bibinfo {author} {\bibfnamefont {Y.}~\bibnamefont {Wu}}, \bibinfo {author} {\bibfnamefont {P.}~\bibnamefont {Zechmann}}, \bibinfo {author} {\bibfnamefont {L.}~\bibnamefont {Zweng}}, \bibinfo {author} {\bibfnamefont {R.~S.~K.}\ \bibnamefont {Mong}}, \bibinfo {author} {\bibfnamefont {M.~P.}\ \bibnamefont {Zaletel}},\ and\ \bibinfo {author} {\bibfnamefont {F.}~\bibnamefont {Pollmann}},\ }\bibfield  {title} {\bibinfo {title} {{Tensor network Python (TeNPy) version 1}},\ }\href {https://doi.org/10.21468/SciPostPhysCodeb.41} {\bibfield  {journal} {\bibinfo  {journal} {SciPost Phys. Codebases}\ ,\ \bibinfo {pages} {41}} (\bibinfo {year} {2024})}\BibitemShut {NoStop}%
\bibitem [{\citenamefont {Rutkevich}(1999)}]{Rutkevich1999}%
  \BibitemOpen
  \bibfield  {author} {\bibinfo {author} {\bibfnamefont {S.~B.}\ \bibnamefont {Rutkevich}},\ }\bibfield  {title} {\bibinfo {title} {Decay of the metastable phase in $d=1$ and $d=2$ ising models},\ }\href {https://doi.org/10.1103/PhysRevB.60.14525} {\bibfield  {journal} {\bibinfo  {journal} {Phys. Rev. B}\ }\textbf {\bibinfo {volume} {60}},\ \bibinfo {pages} {14525} (\bibinfo {year} {1999})}\BibitemShut {NoStop}%
\bibitem [{\citenamefont {Rutkevich}(2005)}]{Rutkevich2005}%
  \BibitemOpen
  \bibfield  {author} {\bibinfo {author} {\bibfnamefont {S.~B.}\ \bibnamefont {Rutkevich}},\ }\bibfield  {title} {\bibinfo {title} {Large-$n$ excitations in the ferromagnetic ising field theory in a weak magnetic field: Mass spectrum and decay widths},\ }\href {https://doi.org/10.1103/PhysRevLett.95.250601} {\bibfield  {journal} {\bibinfo  {journal} {Phys. Rev. Lett.}\ }\textbf {\bibinfo {volume} {95}},\ \bibinfo {pages} {250601} (\bibinfo {year} {2005})}\BibitemShut {NoStop}%
\bibitem [{\citenamefont {Lagnese}\ \emph {et~al.}(2021)\citenamefont {Lagnese}, \citenamefont {Surace}, \citenamefont {Kormos},\ and\ \citenamefont {Calabrese}}]{Lagnese2021}%
  \BibitemOpen
  \bibfield  {author} {\bibinfo {author} {\bibfnamefont {G.}~\bibnamefont {Lagnese}}, \bibinfo {author} {\bibfnamefont {F.~M.}\ \bibnamefont {Surace}}, \bibinfo {author} {\bibfnamefont {M.}~\bibnamefont {Kormos}},\ and\ \bibinfo {author} {\bibfnamefont {P.}~\bibnamefont {Calabrese}},\ }\bibfield  {title} {\bibinfo {title} {False vacuum decay in quantum spin chains},\ }\href {https://doi.org/10.1103/PhysRevB.104.L201106} {\bibfield  {journal} {\bibinfo  {journal} {Phys. Rev. B}\ }\textbf {\bibinfo {volume} {104}},\ \bibinfo {pages} {L201106} (\bibinfo {year} {2021})}\BibitemShut {NoStop}%
\bibitem [{\citenamefont {Maertens}\ \emph {et~al.}(2025)\citenamefont {Maertens}, \citenamefont {Haegeman},\ and\ \citenamefont {Van~Acoleyen}}]{maertens2025real}%
  \BibitemOpen
  \bibfield  {author} {\bibinfo {author} {\bibfnamefont {D.}~\bibnamefont {Maertens}}, \bibinfo {author} {\bibfnamefont {J.}~\bibnamefont {Haegeman}},\ and\ \bibinfo {author} {\bibfnamefont {K.}~\bibnamefont {Van~Acoleyen}},\ }\bibfield  {title} {\bibinfo {title} {Real-time bubble nucleation and growth for false vacuum decay on the lattice},\ }\href {https://doi.org/10.48550/arXiv.2508.13645} {\bibfield  {journal} {\bibinfo  {journal} {arXiv preprint arXiv:2508.13645}\ } (\bibinfo {year} {2025})}\BibitemShut {NoStop}%
\bibitem [{\citenamefont {Johansen}\ \emph {et~al.}(2025)\citenamefont {Johansen}, \citenamefont {Recati}, \citenamefont {Carusotto},\ and\ \citenamefont {Biella}}]{johansen2025many}%
  \BibitemOpen
  \bibfield  {author} {\bibinfo {author} {\bibfnamefont {C.}~\bibnamefont {Johansen}}, \bibinfo {author} {\bibfnamefont {A.}~\bibnamefont {Recati}}, \bibinfo {author} {\bibfnamefont {I.}~\bibnamefont {Carusotto}},\ and\ \bibinfo {author} {\bibfnamefont {A.}~\bibnamefont {Biella}},\ }\bibfield  {title} {\bibinfo {title} {Many-body theory of false vacuum decay in quantum spin chains},\ }\bibfield  {journal} {\bibinfo  {journal} {arXiv preprint arXiv:2508.13780}\ }\href {https://doi.org/https://doi.org/10.48550/arXiv.2508.13780} {https://doi.org/10.48550/arXiv.2508.13780} (\bibinfo {year} {2025})\BibitemShut {NoStop}%
\bibitem [{\citenamefont {Yin}\ \emph {et~al.}(2025)\citenamefont {Yin}, \citenamefont {Surace},\ and\ \citenamefont {Lucas}}]{Yin2025}%
  \BibitemOpen
  \bibfield  {author} {\bibinfo {author} {\bibfnamefont {C.}~\bibnamefont {Yin}}, \bibinfo {author} {\bibfnamefont {F.~M.}\ \bibnamefont {Surace}},\ and\ \bibinfo {author} {\bibfnamefont {A.}~\bibnamefont {Lucas}},\ }\bibfield  {title} {\bibinfo {title} {Theory of metastable states in many-body quantum systems},\ }\href {https://doi.org/10.1103/PhysRevX.15.011064} {\bibfield  {journal} {\bibinfo  {journal} {Phys. Rev. X}\ }\textbf {\bibinfo {volume} {15}},\ \bibinfo {pages} {011064} (\bibinfo {year} {2025})}\BibitemShut {NoStop}%
\bibitem [{\citenamefont {Christandl}\ \emph {et~al.}(2004)\citenamefont {Christandl}, \citenamefont {Datta}, \citenamefont {Ekert},\ and\ \citenamefont {Landahl}}]{Christandl2004}%
  \BibitemOpen
  \bibfield  {author} {\bibinfo {author} {\bibfnamefont {M.}~\bibnamefont {Christandl}}, \bibinfo {author} {\bibfnamefont {N.}~\bibnamefont {Datta}}, \bibinfo {author} {\bibfnamefont {A.}~\bibnamefont {Ekert}},\ and\ \bibinfo {author} {\bibfnamefont {A.~J.}\ \bibnamefont {Landahl}},\ }\bibfield  {title} {\bibinfo {title} {Perfect state transfer in quantum spin networks},\ }\href {https://doi.org/10.1103/PhysRevLett.92.187902} {\bibfield  {journal} {\bibinfo  {journal} {Phys. Rev. Lett.}\ }\textbf {\bibinfo {volume} {92}},\ \bibinfo {pages} {187902} (\bibinfo {year} {2004})}\BibitemShut {NoStop}%
\end{thebibliography}%

\appendix

\section{Exact solution for the single-particle model}
\label{app:sp}
The eigenstates of the Hamiltonian in Eq.~(\ref{eq:Hwv}) can be obtained using the ansatz
\begin{equation}
\label{eq:ansatz}
    \psi_k(j)\equiv \braket{j|\psi_k}=\begin{cases}
        A_k e^{ikj}+B_ke^{-ikj}&\qquad \text{for }j>0,\\
        C_k&\qquad \text{for }j=0,
    \end{cases}
\end{equation}
where $k\in [0,\pi]$. For $j>1$ this state satisfies the Schroedinger equation
\begin{equation}
    -w\Big(\psi_k(j-1)+\psi_k(j+1)\Big)=E_k \psi_k(j),
\end{equation}
with $E_k=-2w\cos k$. We then have to impose that it also satifies the Schroedinger equation for $j=0,1$:
\begin{align}
    &V_0\psi_k(0)-w'\psi_k(1)=-2w\cos(k)\psi_k(0)\\
    &-w' \psi_k(0)-w\psi_k(2)=-2w\cos(k)\psi_k(1),
\end{align}
from which we get 
\begin{align}
    &\Big(V_0+2w\cos(k)\Big)C_k-w'(A_ke^{ik}+B_ke^{-ik})=0 \label{eq:cond1}\\
    &-w'C_k+w(A_k+B_k)=0.\label{eq:cond2}
\end{align}
To solve this system of equations, we note that the conservation of probability implies $|A_k|^2=|B_k|^2$, since $B_k$ and $A_k$ represent the amplitudes of an incoming and a reflected wave, respectively. We can then set $|A_k|=|B_k|=1$ (fixing the normalization of the propagating wave) and define $z_k$ such that $A_k=-i e^{iz_k}$ and $B_k=A_k^*$ (fixing the global phase of the wavefunction). From Eq.~(\ref{eq:cond2}) we obtain
\begin{equation}
    C_k=2\frac{w}{w'}\sin z_k.
\end{equation}
The condition in Eq.~(\ref{eq:cond1}) then results in the following equation for $z_k$:
\begin{equation}
    z_k =\tan^{-1}\left(\frac{(w')^2 \sin k}{wV_0-[(w')^2-2w^2] \cos k }\right).
\end{equation}

Additionally, for $|V_0|\le 2|w|-\frac{(w')^2}{|w|}$ there are no bound states (i.e., no eigenstates that decay as $e^{-\kappa j}$ for large $j$), so all energy eigenstates are of the form in Eq.~(\ref{eq:ansatz}).

We can use the fact that $\ket{\psi_k}$ form a complete orthogonal set and $\braket{\psi_{k'}|\psi_k}=2\pi\delta(k-k')$ to write the initial state $\ket{\psi(t=0)}=\ket{0}$ as 

\begin{equation}
    \ket{0}=\int_0^{\pi} \frac{dk}{2\pi} C_k^* \ket{\psi_k}.
\end{equation}
The distribution $|C_k|^2=4\left(\frac{w}{w'}\right)^2 \sin^2 z_k$ then corresponds to the expected momentum distribution at long times.

\section{Dispersion relations and kinematic analysis}
\label{app:disp}

We here show the numerical results about the dispersion relations in the Rydberg atom chain and the mixed-field quantum Ising chain.

\begin{figure}[t]
    \centering
    \begin{minipage}[t]{0.50\linewidth}
    \vspace{0pt}
    \includegraphics[width=\linewidth]{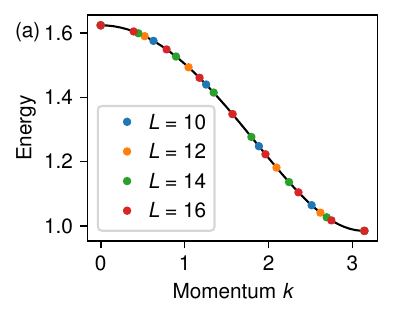}
    
    \end{minipage}
    \begin{minipage}[t]{0.48\linewidth}
    \vspace{2pt}
    \includegraphics[width=\linewidth]{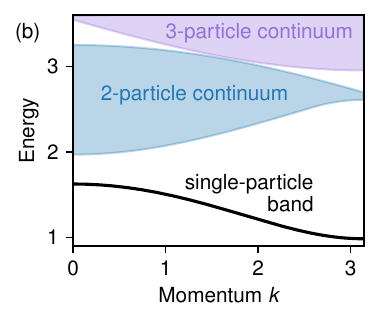}
    \end{minipage}
    \caption{(a) Dispersion relation of the Rydberg atom chain for the same parameters as in Sec.~\ref{sec:MB}. The low-energy spectrum is obtained by exact diagonalization for several system sizes $L$ (colored dots) and is already well converged for $L\geq 10$. The solid line shows a cubic-spline interpolation of the numerical data. (b) Two-particle and three-particle continua constructed from the single-particle dispersion by summing the corresponding energies and momenta. For the chosen parameters, the two continua do not overlap, indicating that inelastic scattering processes are kinematically forbidden.}
    \label{fig:spec}
\end{figure}

\subsection{Rydberg atom chain}
We used exact diagonalization to compute the momentum-resolved low-energy spectrum of the Hamiltonian in Eq.~(\ref{eq:Ryd}) with periodic boundary conditions. The results are shown in Fig.~\ref{fig:spec}a for different system sizes $L$, and appear to be well converged with $L$, exhibiting no visible finite size effects. We used the single-particle band obtained with this method to further compute the 2-particle and 3-particle continua (Fig.~\ref{fig:spec}b): these data show that, for the parameters chosen in the numerical simulation, the 2-particle and 3-particle continua do not overlap, so no inelastic channel is available in the collision of two particles.

\subsection{Mixed field quantum Ising chain}

Together with exact diagonalization, the dispersion relations of the meson bands in the mixed-field quantum Ising chain can be computed with high accuracy using a two-fermion approximation. This approach is expected to be reliable in the regime of weak longitudinal field $h$. Here we briefly summarize the method and refer to Refs.~\cite{Rutkevich1999,Rutkevich2005,Lagnese2021,maertens2025real,johansen2025many} for a detailed derivation.

The key observation is that the Hamiltonian is exactly solvable when $h=0$. Applying a Jordan--Wigner transformation followed by a Bogoliubov transformation maps the model to free fermions,
\begin{equation}
H_0=\int_{-\pi}^{\pi}\frac{d\theta}{2\pi}\,\omega(\theta)\,\gamma^\dagger(\theta)\gamma(\theta),
\end{equation}
where $\gamma(\theta)$ are fermionic quasiparticle operators and $\omega(\theta)$ is the single-particle dispersion relation. We then introduce the real-space Bogoliubov fermions
\begin{equation}
b_j=\int_{-\pi}^{\pi}\frac{d\theta}{2\pi}\,\gamma(\theta)e^{ij\theta}.
\end{equation}

Within the two-fermion approximation, the $\ell$-th meson state with momentum $k$ is described by the ansatz
\begin{equation}
\ket{\Psi_{k,\ell}}
=
\sum_{j,n>0}
e^{ik(j+\frac{n}{2})}
f_{\ell}(n)\,
b_j^\dagger b_{j+n}^\dagger
\ket{0},
\end{equation}
where $\ket{0}$ denotes the ground state of $H_0$. The function $f_{\ell}(n)$ describes the relative wavefunction of the two fermions, while the index $\ell$ labels the different meson bands. It is determined by requiring that $\ket{\Psi_{k,\ell}}$ be an eigenstate of the full Hamiltonian projected onto the two-fermion sector.

This condition leads to the effective eigenvalue equation
\begin{widetext}
\begin{equation}
2hMn\,f_{\ell}(n)
+
2\sum_{m>0}
f_{\ell}(m)
\left[
\cos\!\left(\frac{k(n-m)}{2}\right)K_{n-m}
-
\cos\!\left(\frac{k(n+m)}{2}\right)K_{n+m}
\right]
=
E_{\ell}(k)\,f_{\ell}(n),
\end{equation}
\end{widetext}
where
\begin{equation}
K_n
=
\int_{-\pi}^{\pi}
\frac{d\theta}{2\pi}\,
\omega(\theta)e^{in\theta},
\end{equation}
and $M=(1-g^2)^{1/8}$ is the spontaneous magnetization of the transverse-field Ising model in the limit $h\to 0$. The term proportional to $hMn$ acts as a linear confining potential between the two fermions, leading to the formation of mesonic bound states. Solving the above eigenvalue problem yields the meson dispersion relations $E_\ell(k)$.

\begin{figure}[b]
    \centering
    \begin{minipage}[t]{0.50\linewidth}
    \vspace{0pt}
    \includegraphics[width=\linewidth]{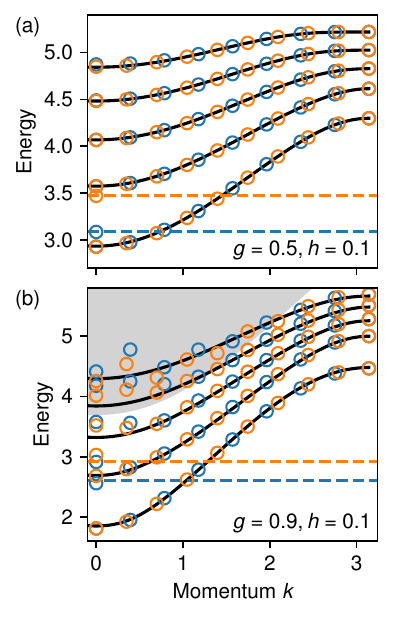}
    \end{minipage}
    \begin{minipage}[t]{0.48\linewidth}
    \vspace{2pt}
    \includegraphics[width=\linewidth]{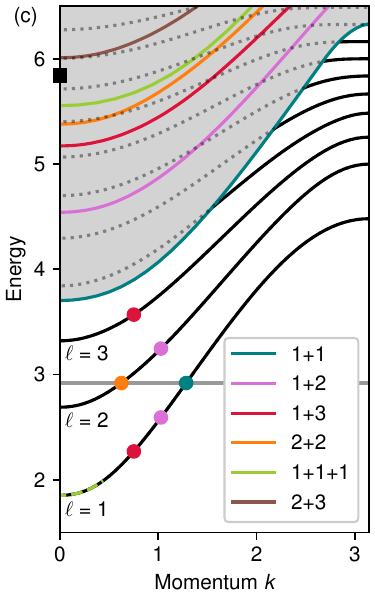}
    \end{minipage}
    \caption{(a,b) Dispersion relations of the five lowest meson bands in the mixed-field quantum Ising chain, computed for two different parameter sets. Exact diagonalization results are shown for system sizes $L=16$ (blue symbols) and $L=18$ (orange symbols). The solid lines correspond to predictions from the two-fermion approximation. The onset of the continuum spectrum is indicated by the grey shaded region. The dashed horizontal lines mark the estimated energy of the false vacuum state $E\approx 2hML$ for the two system sizes. (c) Low-energy spectrum of the mixed-field quantum Ising chain for $g=0.9$ and $h=0.1$. The solid black lines show the meson bands obtained within the two-fermion approximation. The same bands are continued as dotted grey lines in regions where the mesons become unstable due to their overlap with the continuum (grey shaded area). Colored curves indicate the thresholds of the various multi-particle continua; for example, ``1+1'' denotes the continuum formed by two particles from the $\ell=1$ meson band.
The total energy and momentum of the collision considered in Fig.~\ref{fig:inelastic_scatt} are marked by a black square. This point lies above several continuum thresholds, implying that inelastic scattering is kinematically allowed. In particular, the collision can decay into final states of the types $1+1$, $1+2$, $1+3$, $2+2$, and $1+1+1$. The energies and momenta of the individual particles corresponding to these final-state channels are indicated by colored circles.}
    \label{fig:specIsing}
\end{figure}

The results of the two-fermion approximation are compared with exact diagonalization in Fig.~\ref{fig:specIsing}a,b. In the regions where the exact diagonalization spectrum appears converged with system size, the agreement between the two methods is excellent. This confirms that the low-energy excitations are accurately captured by the two-fermion description.

At $k=0$, the exact diagonalization spectrum contains an additional state whose energy lies within the range spanned by the single-meson bands. This state can be interpreted as a false vacuum state \cite{Rutkevich1999,Lagnese2021,Yin2025}, whose energy (approximately $2hML$) grows proportionally to the system size, indicating that it does not belong to the low-energy spectrum in the thermodynamic limit.

Additional finite-size effects are visible in regions where the meson bands overlap with the continuum. This behavior is expected, since mesons are no longer stable in this regime and can hybridize with multi-particle states. As a result, the finite-size spectrum exhibits stronger size dependence, and the assumptions underlying the two-fermion approximation become less accurate.

\section{Quantum slide in a single-particle model}
\label{app:slide}

\begin{figure}
    \centering
    \includegraphics[width=\linewidth]{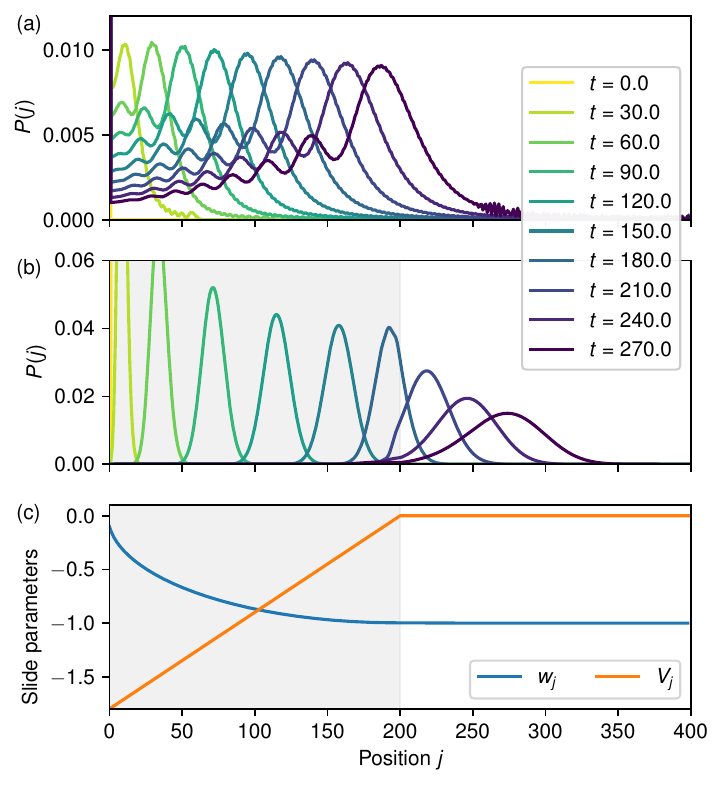}
    \caption{(a) Probability distribution $P(j)$ for a wave packet prepared as in Fig.~\ref{fig:single_x}. (b) Probability distribution $P(j)$ for a wave packet with the same mean energy and energy variance prepared using a quantum slide. Compared to the standard preparation, the quantum slide strongly suppresses oscillatory tails and produces a wave packet with a nearly Gaussian profile. (c) Spatial profiles of the hopping amplitudes $w_j$ and local potential $V_j$ defining the slide. The grey shaded region indicates the slide area.}
    \label{fig:slide}
\end{figure}
We now illustrate how to prepare a nearly ideal Gaussian wave packet in the single-particle regime using a quantum slide. The construction follows the approach of Refs.~\cite{Wang2020Integrated,wang2022quantum}. Both the hopping amplitudes and the local potential are made site dependent, with profiles $w_j=\sqrt{(j+1)(2R-j-4)}$ and $V_j=V_0\left(1-\frac{j}{R}\right)$ (Fig.~\ref{fig:slide}c).

The hopping profile is inspired by perfect-state-transfer protocols \cite{Christandl2004}, in which the hopping Hamiltonian can be identified with the operator $H\propto S_x$ of a fictitious spin $S=(L-1)/2$. In such protocols, the state evolves into an approximately Gaussian distribution centered at the middle of the chain at half of the state-transfer time. The quantum-slide construction adapts this idea by engineering a hopping matrix that resembles one half of the spin Hamiltonian. As shown in Refs.~\cite{Wang2020Integrated,wang2022quantum}, this modification still produces wave packets with an approximately Gaussian spatial profile.

We test this protocol using the parameters of Fig.~\ref{fig:single_x}. To obtain a wave packet with the same variance as the one considered there, the slide must contain $R=200$ sites. The resulting wave packet dynamics, with and without the quantum slide, are shown in Fig.~\ref{fig:slide}a,b. Without the slide, the wave packet displays large oscillatory tails and significant deviations from a Gaussian profile. In contrast, the slide produces a wave packet that remains approximately Gaussian throughout its evolution, undergoing only a gradual broadening as it moves along the chain.

\section{Numerical estimation of the decay rate}
\label{app:decay}
\begin{figure}
    \centering
    \includegraphics[width=\linewidth]{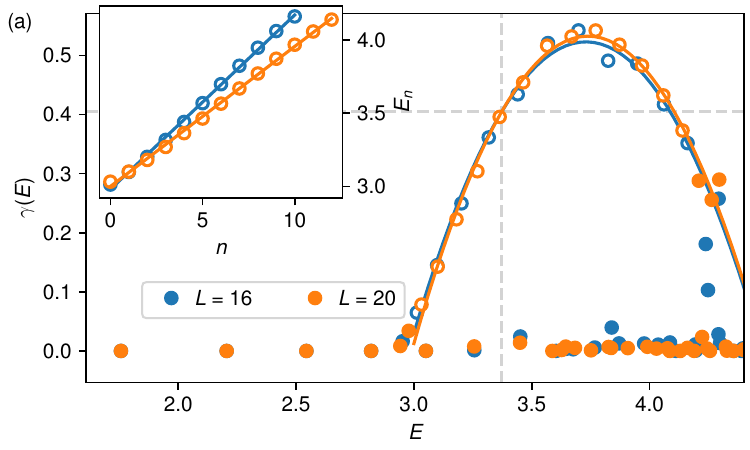}
    \includegraphics[width=\linewidth]{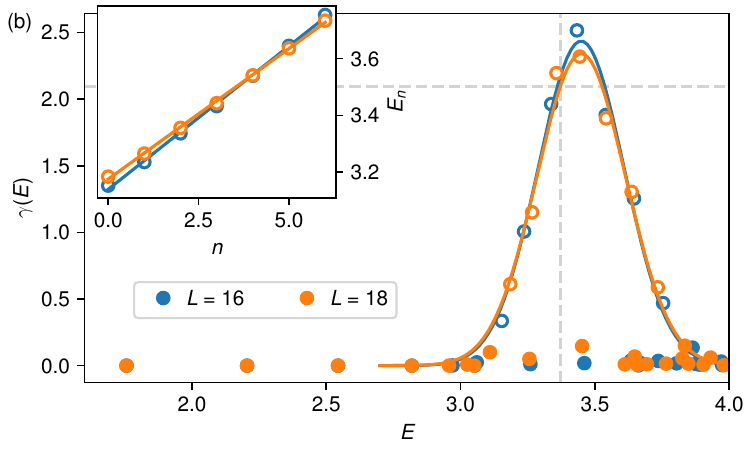}
    \caption{Estimates of the decay factor (a) in the absence of the slide and (b) with a slide of $R=5$ sites. The overlaps $|\langle \epsilon_\downarrow|GS_\uparrow\rangle|^2$ are computed with exact diagonalization for two different system sizes. The values rescaled by the density of states $\gamma=\rho|\langle \epsilon_\downarrow|GS_\uparrow\rangle|^2$ are plotted vs the energy $E=\epsilon_\downarrow-h_0$. A subset of the data marked by white dots is fitted with (a) a parabolic function or (b) a Gaussian function. The vertical and horizontal dashed lines mark the energy $E_0$ and decay factor $\gamma(E_0)$ corresponding to the target particle in the state preparation protocol, as discussed in Sec.~\ref{sec:slide}. The density of states is computed with a linear fit (inset). }
    \label{fig:gamma}
\end{figure}
We now discuss how the decay factor $\gamma(E)$ introduced in Sec.~\ref{sec:slide} is estimated numerically.

Using exact diagonalization, we compute the low-energy spectrum of the Hamiltonian $H_\downarrow = H + J\sigma_1^z$. We then evaluate the overlaps between $\ket{GS_\uparrow}$—the ground state of $H - J\sigma_1^z$—and the eigenstates $\ket{\epsilon_\downarrow}$ of $H_\downarrow$. The low-energy excited states exhibiting large overlap with $\ket{GS_\uparrow}$ (shown as white dots in Fig.~\ref{fig:gamma}) correspond to the single-particle excitations populated during the state-preparation protocol.

The energies of these states, displayed in the insets, are used to estimate the density of states, $\rho=(dE/dn)^{-1}$, through a linear fit of the excitation energies as a function of their index. This procedure assumes that the density of single-particle states is approximately constant within the relevant energy window. The validity of this assumption is supported by the quality of the linear fit. Under this approximation, $\rho$ depends only on the system size $L$.

We then plot the quantity $\rho\,|\langle \epsilon_\downarrow | GS_\uparrow \rangle|^2$ as a function of the excitation energy $E=\epsilon_\downarrow-h_0$ (Fig.~\ref{fig:gamma}). The resulting data are fitted with a parabolic function for $R=0$ and with a Gaussian function for $R=5$. The fitted curves exhibit only weak dependence on system size, suggesting that the extracted estimates are representative of the thermodynamic-limit behavior.

\section{Momentum detection in 2D}
\label{app:2Ddet}

\begin{figure}
    \centering
    \includegraphics[width=\linewidth]{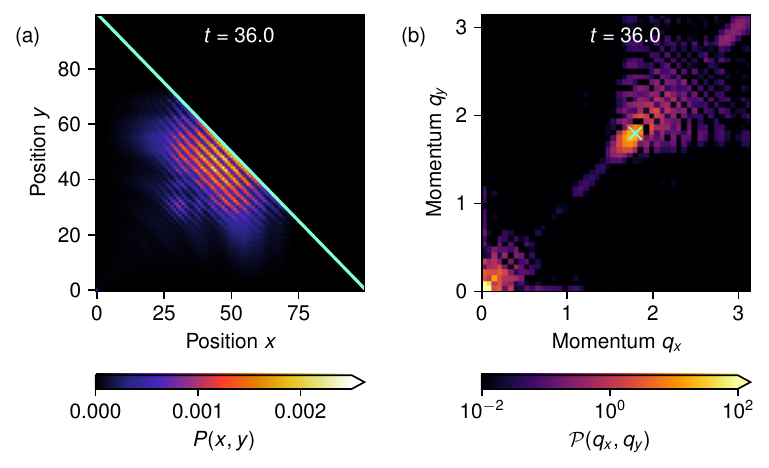}
    \includegraphics[width=\linewidth]{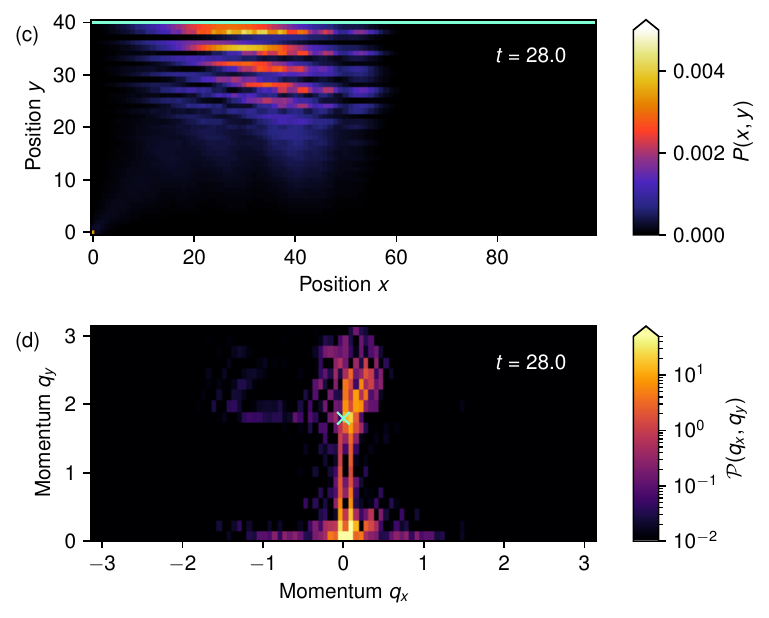}
    \caption{Momentum detection in two dimensions. (a) Probability distribution $P(x,y)$ at time $t=36.0$, showing a wave packet reflected by a diagonal boundary (light-blue line). The interference between the incoming and reflected wave packets produces a characteristic fringe pattern. (b) Fourier transform of $P(x,y)$. A peak at $\vec q \approx 2k_* \hat n$ (light-blue cross), where $\hat n$ is the unit vector normal to the boundary, reveals the momentum component perpendicular to the wall. (c) Probability distribution $P(x,y)$ for a wave packet reflected by a horizontal boundary. (d) Fourier transform of the corresponding probability distribution. The peak at $\vec q \approx 2k_y \hat y$ (light-blue cross) provides an estimate of the vertical momentum component $k_y$. }
    \label{fig:2Ddetect}
\end{figure}

We now present numerical results for the state-detection protocol in two dimensions. As in the one-dimensional case, the method relies on observing the interference between the incoming and reflected wave packet after reflection from a boundary. In two dimensions, however, an additional complication arises: depending on the orientation of the boundary, the interference pattern probes different projections of the momentum vector $\vec{k}$.

We consider a wave packet prepared as described in Fig.~\ref{fig:2D}a of the main text. Figure~\ref{fig:2Ddetect}a shows its reflection from a diagonal wall. Upon reflection, the component of the momentum perpendicular to the wall changes sign, generating a characteristic interference pattern. Taking the Fourier transform of the probability distribution $P(x,y)$ along both spatial directions reveals a peak at $\vec{q}\approx 2k_*\hat n$,where $\hat n$ is the unit vector normal to the wall (Fig.~\ref{fig:2Ddetect}b). More generally, this approach provides a measurement of the momentum component $k_\perp=\vec{k}\cdot \hat n$ perpendicular to the reflecting boundary. In the present example, the wave packet is emitted predominantly along the diagonal direction, so the observed peak provides a good estimate of the momentum magnitude $k_*$.

Other boundary orientations can be used to probe different momentum components. As an illustration, Fig.~\ref{fig:2Ddetect}c shows the interference pattern produced by reflection from a wall normal to the $y$ direction. In this case, the Fourier-space signature is less pronounced, but a peak can still be identified near $\vec q\approx 2(\vec k_*\cdot \hat y) \hat y$ (Fig.~\ref{fig:2Ddetect}d). This behavior is consistent with the fact that the interference pattern is sensitive only to the component of the momentum perpendicular to the reflecting boundary.

\end{document}